\documentclass[
aps,
pra,
reprint,
amsmath,
amssymb,
superscriptaddress,
longbibliography]{revtex4-2}
\usepackage{amssymb}
\usepackage{amsmath}
\usepackage{amsthm}
\usepackage[english]{babel}
\usepackage{bm}
\usepackage{bbold}
\usepackage{hyperref}
\usepackage{graphicx}
\usepackage{subfigure}
\usepackage{xcolor}
\usepackage{booktabs}
\usepackage{blindtext}
\usepackage{xurl}
\newcommand{\ket}[1]{|#1\rangle}
\newcommand{\bra}[1]{\langle #1|}
\newcommand{\inp}[2]{\langle #1|#2\rangle}

\def\<{\langle}  %% overiding the original command \<
\def\>{\rangle}  %% overiding the original command \>

\begin{document}
	\title{Tight tradeoff relation and optimal measurement for multiparameter quantum estimation}
	
	\author{Lingna Wang}
	\email{lnwang@mae.cuhk.edu.hk}
	\affiliation{Department of Mechanical and Automation Engineering, The Chinese University of Hong Kong, Shatin, Hong Kong}
	\affiliation{The Hong Kong Institute of Quantum Information Science and Technology, The Chinese University of Hong Kong, Shatin, Hong Kong SAR, China}
    \affiliation{State Key Laboratory of Quantum Information Technologies and Materials, The Chinese University of Hong Kong, Shatin, Hong Kong SAR, China}
    \author{Hongzhen Chen}
    \email{hzchen@szu.edu.cn}
    \affiliation{Institute of Quantum Precision Measurement, State Key Laboratory of Radio Frequency Heterogeneous Integration, College of Physics and Optoelectronic Engineering, Shenzhen University, Shenzhen, China}
	\author{Haidong Yuan}
	\email{hdyuan@mae.cuhk.edu.hk}
	\affiliation{Department of Mechanical and Automation Engineering, The Chinese University of Hong Kong, Shatin, Hong Kong}
	\affiliation{The Hong Kong Institute of Quantum Information Science and Technology, The Chinese University of Hong Kong, Shatin, Hong Kong SAR, China}
    \affiliation{State Key Laboratory of Quantum Information Technologies and Materials, The Chinese University of Hong Kong, Shatin, Hong Kong SAR, China}
	
	\date{\today}
	
	\begin{abstract}
In multiparameter quantum estimation, the optimal measurements for different parameters encoded in a quantum state are in general incompatible, giving rise to nontrivial tradeoffs between their attainable precisions. 
Understanding and characterizing such tradeoffs is essential for determining the ultimate precision limits in multiparameter quantum estimation and is therefore a central topic in quantum metrology.
In this article, we present an approach that precisely quantifies the tradeoff resulting from incompatible optimal measurements in multiparameter estimation. 
We derive a tight analytical tradeoff relation that determines the ultimate precision limits for estimating an arbitrary number of parameters encoded in pure quantum states. Additionally, we provide a systematic methodology for constructing optimal measurements that saturate this tight bound in an analytical and structured manner. To demonstrate the power of our findings, we apply our methodology to quantum radar, resulting in a refined Arthurs-Kelly relation that characterizes the ultimate performance for the simultaneous estimation of range and velocity.
	\end{abstract}
 \maketitle

\section{Introduction}
Quantum metrology harnesses quantum mechanical phenomena, such as superposition and entanglement, to surpass the precision limits achievable in classical metrology. There is now a good understanding of the local precision limit for single-parameter quantum estimation~\cite{Hole82book,Hels76book,Liu2022,Qiushi2023,Kurdzialek22}, where the precision limit can be quantified by the quantum Cram\'er-Rao bound~\cite{Hels76book}. However, practical applications frequently involve multiple parameters, posing a more complex challenge.  Advancing our understanding of multiparameter quantum estimation is crucial, especially in applications such as quantum sensing, quantum imaging, and quantum communication, where accurate estimation of multiple parameters is required.

An estimation scheme typically consists of three stages: probe-state preparation, the application of coherent controls during the evolution, and a final measurement.
A key feature that distinguishes multiparameter quantum estimation from the single-parameter case is that schemes that are individually optimal for different parameters can be mutually incompatible. 
At each stage, a design tailored to enhance sensitivity for one subset of parameters may be suboptimal or even detrimental for another, giving rise to probe incompatibility, control incompatibility, and measurement incompatibility. 
These incompatibilities lead to nontrivial tradeoffs in the achievable precision for different parameters, and quantifying such tradeoffs has become a central focus in quantum metrology~\cite{MatsumotoThesis,HongzhenPra,HongzhenPRL,GillM00,HayaM05,Zhu2018universally,Lu2021,Suzuki2016,Sidhu2021,Nagaoka1,Nagaoka2,Conlon2021,ALBARELLI2020126311,Federico2021,Carollo_2019,Ragy2016,Chen_2017,Liu_2019,ChenHZ2019,Rafal2020,Kok2020,vidrighin2014,crowley2014,Yue2014,Zhang2014,Liu2017,Roccia_2017,e22111197,Candeloro_2021,Koichi2013,Kahn2009,Yuxiang2019,Hou20minimal,HouSuper2021,Szczykulska2016,FrancescoPRL,Conlon2023,hu2024control,imai2026semiclassical}, as it is essential for identifying the ultimate precision limits in multiparameter estimation tasks. In this work, we focus on measurement incompatibility-the absence of a single measurement that is simultaneously optimal for all parameters. This directly reflects the fundamental restriction imposed by noncommuting observables in quantum mechanics.

When measurements can be applied in a collective manner on infinite copies of quantum states, the Holevo bound serves as a theoretical limit for the precision that can be achieved in multiparameter quantum estimation~\cite{Hole82book,Koichi2013,Kahn2009,Yuxiang2019, MatsumotoThesis, Matsumoto_2002}. However, evaluating the Holevo bound typically requires numerical calculations~\cite{FrancescoPRL, MatsumotoThesis, Matsumoto_2002,Suzuki2016,Sidhu2021}, which limits the insights gained from it. Furthermore, practical measurements can only be performed collectively on a finite number of quantum states, under which the Holevo bound may not be achievable. 
For practical separable measurements, which are performed on each copy of the quantum state separately, Nagaoka introduced a bound for two-parameter estimation that is tighter than the Holevo bound~\cite{Nagaoka1,Nagaoka2}, later generalized by Conlon et al. to the Nagaoka–Hayashi bound for an arbitrary number of parameters~\cite{Conlon2021}. However, these bounds generally require numerical optimization for evaluation and are not tight in most cases. Alternatively, analytical bounds such as the Gill–Massar bound~\cite{GillM00} and the Lu-Wang bound~\cite{Lu2021} provide explicit expressions that quantify the precision tradeoffs arising from the incompatibility of separable measurements in multiparameter quantum estimation. Although these bounds are insightful, they are only tight in specific scenarios. Furthermore, systematic procedures for constructing optimal measurements that achieve these bounds are lacking.

Here, we present an analytical tight tradeoff relation for the estimation of an arbitrary number of parameters encoded in pure quantum states. Additionally, we introduce a systematic approach that enables the analytical construction of optimal separable measurements capable of saturating this relation. The structured approach offers significant insights into the interplay between the estimation of various parameters, making it possible to optimize the performance of various quantum technologies that are related to multiparameter quantum estimation. As a demonstration, we apply the approach to quantum radar and quantify the fundamental limit for simultaneous estimation of the range and velocity. We consider two distinct scenarios: one involving separable photon sources and the other utilizing entangled biphoton sources. In the case of separable photon sources, we provide systematic constructions of the optimal measurement that saturate the Arthurs-Kelly relation~\cite{arthurs1965}. This allows us to achieve the fundamental limit for simultaneously estimating the range and velocity in quantum radar with separable photons. For the case of entangled biphoton sources, we derive a refined Arthurs-Kelly relation that quantifies the ultimate precision for simultaneous range and velocity estimation using biphoton states with any given amount of entanglement. This refined relation precisely quantifies the gain provided by the entanglement in the quantum radar. 

The article is organized as follows: in Sec.~\ref{sec:incompatibility} we introduce the measure of the tradeoff induced by the incompatibility of optimal measurements; in Sec.~\ref{sec:bound} we present the analytical tight tradeoff relation and optimal measurement that saturates the tight bound for the estimation of an arbitrary number of parameters in pure states;
in Sec.~\ref{sec:mix} we extend the bound to mixed states; 
in Sec.~\ref{Sec:example} we illustrate the general procedure with three examples: 1) estimating multiple parameters encoded in squeezed coherent states, 2) simultaneous estimation of the range and velocity with both separable photons and entangled biphoton states in quantum radar, and 3) an implementation of multiparameter estimation on a cloud quantum superconducting platform. This paper is an extended companion to \cite{prl}.

\section{Measurement incompatibility in multiparameter quantum estimation}\label{sec:incompatibility}
In general, to estimate a parameter, $x$, encoded in a quantum state, $\rho_x$, a positive operator-valued measurement (POVM), denoted by $\{M_m\geq 0|\sum_m M_m=I\}$, needs to be performed on the state whereby the probability of obtaining the measurement result $m$ is given by $p(m|x)=\operatorname{Tr}(M_m\rho_x)$. An estimator can then be constructed from the results obtained. In the case of estimating a single parameter, the variance of any locally unbiased estimator is limited by the Cram\'er-Rao bound~\cite{Cram46} as
%\begin{equation}
$\delta\hat{x}^2\geq \frac{1}{\nu F_C},$
%\end{equation}
here $\delta\hat{x}^2=E[(\hat{x}-x)^2]$ is the variance of the locally unbiased estimator, $\nu$ is the number of copies, $F_C=\sum_{m} \frac{(\partial_x p(m|x))^2}{p(m|x)}$ is the Fisher information~\cite{Fish22}. Regardless of the choice of measurement, the variance is always further bounded by the quantum Cram\'er-Rao bound(QCRB)~\cite{Hole82book,Hels76book} as
\begin{equation}
\delta\hat{x}^2\geq \frac{1}{\nu F_C} \geq \frac{1}{\nu F_Q},
\end{equation}
here $F_Q=\operatorname{Tr}(\rho_x L^2)$ is the quantum Fisher information with $L$ as the symmetric logarithmic derivative(SLD) operator which can be obtained implicitly from the equation $\partial_x \rho_x=\frac{1}{2}(\rho_xL+L\rho_x)$. In the instance of estimating a single parameter, it is always possible to find a measurement under which $F_C=F_Q$, which makes the quantum Cram\'er-Rao bound (QCRB) attainable. The projective measurement on the eigenspaces of the SLD is an exemplary method for the optimal measurement that attains the QCRB. Thus, the SLD is an optimal observable for the estimation of the corresponding parameter.

When estimating multiple parameters, where $x=(x_1,\cdots, x_n)$ is a vector, the QCRB generalizes to
\begin{equation}
\operatorname{Cov}(\hat{x})\geq \frac{1}{\nu} F_C^{-1} \geq \frac{1}{\nu} F_Q^{-1},
\end{equation}
here $\operatorname{Cov}(\hat{x})$ is the covariance matrix for locally unbiased estimators, $\hat{x}=(\hat{x}_1,\cdots,\hat{x}_n)$, with $\operatorname{Cov}(\hat{x})_{jk}=E[(\hat{x}_j-x_j)(\hat{x}_k-x_k)]$, $F_C$ is the Fisher information matrix with the $jk$-th entry given by $(F_C)_{jk}=\sum_{m}\frac{\partial_{x_j}p(m|x)\partial_{x_k}p(m|x)}{p(m|x)}$, $F_Q$ is the quantum Fisher information matrix(QFIM) with $(F_Q)_{jk}=\frac{1}{2}\operatorname{Tr}(\rho_x\{L_j,L_k\})$, where $L_{q}$ is the SLD for $x_{q}$ with $\partial_{x_{q}}\rho_x=\frac{1}{2}(\rho_xL_{q}+L_{q}\rho_x)$. In this article, we assume all parameters are locally identifiable, so that the quantum Fisher information matrix is invertible. The study of singular QFIMs~\cite{singularqfim1,singularqfim2,singularqfim3,singularqfim4,singularqfim5} lies beyond the scope of this work. 

In contrast to the single-parameter scenario, the multiparameter QCRB is generally not saturable due to the incompatibility of optimal measurements for different parameters, leading to complex tradeoffs between the precision limits of various parameters. A necessary and sufficient condition for achieving the multiparameter QCRB is the weak commutativity condition: $\operatorname{Tr}(\rho_x[L_j,L_k])=0$, $\forall j, k$. When this condition holds, the optimal measurement that saturates the QCRB for pure states has been identified~\cite{MatsumotoThesis,Luca2017}. However, scenarios where the weak commutativity condition is violated remain far less understood. Elucidating the gap between $F_C$ and $F_Q$ in such general settings, along with constructing the optimal measurement that minimizes this gap, are critical for determining the ultimate precision limits in multiparameter quantum estimation. These challenges currently represent a central focus in multiparameter quantum metrology~\cite{MatsumotoThesis,HongzhenPra,HongzhenPRL,GillM00,HayaM05,Zhu2018universally,Lu2021,Suzuki2016,Sidhu2021,Nagaoka1,Nagaoka2,Conlon2021,ALBARELLI2020126311,Federico2021,Carollo_2019,Ragy2016,Chen_2017,Liu_2019,ChenHZ2019,Rafal2020,Kok2020,vidrighin2014,crowley2014,Yue2014,Zhang2014,Liu2017,Roccia_2017,e22111197,Candeloro_2021,Koichi2013,Kahn2009,Yuxiang2019,Yuan2016,Hou20minimal,HouSuper2021,Ragy2016,Szczykulska2016,FrancescoPRL}.

To assess the disparity between the classical Fisher information matrix and the quantum Fisher information matrix caused by measurement incompatibility, two metrics are commonly employed: $\operatorname{Tr}(F_Q^{-1}F_C)$ and $\operatorname{Tr}(F_QF_C^{-1})$~\cite{GillM00,ALBARELLI2020126311,Federico2021,Carollo_2019,HongzhenPra,HongzhenPRL, MatsumotoThesis,Zhu2018universally}. These metrics are invariant under reparametrization and quantify the similarity between the classical and quantum Fisher information matrices. Since $F_C$ is always less than or equal to $F_Q$ (i.e., $F_C\leq F_Q$), it follows that $\operatorname{Tr}(F_Q^{-1}F_C) \leq n$, where $n$ is the number of parameters. The equality $\operatorname{Tr}(F_Q^{-1}F_C)=n$ holds only when there is no incompatibility, meaning there exists a measurement such that $F_C=F_Q$. The discrepancy between $\operatorname{Tr}(F_Q^{-1}F_C)$ and $n$ quantifies the tradeoff~\cite{HongzhenPra,HongzhenPRL}. Similarly, for the second metric we always have $\operatorname{Tr}(F_QF_C^{-1})\geq n$, and the difference between $\operatorname{Tr}(F_QF_C^{-1})$ and $n$ also quantifies the tradeoff.
The two quantities are related through the Cauchy–Schwarz inequality as $\operatorname{Tr}(F_QF_C^{-1})\geq \frac{n^2}{\operatorname{Tr}(F_Q^{-1}F_C)}$. 
An upper bound of $\operatorname{Tr}(F_Q^{-1}F_C)$ can be immediately converted to a lower bound on $\operatorname{Tr}(F_QF_C^{-1})$ through the Cauchy–Schwarz inequality. However, the converse is not true: a lower bound on $\operatorname{Tr}(F_QF_C^{-1})$ cannot be directly transformed 
into an upper bound on $\operatorname{Tr}(F_Q^{-1}F_C)$~\cite{HongzhenPra}. Moreover, an upper bound on $\operatorname{Tr}(F_Q^{-1}F_C)$ can be converted to a lower bound on $\operatorname{Tr}(G\operatorname{Cov}(\hat{x}))$ for any $G\ge0$ via the inequality \begin{equation}
    \nu \operatorname{Tr}[G\operatorname{Cov}(\hat{x})]\geq \frac{\left(\operatorname{Tr} \sqrt{F_Q^{-\frac{1}{2}} G F_Q^{-\frac{1}{2}}}\right)^2}{\operatorname{Tr}(F_Q^{-1}F_C)},
\end{equation} which can be further used to bound other incompatibility measures defined in~\cite{Federico2021} (see Appendix~\ref{sec:appendix_measures} for discussion).
For these reasons, in this article we shall use $\operatorname{Tr}(F_Q^{-1}F_C)$ as the primary metric.

\section{Tight tradeoff relation and optimal measurement}\label{sec:bound}
Here, we present a tight analytical tradeoff relation for estimating an arbitrary number of parameters encoded in pure quantum states, addressing the general scenario where the weak commutativity condition may not hold. In addition, we provide an analytical construction of the optimal measurement that saturates this tradeoff relation.

We first present the results for the estimation of two parameters encoded in a pure state, then extend it to an arbitrary number of parameters. Note that since $\operatorname{Tr}(F_Q^{-1}F_C)$ is invariant under reparametrization, we can, without loss of generality, assume assume that the QFIM is the identity matrix, $F_Q=I$. This can be achieved by the reparametrization with $\tilde{x}=F_Q^{\frac{1}{2}}x$ under which $\tilde{F}_Q=I$ and $\tilde{F}_C=F_Q^{-\frac{1}{2}}F_CF_Q^{-\frac{1}{2}}$, where the metric $\operatorname{Tr}(\tilde{F}_Q^{-1}\tilde{F}_C)=\operatorname{Tr}(\tilde{F}_C)=\operatorname{Tr}(F_Q^{-1}F_C)$ is invariant.

\subsection{Tight tradeoff relation and optimal measurement for two parameters}
Now consider the estimation of two parameters, $x=(x_1,x_2)$, encoded in a pure state $|\Psi_x\rangle$, with associated symmetric logarithmic derivatives (SLDs) $L_1$ and $L_2$. In general, the SLDs associated with different parameters may not commute. As a result, the optimal measurement cannot be directly obtained from the eigenvectors of the SLDs. One approach to address this challenge is to utilize the measurement uncertainty relation~\cite{arthurs1965,Arthurs1988,Ozawa2003,OZAWA2004367,OZAWA2004350,OZAWA200321,Ozawa_2014,Hall2004,Branciard2013,Branciard2014,2014Error,Lu2014,chen2024simultaneous}. This involves constructing commuting observables that can approximate the SLDs. These approximations inevitably introduce residual errors, a direct manifestation of the tradeoff induced by the incompatibility of the SLDs~\cite{Lu2021}. 

\begin{figure}[t]
  \centering
  \includegraphics[width=0.45\textwidth]{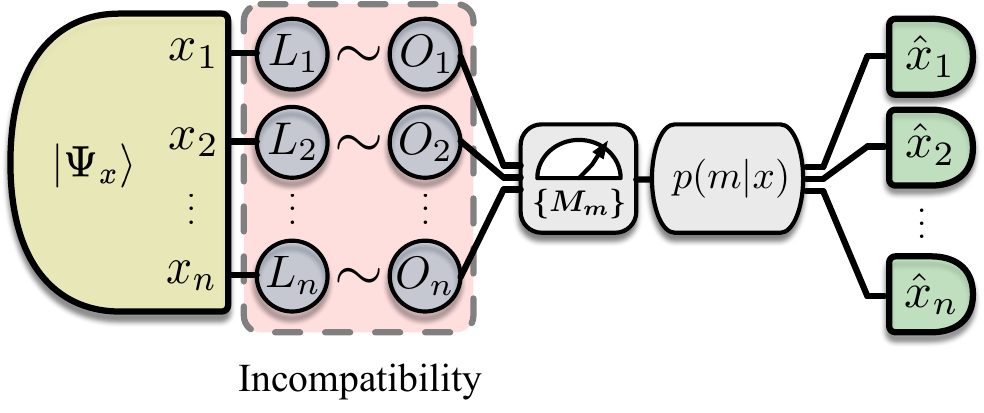}
  \caption{Incompatibility in multiparameter quantum estimation. For a single parameter $x_j$, the optimal measurement is given by the eigenvectors of the SLD operator $L_j$. However, multiple SLDs for different parameters in general cannot be implemented via a single measurement $\{M_m\}$, necessitating approximations of $\{L_j\}$ with another set of observables $\{O_j\}$.}
  \label{MPQE}
\end{figure}

For any positive operator-valued measure(POVM) $\{M_m\}$ acting on $|\Psi_x\rangle$, it is always possible to realize it as a projective measurement $\{|m\rangle\langle m|\}$ on an extended state $|\Psi_x\rangle|\xi\rangle$ with $|\xi\rangle$ as an ancillary state. Based on the projective measurement we construct a set of commuting observables $\{O_j=\sum_m f_j(m)|m\rangle\langle m|\}$ with real coefficients $f_j(m)\in \mathbb{R}$. These commuting observables serve as approximations to the SLDs of the extended state, $\{L_j\otimes I\}$, where $I$ is the Identity operator on the ancilla (see Fig.\ref{MPQE}). The mean squared error of the approximation is given by~\cite{arthurs1965,Arthurs1988,Ozawa2003,OZAWA2004367,OZAWA2004350,OZAWA200321,Ozawa_2014,Hall2004,Branciard2013,Branciard2014,2014Error,Lu2014}
 \begin{eqnarray}\label{eq:error_1}
 \aligned
 \epsilon_j^2&=\bra{\xi}\bra{\Psi_x}(O_j-L_j\otimes I)^2\ket{\Psi_x}\ket{\xi}.
%&=\||o_j\rangle-|l_j\rangle\|_2,\\
 \endaligned
 \end{eqnarray}
Under a given measurement, $\{|m\rangle\langle m|\}$, the probability of obtaining the measurement result, $m$, is given by $p_m(x)=|\langle m|\Psi_x\rangle|\xi\rangle|^2 $. The optimal choice of $f_j(m)$ that minimizes $\epsilon_j^2$ is given by $f_j(m)=\frac{\partial_{x_j}p_m(x)}{p_m(x)}$(see Appendix~\ref{sec:f} for detailed derivation). 
We then define the vectors
\begin{eqnarray}
\aligned
|l_j\rangle&=L_j\otimes I|\Psi_x\rangle|\xi\rangle,\\
%|l_2\rangle=L_2\otimes I|\Psi_x\rangle|\xi\rangle,\\
|o_j\rangle&=O_j|\Psi_x\rangle|\xi\rangle.
%|o_2\rangle=O_2-\langle O_2\rangle|\Psi_x\rangle|\xi\rangle,
\endaligned
\end{eqnarray}
The mean squared error can then be expressed as the Euclidean distance between $|l_j\rangle$ and $|o_j\rangle$ as $\epsilon_j^2=\||o_j\rangle-|l_j\rangle\|^2$, here $\||v\rangle\|^2=\langle v|v\rangle$. With the optimal choice $f_j(m)=\frac{\partial_{x_j}p_m(x)}{p_m(x)}$, we have
%\begin{eqnarray}
%     \aligned
 $         \langle o_j|o_j\rangle%&=\bra{\xi}\bra{\Psi_x}O_j^2\ket{\Psi_x}\ket{\xi}\\
     =\sum_{m} \frac{[\partial_{x_j}p_m(x)]^2}{p_m(x)}=(F_C)_{jj},$
%     &=(F_C)_{jj},
 %    \endaligned
 %     \end{eqnarray}
which corresponds to the diagonal element of the classical Fisher information matrix. %the $j$-th diagonal term of the classical Fisher information matrix.
Similarly, the off-diagonal terms satisfy $\langle o_j|o_k\rangle=\langle o_k|o_j\rangle=\sum_{m} \frac{\partial_{x_j}p_m(x)\partial_{x_k}p_m(x)}{p_m(x)}=(F_C)_{jk}$. Meanwhile, we also have $\operatorname{Re}\langle l_j|l_k\rangle=(F_Q)_{jk}$, $\operatorname{Re}\langle o_j|l_j\rangle=(F_C)_{jj}$, where $\operatorname{Re}$ denotes the real part (see Appendix~\ref{sec:f} for derivation). From these relations it follows that $\epsilon_j^2=(F_Q)_{jj}-(F_C)_{jj}$, showing that the error in approximating the SLDs with commuting observables is precisely the gap between the quantum Fisher information and the classical Fisher information.

We now determine the optimal measurement and the corresponding vectors $\{|o_j\rangle\}$ that minimize the total error
\begin{equation}
\sum_{j=1,2}\epsilon_j^2= \sum_{j=1,2}\||o_j\rangle-|l_j\rangle\|^2=\operatorname{Tr}(F_Q-F_C).    
\end{equation}
Under the parametrization where $F_Q=I$, we have $\operatorname{Tr}(F_Q^{-1}F_C)=\operatorname{Tr}(F_C)$ and 
\begin{equation}\label{eq:mainerrorsum}
\sum_{j=1,2}\epsilon_j^2=\operatorname{Tr}(F_Q-F_C)= 2 -\operatorname{Tr}(F_C).
\end{equation}
A key observation is that, for any measurement, the inner product $\langle o_1|o_2\rangle=\sum_m f_1(m)f_2(m)p_m(x)$ is strictly real. Hence, its imaginary part vanishes: $\operatorname{Im}\langle o_1|o_2\rangle=0$. This provides an intuitive explanation for why the weak commutativity condition is necessary for saturating the quantum Cram\'er-Rao bound. The weak commutativity condition requires $\frac{1}{2i}\langle \xi| \langle \Psi_x|[L_1\otimes I,L_2\otimes I]|\Psi_x\rangle|\xi\rangle=\operatorname{Im} \langle l_1|l_2\rangle=0$. If this condition is violated ($\operatorname{Im} \langle l_1|l_2\rangle\neq 0$), $\{|o_1\rangle, |o_2\rangle\}$ can not perfectly match $\{|l_1\rangle, |l_2\rangle\}$, and the root-mean-squared errors $\epsilon_j$ can not all vanish. Consequently, $F_C$ can not equal $F_Q$ under any measurement.

As $(F_Q)_{jk}=\operatorname{Re}\langle l_j|l_k\rangle$, when $F_Q=I$, the following hold: $\langle l_1|l_1\rangle=\langle l_2|l_2\rangle=1$, $\operatorname{Re}\langle l_1|l_2\rangle=0$. Thus $\langle l_1|l_2\rangle$ is purely imaginary and can be expressed as $\langle l_1|l_2\rangle=i\beta$,  $\beta\in\mathbb{R}$. Applying the Cauchy-Schwarz inequality $|\langle l_1|l_2\rangle|^2\leq \langle l_1|l_1\rangle\langle l_2|l_2\rangle$ yields $|\beta|\leq 1$. In Appendix~\ref{sec:opt_oj}, we show that for given $|l_1\rangle$ and $|l_2\rangle$ with $\langle l_1|l_2\rangle=i\beta$, the total approximation error is bounded below by $\epsilon_1^2+\epsilon_2^2\geq 1-\sqrt{1-\beta^2}$.
Combining this with Eq.~(\ref{eq:mainerrorsum}), we then have \begin{equation}\label{eq:upperbound2}
    \operatorname{Tr}(F_C)\leq 1+\sqrt{1-\beta^2}.
\end{equation}
This bound is tight, and we will now provide the optimal $\{|o_j\rangle\}$ that saturates the bound (see detailed derivation in Appendix~\ref{sec:opt_oj}).

When $|\beta|< 1$, the optimal $\{|o_j\rangle\}$ are 
\begin{eqnarray}
\aligned
|o_{1}\rangle&=a|l_{1}\rangle-ib|l_{2}\rangle,\\
|o_{2}\rangle&=ib|l_{1}\rangle+a|l_{2}\rangle,
\endaligned
\label{eq:optimalo}
\end{eqnarray}
here $a=\frac{1+\cos\phi}{2\cos\phi}$, $b=-\frac{\sin\phi}{2\cos\phi}$, $\phi=\arcsin \beta$, $\phi\in (-\frac{\pi}{2},\frac{\pi}{2})$. In this case, the classical Fisher information matrix, which can be obtained with $(F_C)_{jk}=\langle o_j|o_k\rangle$, is $F_C=\left(\begin{array}{cc}
\frac{1+\sqrt{1-\beta^2}}{2} & 0\\
0 & \frac{1+\sqrt{1-\beta^2}}{2}
\end{array}\right)=\frac{1+\sqrt{1-\beta^2}}{2}I$.

When $|\beta|=1$, we have $|\langle l_1|l_2\rangle|=1$, $|l_1\rangle$ and $|l_2\rangle$ are linearly dependent. In this case, we can arbitrarily choose a $\ket{l_{\perp}}$ which satisfies $\operatorname{Im} \inp{l_{\perp}}{l_1} = 0$, $\inp{l_{\perp}}{\Psi_x\rangle|\xi} = 0$ and $\inp{l_{\perp}}{l_{\perp}} = 1$. The optimal $\{|o_j\rangle\}$ can then be constructed as
\begin{equation}
  \begin{aligned}
    \ket{o_1} &= \frac{1}{2} (1-\sin 2\varphi) |l_1\rangle + \frac{i}{2} \beta \cos 2\varphi \ket{l_{\perp}}, \\
    \ket{o_2} &= \frac{i}{2}\beta(1+\sin 2\varphi) |l_1\rangle + \frac{1}{2}\cos 2\varphi \ket{l_{\perp}},
  \end{aligned}
  \label{mainbeta_1}
\end{equation}
here $\varphi$ can take any real value. In this case,
\begin{equation}\label{eq:FCbeta1}
F_C = \left(\begin{matrix}
    \frac{1}{2}(1-\sin 2\varphi) & \frac{1}{2} \cos 2\varphi \operatorname{Re} \inp{l_{\perp}}{l_1} \\
    \frac{1}{2} \cos 2\varphi \operatorname{Re} \inp{l_{\perp}}{l_1} & \frac{1}{2}(1+\sin 2\varphi)
\end{matrix}\right).
\end{equation}
The bound is saturated for any $\ket{l_{\perp}}$ and $\varphi$. If we choose a $\ket{l_{\perp}}$ that is orthogonal to $\ket{l_1}$ (which is always possible—for instance, by taking $\ket{l_{\perp}}= |\Phi\rangle|\xi_{\perp}\rangle$ where $|\xi_{\perp}\rangle$ is orthogonal to $|\xi\rangle$ and $|\Phi\rangle$ is an arbitrary state), then $F_C$ becomes diagonal. Further setting $\varphi=0$ yields $F_C=\frac{1}{2}I$. Notably, when $|\beta|=1$, this can equivalently be expressed as $\frac{1+\sqrt{1-\beta^2}}{2}I$. Thus for all $\beta$, there exists a classical Fisher information matrix as $F_C=\frac{1+\sqrt{1-\beta^2}}{2}I$ that saturates the bound in Eq.~(\ref{eq:upperbound2}).

We now construct the optimal measurement, $\{|m\rangle\langle m|\}$, that saturates the bound. First note that in the measurement basis, $\{O_j=\sum_m f_j(m)|m\rangle\langle m|\}$ are diagonal with real diagonal entries. We can write the state in the measurement basis as $|\Psi_x\rangle|\xi\rangle=\sum_m c_m|m\rangle$. If $c_m\notin \mathbb{R}$, we can write $c_m=r_me^{i\phi_m}$ with $r_m\in \mathbb{R}$ and let $|\tilde{m}\rangle=e^{i\phi_m}|m\rangle$, then $|\Psi_x\rangle|\xi\rangle=\sum_m r_m|\tilde{m}\rangle$ where $\{\ket{\tilde{m}}\bra{\tilde{m}}\}$ are the same projective measurement as $\{|m\rangle\langle m|\}$. Thus without loss of generality, we assume $c_m\in \mathbb{R}$, 
then
\begin{equation}
|o_j\rangle=\sum_m f_j(m)|m\rangle\langle m|\Psi_x\rangle|\xi\rangle=\sum_m f_j(m)c_m|m\rangle
\end{equation}
with $f_j(m)c_m\in \mathbb{R}$. $\{|o_j\rangle\}$ are thus real vectors in the measurement basis.

To get the measurement basis, $\{|m\rangle\langle m|\}$, under which $\{|\Psi_x\rangle|\xi\rangle, |o_1\rangle, |o_2\rangle\}$ are all real, we first perform the Gram-Schmidt orthonormalization on $\{|\Psi_x\rangle|\xi\rangle, |o_1\rangle, |o_2\rangle\}$ and let
 \begin{eqnarray}
 \aligned
|a_0\rangle&=|\Psi_x\rangle|\xi\rangle,\\
 |a_1\rangle&=\frac{|o_1\rangle}{\sqrt{\langle o_1|o_1\rangle}},\\
 |a_2\rangle&=\frac{|o_2\rangle-\langle a_1|o_2\rangle |a_1\rangle}{\sqrt{\langle o_2|o_2\rangle-|\langle a_1|o_2\rangle|^2}},
 \endaligned
 \end{eqnarray}
here the fact that $|o_1\rangle$ and $|o_2\rangle$ are orthogonal to $|\Psi_x\rangle|\xi\rangle$ has been used. These vectors can be expanded into a complete basis by adding additional orthonormal vectors $\{|a_j\rangle|3\leq j\leq d-1\}$, where $d$ is the total dimension. Note that if $\{|\Psi_x\rangle|\xi\rangle, |o_1\rangle,|o_2\rangle\}$ have real entries in a basis, the entries of $\{|a_0\rangle, |a_1\rangle, |a_2\rangle\}$ are also real in that basis. 

To find a basis in which $\{|a_j\rangle\}$ are real, we require a unitary transformation that maps $\{|a_j\rangle\}$ to a set of real orthonormal vectors $\{|b_j\rangle\}$. Let the columns of matrix  $A$ be the vectors $\{|a_j\rangle\}$ and let the columns of a real orthogonal matrix $B$ be the vectors $\{|b_j\rangle\}$. The unitary that performs the change of basis is then $U=BA^{-1}$. The measurement basis can then be taken as the rows of $U$, i.e., if the $m$-th row of $U$ is $\langle m|$, $|b_j\rangle$ is precisely the representation of $|a_j\rangle$ in the basis of $\{|m\rangle\}$. This follows because $B=UA$, so that $|b_j\rangle =U|a_j\rangle$, the $m$-th entry of $|b_j\rangle$ is then exactly $\langle m|a_j\rangle$, $0\leq j\leq d-1$. 

The matrix $B$ can be taken as any real orthogonal matrix whose first column, i.e., $|b_0\rangle$, corresponds to the representation of $|\Psi_x\rangle|\xi\rangle$ in the basis of $\{|m\rangle\}$, has no zero entries. As in this case, we can always choose proper $f_j(m)$ to get the optimal $|o_j\rangle=\sum_m f_j(m)|m\rangle\langle m|\Psi_x\rangle|\xi\rangle$, specifically we can take $f_j(m)=\frac{\langle m|o_j\rangle}{\langle m|\Psi_x\rangle|\xi\rangle}$ when $\langle m|\Psi_x\rangle|\xi\rangle\neq 0$. 
Therefore, for any real orthogonal matrix, $B$, whose first column has no zero entries, the rows of $U=BA^{-1}$ form the basis for the optimal projective measurement that saturates the tight tradeoff relation. In special cases, the requirement on $|b_0\rangle$ may be relaxed: if the $m$-th entry of both $|o_1\rangle$ and $|o_2\rangle$ in the measurement basis are zero, then $\langle m|\Psi_x\rangle|\xi\rangle$ may also be zero because the relation $|o_j\rangle=\sum_m f_j(m)|m\rangle\langle m|\Psi_x\rangle|\xi\rangle$ can still hold with the $m$-th entry of both sides vanishing.

\subsection{Tight tradeoff relation and optimal measurement for an arbitrary number of parameters}
For an arbitrary number of parameters, $x=(x_1,\cdots, x_n)$, encoded in a pure state $|\Psi_x\rangle$, we denote the corresponding SLDs as $\{L_j\}$. For any projective measurement $\{|m\rangle\langle m|\}$ on the system+ancilla, we can similarly define $O_j=\sum_m f_j(m)|m\rangle \langle m|$ and
\begin{eqnarray}
    |l_j\rangle=L_j\otimes I |\Psi_x\rangle |\xi\rangle,\\
    |o_j\rangle=O_j |\Psi_x\rangle |\xi\rangle,
\end{eqnarray}
where $j\in\{1,\cdots, n\}$ and $|\xi\rangle$ is an  ancillary state. The mean squared error between $O_j$ and $L_j\otimes I$ is then
\begin{eqnarray}\label{eq:error_2}
 \aligned
 \epsilon_j^2&=\bra{\xi}\bra{\Psi_x}(O_j-L_j\otimes I)^2\ket{\Psi_x}\ket{\xi}\\
 &=\||o_j\rangle-|l_j\rangle\|^2.
 %&=(F_Q)_{jj}-(F_C)_{jj}.
  \endaligned
 \end{eqnarray}
As in the two-parameter case, it can be similarly shown that under any given measurement, with the optimal choice of $\{f_j(m)\}$ we have $\epsilon_j^2=(F_Q)_{jj}-(F_C)_{jj}$.

We then define a matrix $F$ whose $jk$-th entry is $F_{jk}=\langle l_j|l_k\rangle$, which is the geometric tensor of $|\Psi_x\rangle$ ~\cite{bengtsson2017geometry,provost1980riemannian,carollo2020geometry}. The real part of $F$ is the QFIM, and the imaginary part, denoted as $F_{\operatorname{Im}}$, is proportional to the Berry curvature~\cite{Berry1984}. %so $F=F_Q+iF_{\operatorname{Im}}$. 
Since $\operatorname{Tr}(F_Q^{-1}F_C)$ is invariant under reparametrization, we can
choose a parametrization under which $F_Q=I$ and $F_{\operatorname{Im}}$ takes the canonical block diagonal form 
\begin{eqnarray}\label{eq:block_main}
F_{\operatorname{Im}}=\left[\begin{array}{ccccccccccc}
0 & \beta_1 &  0 &  \ldots & &   & & & \\
-\beta_1 & 0 &  & & & & & & \\
 0 & \cdots & 0 & \beta_2 & & & & & & \\
& & -\beta_2 & 0 & & & & & \\
 \vdots &  & & & \ddots & \vdots &  & & & \\
 0 & &\cdots & 0 &  \cdots & 0 & \beta_r & & & \\
& & & & &  -\beta_r & 0 & & & \\
& & & & & & & 0 & & \\
& & & & & & & & \ddots & \\
& & & & & & & & & 0
\end{array}\right].
\end{eqnarray}
If the original parametrization does not exhibit this structure, we first make a reparametrization with $\tilde{x}=F_Q^{-\frac{1}{2}}x$, under which $\tilde{F}=I+i\tilde{F}_{\operatorname{Im}}$ with $\tilde{F}_{\operatorname{Im}}=F_Q^{-\frac{1}{2}}F_{\operatorname{Im}}F_Q^{-\frac{1}{2}}$. Because $\tilde{F}_{\operatorname{Im}}$ is anti-symmetric, there exists an orthogonal matrix $P$ such that $P\tilde{F}_{\operatorname{Im}}P^T$ takes the block diagonal form given in Eq.~(\ref{eq:block_main}). A second  reparametrization $\breve{x}=P\tilde{x}$ then gives $\breve{F}_Q=PIP^T=I$ and $\breve{F}_{\operatorname{Im}}=P\tilde{F}_{\operatorname{Im}}P^T$, which has the desired canonical structure. Note that the eigenvalues of $\breve{F}_{\operatorname{Im}}$ are the same as those of $\tilde{F}_{\operatorname{Im}}=F_Q^{-\frac{1}{2}}F_{\operatorname{Im}}F_Q^{-\frac{1}{2}}$. 

Hence, without loss of generality we assume $F_Q=I$ and $F_{\operatorname{Im}}$ as given in Eq.~(\ref{eq:block_main}). Under this parametrization, $\operatorname{Tr}(F_Q^{-1}F_C)=\operatorname{Tr}(F_C)$ and $\sum_j\epsilon_j^2=n-\operatorname{Tr}(F_C)$. 
For each pair $1\leq j\leq r$ we have
\begin{equation}
\epsilon_{2j-1}^2+\epsilon_{2j}^2\geq %\frac{\beta_j^2}{1+\sqrt{1-\beta_j^2}}=
1-\sqrt{1-\beta_j^2},
\end{equation}
where the equality can be saturated by the optimal choices of $\{|o_{2j-1}\rangle,|o_{2j}\rangle\}$, which are in the subspace spanned by $|l_{2j-1}\rangle$ and $|l_{2j}\rangle$(if $|\beta_j|=1$ we replace $|l_{2j}\rangle$ with a $|l_{j\perp}\rangle$ that is orthogonal to all $\{|l_j\rangle\}$).  For indices $k> 2r$ we have $\epsilon_{k}^2\geq 0$, and the equality can be attained by choosing $|o_k\rangle=|l_k\rangle$. Consequently $\sum_{q=1}^n \epsilon_q^2\geq \sum_{j=1}^r (1-\sqrt{1-\beta_j^2})$, which implies
\begin{eqnarray}
\aligned
\operatorname{Tr}(F_C)&\leq n-\sum_{j=1}^r (1-\sqrt{1-\beta_j^2}).
%&=\sum_{j=1}^r (1+\sqrt{1-\beta_j^2})+n-2r\\
%&=n-\sum_{j=1}^r(1-\sqrt{1-\beta_j^2})\\
%&=n-\frac{1}{2}\sum_{q=1}^n(1-\sqrt{1-|\lambda_q|^2}).
\endaligned
\end{eqnarray}
The bound can be expressed compactly using the eigenvalues of $F_{\operatorname{Im}}$, $\{\lambda_1,\cdots, \lambda_n\}=\{\pm i\beta_1,\cdots,\pm i\beta_r, 0,\cdots, 0\}$, as \begin{eqnarray}
\aligned
\operatorname{Tr}(F_C)&\leq %n-\sum_{j=1}^r (1-\sqrt{1-\beta_j^2})\\
%&=\sum_{j=1}^r (1+\sqrt{1-\beta_j^2})+n-2r\\
%&=n-\sum_{j=1}^r(1-\sqrt{1-\beta_j^2})\\
&n-\frac{1}{2}\sum_{q=1}^n(1-\sqrt{1-|\lambda_q|^2}).
\endaligned
\end{eqnarray}

For arbitrary parametrization with general $F=F_Q+iF_{\operatorname{Im}}$, the bound can be written as
\begin{eqnarray}\label{eq:mainbound}
\aligned
\operatorname{Tr}(F_Q^{-1}F_C)&\leq %\sum_{j=1}^r (1+\sqrt{1-\beta_j^2})+n-2r\\
%&=n-\sum_{j=1}^r(1-\sqrt{1-\beta_j^2})\\
%&=
n-\frac{1}{2}\sum_{q=1}^n(1-\sqrt{1-|\lambda_q|^2}),
\endaligned
\end{eqnarray}
where $\{\lambda_q\}$ are eigenvalues of $F_Q^{-\frac{1}{2}}F_{\operatorname{Im}}F_Q^{-\frac{1}{2}}$.

We note that different $|l_j\rangle$ that correspond to different blocks are orthogonal to each other, the obtained set of $\{|o_j\rangle\}$ satisfies $\langle o_j|o_k\rangle\in \mathbb{R}$, $\forall j, k\in \{1,\cdots, n\}$. The optimal measurement can then be constructed from $\{|\Psi_x\rangle|\xi\rangle, |o_1\rangle,\cdots, |o_n\rangle\}$ in a similar way. We first perform Gram-Schmidt orthonormalization on these vectors and obtain a set of orthonormal vectors $\{|a_j\rangle\}$ with $0\leq j\leq n$, here $|a_0\rangle=|\Psi_x\rangle|\xi\rangle$. These vectors are then extended into a complete basis of the system–ancilla Hilbert space by adding additional orthonormal vectors $\{|a_{n+1}\rangle,\cdots, |a_{d-1}\rangle\}$, with $d$ the total dimension. 

For any measurement basis, $\{|m\rangle\langle m|\}$, we may assume without loss of generality that $|\Psi_x\rangle|\xi\rangle$ is a real vector in the measurement basis. Consequently, the vectors $\{|o_j\rangle=\sum_m f_j(m)|m\rangle\langle m|\Psi_x\rangle|\xi\rangle\}$ are also real in this basis. Let $A$ be the matrix whose columns are $\{|a_j\rangle\}$ and $B$ be any real orthogonal matrix whose first column has no zero entries, then the rows of $U=BA^{-1}$ form the basis for the optimal projective measurement that saturates the bound.

In Appendix~\ref{sec:appendixverify},  we explicitly verify the optimality of our constructed measurement by computing $F_C$ from the measurement outcomes and confirming that it saturates the derived tradeoff relation. In Appendix~\ref{sec:appendixPre}, we provide detailed comparisons between our bound and existing bounds~\cite{HongzhenPra,HongzhenPRL,GillM00,MatsumotoThesis}.

A notable special case occurs in the estimation of $2d-2$ parameters encoded in a $d$-dimensional pure state. In this case the eigenvalues of $F_Q^{-\frac{1}{2}}F_{\operatorname{Im}}F_Q^{-\frac{1}{2}}$ always satisfy $|\lambda_j|=1$, $\forall j=1,\cdots, 2d-2$ (see Appendix~\ref{sec:appendixPre} for detail). Our general tradeoff relation then reduces to $\text{Tr}(F_Q^{-1}F_C)\leq 2d-2-\frac{1}{2}(2d-2)=d-1$, which coincides with the Gill-Massar bound that is known to be tight for this special case~\cite{GillM00}. However, in general, the Gill-Massar bound is not saturable.

\section{Tradeoff relation for mixed states and collective measurement}\label{sec:mix}
The tradeoff relation can be extended to mixed states via purification. For a mixed state, $\rho_x=\sum_j \lambda_j |\Psi_j\rangle\langle \Psi_j|$, we consider a purification, $|\Psi_x\rangle=\sum_j \sqrt{\lambda_j}|j_E\rangle|\Psi_j\rangle$, where $\{|j_E\rangle\}$ are orthonormal vectors in the Hilbert space, $H_E$, and $\operatorname{Tr}_E(|\Psi_x\rangle\langle\Psi_x|)=\rho_x$. A general positive operator-valued measure(POVM), $\{M_m\}$, on the state $\rho_x$ can always be realized as a projective measurement, $\{|m\rangle\langle m|\}$, on an extended state $\rho_x\otimes |\xi\rangle\langle \xi|$ with $|\xi\rangle$ as an ancillary state in the Hilbert space, $H_A$. Commuting observables, $O_j=\sum_m f_j(m)|m\rangle\langle m|$ with $f_j(m)\in \mathbb{R}$, can then be constructed from the projective measurement to approximate the SLDs of the extended state, $L_j\otimes I_A$, here $I_A$ denotes the Identity operator on the ancilla. The mean squared error of the approximation is given by~\cite{arthurs1965,Arthurs1988,Ozawa2003,OZAWA2004367,OZAWA2004350,OZAWA200321,Ozawa_2014,Hall2004,Branciard2013,Branciard2014,2014Error,Lu2014}
 \begin{eqnarray}
 \aligned
 \epsilon_j^2&=\operatorname{Tr}[(O_j-L_j\otimes I_A)^2\rho_x\otimes\ket{\xi}\bra{\xi}].
%&=\||o_j\rangle-|l_j\rangle\|_2,\\
 \endaligned
 \end{eqnarray}
With the purified state, this can be written as
\begin{eqnarray}
 \aligned
 \epsilon_j^2&=\bra{\xi}\bra{\Psi_x}(I_E\otimes O_j-I_E\otimes L_j\otimes I_A)^2|\Psi_x\rangle\ket{\xi}.
%&=\||o_j\rangle-|l_j\rangle\|_2,\\
 \endaligned
 \end{eqnarray}
Again we first consider two parameters where $x=(x_1,x_2)$ and choose a parameterization under which $F_Q=I$. Let $|l_j\rangle=I_E\otimes L_j\otimes I_A |\Psi_x\rangle|\xi\rangle$, we can then get $\epsilon_1^2+\epsilon_2^2\geq 1-\sqrt{1-\beta^2}$ where $\beta=\text{Im}\langle l_1|l_2\rangle=\text{Im} \operatorname{Tr}(\rho_xL_1L_2)$. From which we can similarly get $\operatorname{Tr}(F_C)\leq 1+\sqrt{1-\beta^2}$, which can be written as $\operatorname{Tr}(F_Q^{-1}F_C)\leq 1+\sqrt{1-\beta^2}$ under an arbitrary parametrization. For an arbitrary finite number of parameters, we can similarly obtain  
\begin{eqnarray}\label{eq:mixed}
\aligned
\operatorname{Tr}(F_Q^{-1}F_C)&\leq %\sum_{j=1}^r (1+\sqrt{1-\beta_j^2})+n-2r\\
%&=n-\sum_{j=1}^r(1-\sqrt{1-\beta_j^2})\\
%&=
n-\frac{1}{2}\sum_{q=1}^n(1-\sqrt{1-|\lambda_q|^2}),
\endaligned
\end{eqnarray}
where $\{\lambda_q\}$ are eigenvalues of $F_Q^{-\frac{1}{2}}F_{\operatorname{Im}}F_Q^{-\frac{1}{2}}$, here $F_{\operatorname{Im}}$ is the imaginary part of $F=F_Q+iF_{\operatorname{Im}}$ with $F_{jk}=\operatorname{Tr}(\rho_xL_jL_k)$.
For mixed states, this tradeoff relation is no longer tight since the measurement can only be performed in the restricted space $H_S\otimes H_A$, not the whole space, $H_E\otimes H_S\otimes H_A$. The bound can be further tightened by combining the techniques in previous studies~\cite{2014Error, HongzhenPRL,HongzhenPra}.

The bound also holds for collective measurements performed on $k$ copies of the state, $\rho_x^{\otimes k}$. This can be seen easily by treating $\rho_x^{\otimes k}$ as a single state, whose quantum geometric tensor satisfies
$F^{(k)}=F_Q^{(k)}+iF_{\operatorname{Im}}^{(k)}=k(F_Q+iF_{\operatorname{Im}})$, so that $F_Q^{(k)}=kF_Q$ and $F_{\operatorname{Im}}^{(k)}=kF_{\operatorname{Im}}$. Consequently, $(F_Q^{(k)})^{-\frac{1}{2}} F_{\operatorname{Im}}^{(k)}(F_Q^{(k)})^{-\frac{1}{2}}=F_Q^{-\frac{1}{2}} F_{\operatorname{Im}} F_Q^{-\frac{1}{2}}$. Since the bound depends only on the eigenvalues of this matrix, it remains unchanged, yielding \begin{equation}\label{eq:mainboundN}
\operatorname{Tr}\left(\left(F_Q^{(k)}\right)^{-1}F_C^{(k)}\right)\leq n-\frac{1}{2} \sum_{q=1}^n\left(1-\sqrt{1-\left|\lambda_q\right|^2}\right),  
\end{equation}
where $\{\lambda_q\}$ are eigenvalues of $(F_Q^{(k)})^{-\frac{1}{2}} F_{\operatorname{Im}}^{(k)}(F_Q^{(k)})^{-\frac{1}{2}}=F_Q^{-\frac{1}{2}} F_{\operatorname{Im}}F_Q^{-\frac{1}{2}}$. 

This can be used to obtain a lower bound on $\operatorname{Tr}\left(G \operatorname{Cov}(\hat{x})\right)$ with collective measurements on $k$ copies of state as (see derivation in Appendix~\ref{sec:appendix_measures})
\begin{align}
  \nu\operatorname{Tr}(G\operatorname{Cov}(\hat{x})) %&\geq \operatorname{Tr}\left(G\left(F_C^{(k)}\right)^{-1}\right)\\
&\geq \frac{\left(\operatorname{Tr} \sqrt{F_Q^{-\frac{1}{2}} G F_Q^{-\frac{1}{2}}}\right)^2}{ n-\frac{1}{2} \sum_{q=1}^n\left(1-\sqrt{1-\left|\lambda_q\right|^2}\right)}.  
\end{align}

\section{Examples}\label{Sec:example}
\subsection{Estimating multiple parameters in squeezed coherent states}\label{sec:squeezed}
We illustrate the general procedure with an example involving the simultaneous estimation of multiple parameters encoded in a continuous-variable Gaussian state.

We consider a squeezed coherent state, 
%\begin{equation}
 $ |\eta, r, 0 \rangle = D(\eta) S(r) \ket{0},$
%\end{equation}
where $\ket{0}$ is the vacuum state, $D(\eta)$ is the displacement operator,
\begin{equation}
  D(\eta) = \exp\left(\eta a^{\dagger} - \eta^{*} a \right),
\end{equation}
and $S(r)$ is the squeezing operator
\begin{equation}
  S(r) = \exp\left[\frac{r}{2}(a^2 - a^{\dagger 2})\right],
\end{equation}
with $r$ the squeezing parameter. Here the parameters to be estimated are $\eta$ and $r$. Since $\eta$ is complex, we have three real parameters,
$x_1 = \operatorname{Re}\eta$, $x_2 = \operatorname{Im}\eta$, and $x_3 = r$.

The SLDs are given by
%\end{equation}
\begin{equation}
  \begin{aligned}
    &L_1 = 2 e^{x_3} \ket{\eta, r, 1} \bra{\eta, r, 0} +  2 e^{x_3} \ket{\eta, r, 0} \bra{\eta, r, 1},\\
    &L_2 = 2 i e^{-x_3} \ket{\eta, r, 1} \bra{\eta, r, 0} -  2 i  e^{- x_3} \ket{\eta, r, 0} \bra{\eta, r, 1},\\
    &L_3 = -\sqrt{2} \ket{\eta, r, 2} \bra{\eta, r, 0} -\sqrt{2} \ket{\eta, r, 0} \bra{\eta, r, 2},
  \end{aligned}
\end{equation}
where $\ket{\eta, r, k}=D(\eta)S(r)\ket{k}$.
The geometric tensor $F=F_Q+iF_{\operatorname{Im}}$ can then be obtained with 
\begin{equation}
  F_Q = \left(\begin{matrix}
    4 e^{2 x_3} & 0 & 0 \\ 0 & 4 e^{-2 x_3} & 0 \\ 0 & 0& 2
  \end{matrix}\right), \quad F_{\operatorname{Im}} = \left(\begin{matrix}
  0 & 4   & 0 \\ -4 & 0 & 0 \\ 0 & 0& 0
  \end{matrix}\right).
\end{equation}
Since $F_Q\neq I$, we perform a reparametrization $\left(\begin{matrix}
    x_1^{\prime}\\ x_2^{\prime}\\ x_3^{\prime}
  \end{matrix}\right) = F_Q^{\frac{1}{2}}  \left(\begin{matrix}
    x_1\\ x_2\\ x_3
  \end{matrix}\right)$ under which $\tilde{F}_Q = I$. The SLDs for the new parameters are
  \begin{equation}
      L_{1}^{\prime} =  \frac{L_1}{2 e^{x_3}}, \quad
      L_{2}^{\prime} = \frac{L_2}{2 e^{-x_3}}, \quad
      L_{3}^{\prime} = \frac{L_3}{\sqrt{2}}.
  \end{equation}
The transformed imaginary part becomes
  \begin{equation}
    \tilde{F}_{\operatorname{Im}} =F_Q^{-\frac{1}{2}}F_{\operatorname{Im}}F_Q^{-\frac{1}{2}}= \left(\begin{matrix}
       0 & 1 & 0 \\ -1 & 0 & 0 \\ 0 & 0 & 0
    \end{matrix}\right),
  \end{equation}
whose eigenvalues are $\pm i$ and $0$.
The tradeoff relation is then
  \begin{equation}
     \operatorname{Tr}(F_Q^{-1} F_C) \leq \frac{1}{2} \sum_j(1+\sqrt{1-|\lambda_j|^2})= 2.
  \end{equation}
To construct the optimal measurement, we let
  \begin{equation}
    \begin{aligned}
      &|l_{1}\rangle = L_{1}^{\prime} \ket{\eta, r, 0}  =  \ket{\eta, r, 1}, \\
      &|l_{2}\rangle = L_{2}^{\prime} \ket{\eta, r, 0} = i \ket{\eta, r, 1}, \\
      &|l_{3}\rangle = L_{3}^{\prime} \ket{\eta, r, 0}  = - \ket{\eta, r, 2}, \\
      &|l_\perp \rangle = \ket{\eta, r, 3},
    \end{aligned}
  \end{equation}
where $|l_\perp \rangle$ satisfies $\inp{\eta, r, 0}{l_\perp} = \inp{l_1}{l_\perp} = \inp{l_2}{l_\perp} =0 $ and $\inp{l_\perp}{l_\perp} = 1$.
Note that in this case no ancillary system is required, and we can restrict attention to the subspace spanned by $\{\ket{\eta,r,0},\ket{\eta,r,1}, \ket{\eta,r,2}, \ket{\eta,r,3}\}$ for the construction of the optimal measurement, since projections outside this subspace do not contribute to the Fisher information.

The optimal $\ket{o_1}$ and $\ket{o_2}$ can be obtained from Eq.~(\ref{mainbeta_1}) with $\varphi=0$ as
  \begin{equation}
    \begin{aligned}
      &\ket{o_1}  = \frac{1}{2} \ket{l_1} + \frac{i}{2} \ket{l_\perp} = \frac{1}{2} \ket{\eta, r, 1} + \frac{i}{2}  \ket{\eta, r, 3}, \\
      &\ket{o_2}  = \frac{i}{2} \ket{l_1} + \frac{1}{2} \ket{l_\perp} = \frac{i}{2} \ket{\eta, r, 1} + \frac{1}{2}  \ket{\eta, r, 3}, \\
    \end{aligned}
  \end{equation}
   while $\ket{o_3}$ can be taken as
      $\ket{o_3}  = \ket{l_3} =- \ket{\eta, r, 2}$.

Through Gram–Schmidt we then construct:
  \begin{equation}
    \begin{aligned}
      &\ket{a_0} = \ket{\eta, r, 0},  \\
      &\ket{a_1} = \frac{\sqrt{2}}{2} \ket{\eta, r, 1} + \frac{i\sqrt{2}}{2}  \ket{\eta, r, 3},\\
      &\ket{a_2} = \frac{i\sqrt{2}}{2} \ket{\eta, r, 1} + \frac{\sqrt{2}}{2}  \ket{\eta, r, 3}, \\
      &\ket{a_3} =- \ket{\eta, r, 2},
    \end{aligned}
  \end{equation}
 which form an orthonormal basis for the four-dimensional subspace spanned by $\{\ket{\eta, r, 0}, \ket{\eta, r, 1}, \ket{\eta, r, 2}, \ket{\eta, r, 3}\}$. Within this subspace, we collect the vectors $\{\ket{a_j}\}$ into the columns of a unitary matrix
  \begin{equation}
    A = \left(\begin{matrix}
      1 & 0 & 0 & 0 \\
      0 & \frac{\sqrt{2}}{2} & \frac{i\sqrt{2}}{2} & 0 \\
      0 & 0 & 0 & -1 \\
      0 & \frac{i\sqrt{2}}{2} & \frac{\sqrt{2}}{2} & 0
    \end{matrix}\right),
  \end{equation}
and choose a real orthogonal matrix
  \begin{equation}
    B = \left(\begin{matrix}
    \frac{1}{2} &\frac{1}{2} & \frac{1}{2} & \frac{1}{2}\\
    \frac{1}{2} & -\frac{1}{2}  & \frac{1}{2} & -\frac{1}{2}\\
    \frac{1}{2} &  \frac{1}{2} & -\frac{1}{2} & -\frac{1}{2} \\
    \frac{1}{2} & -\frac{1}{2} & -\frac{1}{2} & \frac{1}{2}
    \end{matrix}\right).
  \end{equation}
The optimal measurement can then be taken as the projective measurement in the basis given by the rows of $U=B A^{-1}$ (see Appendix~\ref{sec:example} for the explicit form of this basis). Similar finite–dimensional projective measurements on continuous–variable modes have been demonstrated in superconducting platforms~\cite{wang2019heisenberg,wang2022quantum,ni2025autonomous}, confirming the feasibility of implementing such measurements.

It can be explicitly verified that the classical Fisher information matrix under this measurement is 
  \begin{equation}
    \begin{aligned}
      F_C = \left(\begin{matrix}
        2 e^{2 x_3} & 0 & 0 \\
        0 & 2 e^{-2 x_3} & 0 \\
        0 & 0 & 2
      \end{matrix}\right),
    \end{aligned}
  \end{equation}
which saturates the tradeoff relation,
  \begin{equation}
    \operatorname{Tr}(F_Q^{-1} F_C) = 2.
  \end{equation}

\subsection{Simultaneous estimation of range and velocity}\label{sec:radar}
%\section{The model}
We now apply our framework to quantum radar, addressing the fundamental challenge of simultaneously estimating the range and velocity of a moving target. While this problem has been extensively investigated~\cite{zhuang2017,zhuang2021,zhuang2022,huang2021,quantumradar2020,Maccone2020,reichert2022quantum,reichert2024heisenberg,li2023entanglement}, the fundamental precision limits have only been established in two extreme scenarios: (i) using separable photons and (ii) assuming perfectly entangled biphoton states~\cite{zhuang2017,huang2021}. In the former case, the fundamental limit is quantified by the Arthurs-Kelly relation~\cite{arthurs1965}. The latter case is unphysical, as real-world entanglement is always finite. The fundamental limits for the general case with any amount of entanglement have remained unknown. This gap stems precisely from the incompatibility between range and velocity estimation, which our method directly addresses. We now derive a refined tradeoff relation for the simultaneous estimation of the range and velocity with any amount of entanglement. This relation provides a clear quantitative link between the attainable precision and the degree of entanglement, thereby bridging the gap between the known extremes and offering a fundamental benchmark for quantum radar. 

\begin{figure}[htbp]
\centering  %图片全局居中
\subfigure[]{
%\label{Fig.sub.1}
\includegraphics[width=0.35\textwidth]{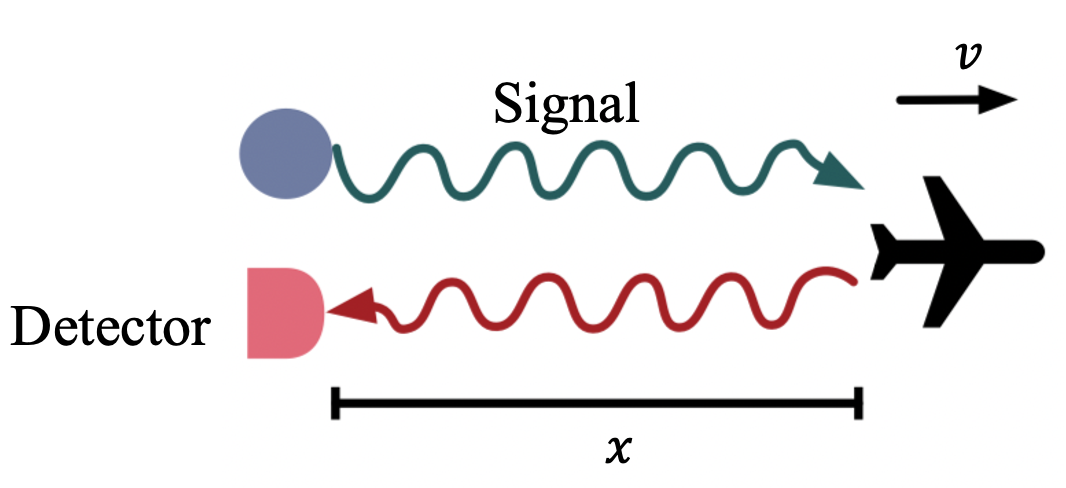}}
\subfigure[]{
%\label{Fig.sub.2}
\includegraphics[width=0.35\textwidth]{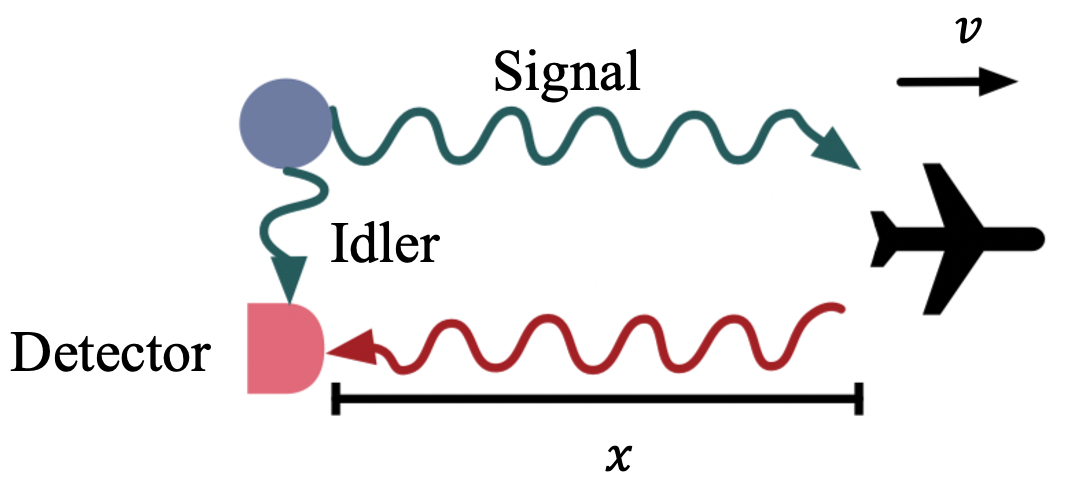}}
\caption{radar sensing of target's range and velocity, using (a) single-photon state and (b) entangled two-photon state.}\label{fig:radar}
\end{figure} 

\subsubsection{Model}
The estimation of range and velocity in the pulsed quantum radar is achieved by first sending a pulsed light to the target, then detecting the light reflected back from the target. The range can be estimated from the time of flight and the velocity can be estimated from the Doppler shift of the frequency of the reflected light. Following~\cite{zhuang2017,huang2021}, we use Gaussian pulses
and adopt an ideal single-target model in the paraxial regime, where the target acts as a perfect mirror with negligible background noise, as illustrated in Fig.\ref{fig:radar}(a) for the single-photon sensing scheme.”

A single photon state in the frequency domain can be described as 
\begin{equation}
  \ket{\psi_0} = \int d\omega \ \tilde{\psi}_0 (\omega) \ket{\omega},
\end{equation}
where
\begin{equation}
  \tilde{\psi}_0 (\omega) = \left(\frac{1}{2 \pi \sigma_0^2}\right)^{1/4} \exp\left\{-\frac{(\omega-\bar{\omega}_0)^2}{4\sigma_0^2} + i \omega \bar{t}_0 \right\},
\end{equation}
here $\bar{\omega}_0$ is the central frequency, $\bar{t}_0$ is the central time, and $\sigma_0$ is the frequency bandwidth. $|\omega\rangle= a_\omega^{\dagger}|\hat{0}\rangle$, where $a_{\omega}^{\dagger}$ is the creation operator of the mode with the frequency $\omega$.
The state can also be represented in the time domain as
\begin{equation}\label{eq:singlephotontime}
  \ket{\psi_0} = \int dt \ \psi_0(t) \ket{t},
\end{equation}
where
\begin{equation}
  \psi_0(t) = \left(\frac{2\sigma_0^2}{\pi}\right)^{1/4} \exp\left\{-(t-\bar{t}_0)^2 \sigma_0^2 - i \bar{\omega}_0 (t-\bar{t}_0)\right\}
\end{equation}
here $|t\rangle=a_t^{\dagger}|\hat{0}\rangle$, where $a_t^{\dagger}$ is the creation operator of the photon at time $t$.

Now assume at $\bar{t}_0$ a target is at a distance $x$ from the radar station and moving away with radial velocity $v$, and a signal photon in the state $\ket{t}$ is sent towards the target at time $t$. As at time $t$ the target is $x+v(t-\bar{t}_0)$ away from the radar station, the signal will reach the target after $\Delta t = \frac{ x+v(t-\bar{t}_0)}{c-v}$, and get back to the radar station at the time
\begin{equation}
  \tau(t) = t + 2 \Delta t = t + \frac{2(x+v(t-\bar{t}_0))}{c-v}
  \label{returntime}
\end{equation}
From the linearity, the returned light of a general single photon state in Eq.~(\ref{eq:singlephotontime}) can be described as
\begin{equation}
  \ket{\psi} = \int dt \ \psi_0(t) \ket{\tau(t)} = \int d\tau \ \psi(\tau) \ket{\tau},
\end{equation}
here
\begin{equation}
  \psi(\tau) = \left(\frac{2\sigma^2}{\pi}\right)^{1/4} \exp\left\{-(\tau-\bar{t})^2 \sigma^2 - i \bar{\omega} (\tau-\bar{t})\right\}
\end{equation}
with 
\begin{equation}
  \sigma = \frac{c-v}{c+v} \sigma_0, \ \bar{t} = \bar{t}_0 + \frac{2x}{c-v}, \ \bar{\omega} = \frac{c-v}{c+v} \bar{\omega}_0.
  \label{btf}
\end{equation}
   
In the classical case, pulses with multiple separable photons are sent and detected. In the quantum case, entangled biphoton states can be used, as shown in Fig. \ref{fig:radar}(b), where one photon acts as the signal which is sent to the target, and the other photon is kept at the station as the reference. In the frequency domain, the two-photon state, which can be generated via the spontaneous parametric down-conversion, can be described as
\begin{equation}
  \ket{\Psi_0} = \int d\omega \int d\omega_i \tilde{\Psi}_0(\omega,\omega_i) \ket{\omega} \ket{\omega_i}
\end{equation}
where
\begin{widetext}
\begin{equation}
 \tilde{\Psi}_0(\omega,\omega_i) = \tilde{\mathcal{N}}_0 e^{i (\omega +\omega_i)\bar{t}_0}\exp\left\{-\frac{1}{2(1-\kappa^2)}\left[\frac{(\omega-\bar{\omega}_0)^2}{2\sigma_0^2}+\frac{(\omega_i-\bar{\omega}_{i0})^2}{2\sigma_{i0}^2}+\frac{\kappa(\omega-\bar{\omega}_0) (\omega_i-\bar{\omega}_{i0})}{\sigma_0\sigma_{i0}}\right]\right\}.
 \end{equation}
\end{widetext}
Here $\tilde{\mathcal{N}}_0 = \frac{1}{\sqrt{2\pi \sigma_0 \sigma_{i0}}(1-\kappa^2)^{1/4}}$ is the normalization factor,
$\kappa \in [0,1)$ quantifies the correlation between the signal and the idler photon.
When $\kappa=0$, the state is separable.  
In the limit $\kappa \to 1$, the state approaches a perfectly entangled state, which, however, is not physical and can not be realized in practice.

The biphoton state can also be written in the time domain as
\begin{equation}
  \ket{\Psi_0} = \int dt \int dt_i \Psi_0(t,t_i) \ket{t} \ket{t_i},
\end{equation}
with
\begin{widetext}
\begin{equation}
  \Psi_0(t,t_i) = \mathcal{N}_0 \exp\{-i \bar{\omega}_0(t-\bar{t}_0)-i\bar{\omega}_{i0}(t_i-\bar{t}_0) -(t-\bar{t}_0)^2 \sigma_0^2 - (t_i-\bar{t}_0)^2 \sigma_{i0}^2 + 2 \kappa (t-\bar{t}_0)(t_i-\bar{t}_0)\sigma_0\sigma_{i0}\},
\end{equation}
\end{widetext}
here $\mathcal{N}_0 = \sqrt{\frac{2\sigma_0 \sigma_{i0}}{\pi}}(1-\kappa^2)^{1/4}$.
%The signal photon is sent toward the target, while the idler photon does not interact with the target.

Similarly, if the signal photon in state $\ket{t}$ is back-scattered by the target, it will return at the time given in Eq.~(\ref{returntime}).
Hence, the returned biphoton state is given by
\begin{equation}
  \ket{\Psi} = \int dt \int dt_i \Psi(t,t_i) \ket{t} \ket{t_i}
\end{equation}
where
%\begin{widetext}
\begin{eqnarray}\label{eq:bientangle}
  \aligned
  \Psi(t,t_i) = &\mathcal{N} \exp\{-i \bar{\omega}(t-\bar{t})-i\bar{\omega}_{i0}(t_i-\bar{t}_0) -(t-\bar{t})^2 \sigma^2 \\
  &- (t_i-\bar{t}_0)^2 \sigma_{i0}^2 + 2 \kappa (t-\bar{t})(t_i-\bar{t}_0)\sigma\sigma_{i0}\},
\endaligned
\end{eqnarray}
%\end{widetext}
here the bandwidth, $\sigma$, the central time, $\bar{t}$, and the frequency, $\bar{\omega}$ are the same as in Eq.~(\ref{btf}), and the normalization factor is given by $\mathcal{N} = \sqrt{\frac{2\sigma \sigma_{i0}}{\pi}}(1-\kappa^2)^{1/4}$.

\subsubsection{Precision limits with separable photons}
We first study the precision limit with separable photons. 
As shown in Eq.~(\ref{btf}), the range and velocity are encoded in the central time $\bar{t}$, frequency $\bar{\omega}$, and bandwidth $\sigma$ of the returned signal photons. The range $x$ and velocity $v$ can be obtained from the central time $\bar{t}$ and the frequency $\bar{\omega}$ in the state of returned photon~\cite{huang2021}, we will characterize the precision limit for the simultaneous estimation of $(\bar{t},\bar{\omega})$ in the following.  For single photon state 
\begin{equation}
  \ket{\psi} = \int dt \ \psi(t) \ket{t},
\end{equation}
with
\begin{equation}
  \psi(t) = \left(\frac{2\sigma^2}{\pi}\right)^{1/4} \exp\left\{-(t-\bar{t})^2 \sigma^2 - i \bar{\omega} (t-\bar{t})\right\}.
\end{equation}
In this case, the SLDs, $L_i = 2\left(\ket{\partial_{x_i} \psi} \bra{\psi} + \ket{\psi} \bra{\partial_{x_i} \psi}\right)$, with $x_1=\bar{t}$ and $x_2=\bar{\omega}$, can be obtained as
\begin{equation}
  \begin{aligned}
    &L_{\bar{t}} = 2\sigma \ket{e_1}\bra{e_2} + 2\sigma \ket{e_2}\bra{e_1},\\
    &L_{\bar{\omega}} = \frac{i}{\sigma}\ket{e_1}\bra{e_2}-\frac{i}{\sigma}\ket{e_2}\bra{e_1},
  \end{aligned}
  \label{eq:sld_sep}
\end{equation}
here $\ket{e_j} = \int dt \ e_j(t) \ket{t}$, for $j = 1, 2$ with
\begin{equation}
  e_1(t) = \psi(t), \quad e_2(t) = 2\sigma (t-\bar{t})\psi(t).
\end{equation}
We can get $F_Q$ and $F_{\operatorname{Im}}$ as
\begin{equation}
  F_Q = \left(\begin{matrix}
    4 \sigma^2 & 0 \\ 0 & \frac{1}{\sigma^2}
  \end{matrix}\right), \quad F_{\operatorname{Im}} = \left(\begin{matrix}
    0 & -2 \\ 2 & 0
  \end{matrix}\right).
\end{equation}
Since $F_{\operatorname{Im}} \neq \boldsymbol{0}$, the incompatibility exists and the quantum Cram\'er-Rao bound is not achievable. If we directly apply the quantum Cram\'er-Rao bound, we will have $\sigma_{\bar{t}}\sigma_{\bar{\omega}}\geq\frac{1}{2\sigma}\sigma=\frac{1}{2}$, where $\sigma_{\bar{t}}$ and $\sigma_{\bar{\omega}}$ denote the standard deviation of the estimators. However, this violates the Arthurs-Kelly relation~\cite{arthurs1965}, $\sigma_{\bar{t}}\sigma_{\bar{\omega}}\geq 1$, which imposes a lower bound on the precision for the simultaneous estimation of $\bar{t}$ and $\bar{\omega}$ with separable photons. In this case, a key question is whether the Arthurs-Kelly uncertainty relation is saturable and what optimal measurement achieves the minimal tradeoff in precision. This can be tackled with our approach. 

We begin by computing the matrix $F_Q^{-1/2}F_{\operatorname{Im}}F_Q^{-1/2}$, which yields
\begin{equation}
    F_Q^{-1/2}F_{\operatorname{Im}}F_Q^{-1/2} = \left(\begin{matrix}
    0 & -1 \\ 1 & 0
  \end{matrix}\right),
\end{equation}
whose eigenvalues are $\{\pm i\}$. From the tradeoff relation in Eq.~(\ref{eq:mainbound}), we get
\begin{equation}
    \operatorname{Tr}(F_Q^{-1}F_C) \leq 1 + \sqrt{1-1} = 1,
\end{equation}
which can be rewritten as $\frac{1}{4\sigma^2}(F_C)_{11}+\sigma^2(F_C)_{22}\leq 1$. Since $\sigma_{\bar{t}}^2\geq \frac{1}{(F_C)_{11}}$ and $\sigma_{\bar{\omega}}^2\geq \frac{1}{(F_C)_{22}}$, we then have 
\begin{equation}
    \frac{1}{4\sigma^2 \sigma_{\bar{t}}^2} + \frac{\sigma^2}{\sigma_{\bar{\omega}}^2} \leq\frac{1}{4\sigma^2}(F_C)_{11}+\sigma^2(F_C)_{22}\leq 1.
    \label{inq_1_sp}
\end{equation}
In Appendix~\ref{sec:opt_separable}, we provide explicit constructions of optimal measurements that saturate the bound.

Since $1\geq \frac{1}{4\sigma^2 \sigma_{\bar{t}}^2} + \frac{\sigma^2}{\sigma_{\bar{\omega}}^2}\geq \frac{1}{\sigma_{\bar{t}}\sigma_{\bar{\omega}}}$, the bound can recover the Arthurs-Kelly relation, but not vice versa. For example, $\sigma_{\bar{t}}^2=\frac{1}{4\sigma^2}$ and $\sigma_{\bar{\omega}}^2=4\sigma^2$ satisfy the Arthurs-Kelly relation but violate the bound. The obtained bound is thus strictly stronger.

\subsubsection{Precision limits with entangled photons}
With an entangled biphoton state, the returned state is given by
\begin{equation}
  \ket{\Psi} = \int dt \int dt_i \Psi(t,t_i) \ket{t} \ket{t_i},
\end{equation}
where $\Psi(t,t_i)$ is given in Eq(\ref{eq:bientangle}).
In this case, the SLDs, $L_i = 2\left(\ket{\partial_{x_i} \Psi} \bra{\Psi} + \ket{\Psi} \bra{\partial_{x_i} \Psi}\right)$, can given by
\begin{widetext}
\begin{equation}
  \begin{aligned}
    &L_{\bar{t}} = \sigma \sqrt{2(1-\kappa)} \ket{e_1}\bra{e_2} + \sigma \sqrt{2(1-\kappa)}  \ket{e_2}\bra{e_1}+ \sigma \sqrt{2(1+\kappa)} \ket{e_1}\bra{e_3} + \sigma \sqrt{2(1+\kappa)}  \ket{e_3}\bra{e_1},\\
    &L_{\bar{\omega}} = \frac{i\sqrt{2}}{2\sigma\sqrt{1-\kappa}} \ket{e_1}\bra{e_2} -\frac{i\sqrt{2}}{2\sigma\sqrt{1-\kappa}} \ket{e_2}\bra{e_1}+\frac{i\sqrt{2}}{2\sigma\sqrt{1+\kappa}} \ket{e_1}\bra{e_3} -\frac{i\sqrt{2}}{2\sigma\sqrt{1+\kappa}}  \ket{e_3}\bra{e_1},
  \end{aligned}
  \label{eq:sld_ent}
\end{equation}
\end{widetext}
where $\ket{e_j} = \int dt \ \int dt_i \ e_j(t, t_i) \ket{t} \ket{t_i}$, $j = 1, 2,3$, are orthonormal with
\begin{equation}
  \begin{aligned}
    &e_1(t,t_i) = \Psi(t,t_i),\\
    &e_2(t,t_i) = \sqrt{2(1-\kappa)}\left(\sigma (t-\bar{t}) + \sigma_i (t_i-\bar{t}_0)\right) \Psi(t,t_i),\\
    &e_3(t,t_i) = \sqrt{2(1+\kappa)}\left(\sigma (t-\bar{t}) - \sigma_i (t_i-\bar{t}_0)\right) \Psi(t,t_i).
  \end{aligned}
\end{equation}
From the SLD operators, we directly obtain $F_Q$ and $F_{\operatorname{Im}}$ as
\begin{equation}
  F_Q = \left(\begin{matrix}
    4 \sigma^2 & 0 \\ 0 & \frac{1}{\sigma^2(1-\kappa^2)}
  \end{matrix}\right), F_{\operatorname{Im}} = \left(\begin{matrix}
    0 & -2 \\ 2 & 0
  \end{matrix}\right).
\end{equation}
If we make a reparametrization with
\begin{equation}
  \left(\begin{matrix}
    \bar{t}^{\prime} \\ \bar{\omega}^{\prime}
  \end{matrix}\right)= F_Q^{1/2} \left(\begin{matrix}
    \bar{t} \\ \bar{\omega}
  \end{matrix}\right),
\end{equation}
under which $\tilde{F}_Q = I$, and 
\begin{equation}
    \tilde{F}_{\operatorname{Im}}  = F_Q^{-\frac{1}{2}}F_{\operatorname{Im}} F_Q^{-\frac{1}{2}}=\left(\begin{matrix}
    0 & -\sqrt{1-\kappa^2}\\ \sqrt{1-\kappa^2} & 0
  \end{matrix}\right).
\end{equation}

From the quantum Cram\'er-Rao bound, previous studies~\cite{huang2021} have obtained a relation as
\begin{equation}\label{eq:relation2}
    \sigma_{\bar{t}} \sigma_{\bar{\omega}} \geq \frac{\sqrt{1-\kappa^2}}{2}.
\end{equation}
This relation is tight in the limit  $\kappa \to 1$ as $\tilde{F}_{\operatorname{Im}}=\boldsymbol{0}$ and the quantum Cram\'er-Rao bound is saturable in this case. But for the practical case where $0 \leq\kappa<1$, this relation is not achievable—indeed, at $\kappa=0$ it violates the Arthurs-Kelly relation. A central challenge in quantum radar is therefore to identify the fundamental bound for arbitrary entanglement and the corresponding optimal measurement. This is exactly what our methodology provides.

By applying the tradeoff relation in Eq.~(\ref{eq:mainbound}), we get
\begin{equation}
    \operatorname{Tr}(F_Q^{-1}F_C) \leq 1 + \sqrt{1-(1-\kappa^2)} = 1+\kappa, 
\end{equation}
which can be equivalently written as $\frac{1}{4\sigma^2}(F_C)_{11}+\sigma^2 (1-\kappa^2)(F_C)_{22}\leq 1+\kappa$. Since $\sigma_{\bar{t}}^2\geq \frac{1}{(F_C)_{11}}$ and $\sigma_{\bar{\omega}}^2\geq \frac{1}{(F_C)_{22}}$, we then obtain the tradeoff relation 
\begin{equation}\label{inq_1_et}
   \frac{1}{4\sigma^2 \sigma_{\bar{t}}^2} + \frac{\sigma^2 (1-\kappa^2)}{\sigma_{\bar{\omega}}^2} \leq 1+\kappa.
    \end{equation}
Since $\frac{\sqrt{1-\kappa^2}}{\sigma_{\bar{t}}\sigma_{\bar{\omega}}}\leq \frac{1}{4\sigma^2 \sigma_{\bar{t}}^2} + \frac{\sigma^2 (1-\kappa^2)}{\sigma_{\bar{\omega}}^2}$, we can also obtain a multiplicative form as
\begin{equation}
\sigma_{\bar{t}}\sigma_{\bar{\omega}} \geq \frac{\sqrt{1-\kappa}}{\sqrt{1+\kappa}}.
\end{equation}
This represents a refined Arthurs-Kelly relation applicable for general $\kappa$. It reduces to the classical Arthurs-Kelly relation when $\kappa=0$, and aligns with the relation in Eq.~({\ref{eq:relation2}) at $\kappa=1$. Thus, it unifies previous extreme‑case results and characterizes the ultimate performance of quantum radar with arbitrary entangled biphoton states. 
% With this refined Arthurs-Kelly relation, we can further confirm that the heuristic measurement proposed previously by Zhuang et.al~\cite{zhuang2017} is indeed optimal as it saturates the refined relation. Previously, this measurement was only known to be optimal for $\kappa=1$, where the QCRB is saturable. In Appendix~\ref{sec:opt_entangled}, we demonstrate the constructions of alternative optimal measurements.
Using this refined relation, we can rigorously confirm that the heuristic continuous‑spectrum measurement proposed by Zhuang et al.~\cite{zhuang2017,huang2021} is indeed optimal, as it saturates the refined relation. Previously, this measurement was only known to be optimal in the perfectly entangled limit ($\kappa=1$) where the QCRB is saturable. In Appendix~\ref{sec:opt_entangled}, we show how this measurement fits into our framework and further show that continuous outcomes are not essential: all extractable information can be obtained via finite‑outcome projective measurements on a chosen subspace, which provides a route to optimality in platforms where continuous-spectrum detection is challenging. The optimal measurements constructed here are parameter-dependent, and their practical implementation therefore requires adaptive updates when the target parameters are not known in advance.

We note that the characterization of the fundamental limit is based on the ideal scenario without any noise. However, the analysis can be extended to cases where photon loss is the dominant source of noise; in such settings, performance can be quantified by scaling the bound with the loss rate. For more general noise models, the bound for the mixed state can be employed, which is valid but is generally not tight, serving as a loose benchmark.

\subsection{Multiparameter estimation on a superconducting quantum processor}\label{sec:exp}
In this section, we apply our general framework to a cloud-based superconducting quantum processor. We consider two multiparameter estimation tasks implemented on this platform, which provide an independent demonstration of the theory and complement the theoretical examples presented in Secs.~\ref{sec:squeezed} and \ref{sec:radar}.
The cloud superconducting processor, namely ``ScQ-P136''(``Baiwang''), is accessed through the \textit{Quafu} cloud quantum computing platform whose single-qubit gate fidelity exceeds 99\%~\cite{Quafu1,Quafu2,Quafu3}.
The layout and additional details of the processor are provided in the Appendix~\ref{sec:exp_theo_circuit}.

\begin{figure}[htbp]
  \centering
\includegraphics[width=0.5\textwidth]{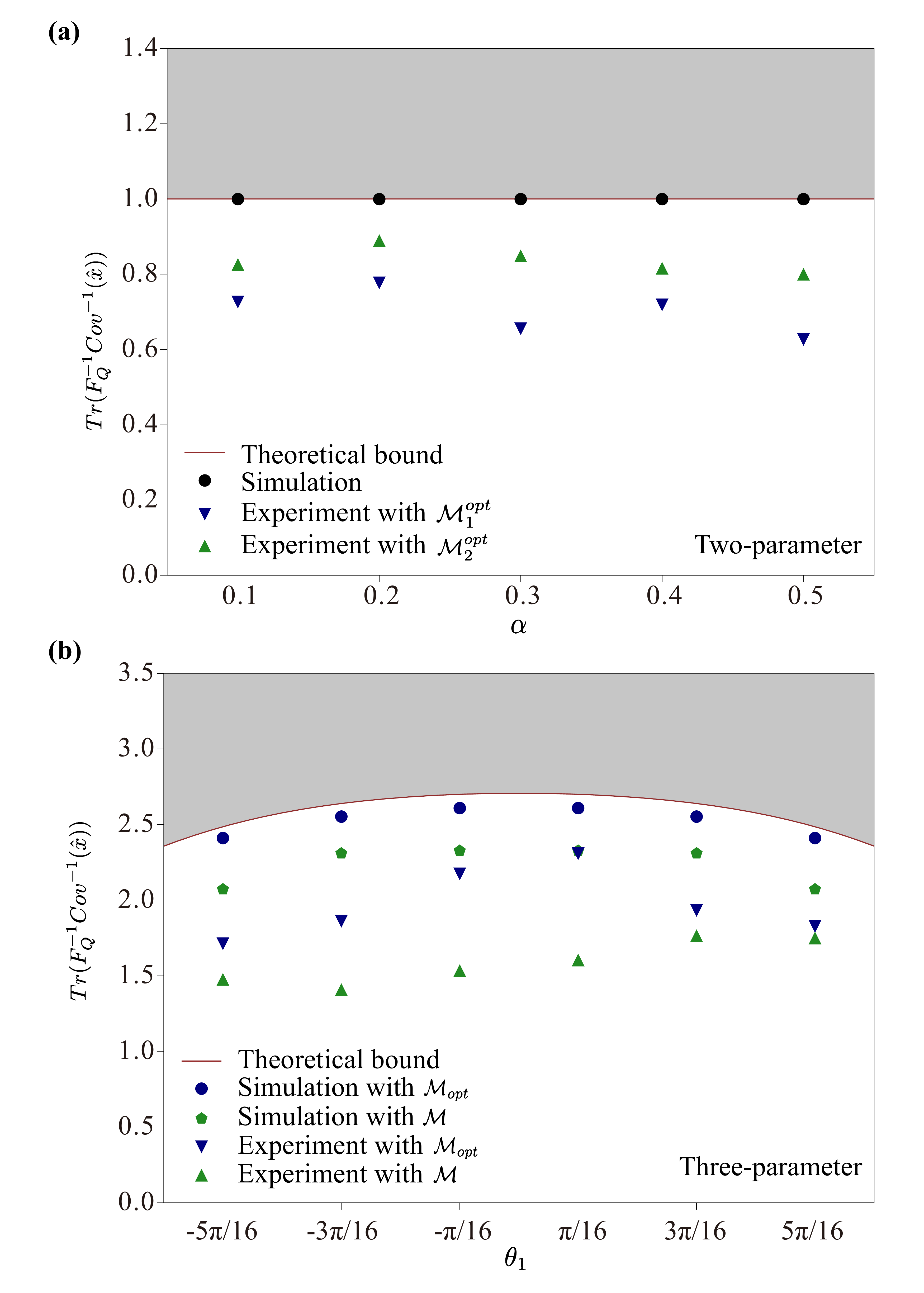}
  \caption{Approaching fundamental tradeoff relation on quantum computing platforms.
  (a) Tradeoff relation for estimating two rotational parameters encoded in $|\psi (\alpha, \theta)\rangle$, with $\theta=\frac{\pi}{6}$. (b) Tradeoff relation for estimating three parameters encoded in $|\psi (\alpha_1, \theta_1, \gamma_1)\rangle$, with $\alpha_1 =\frac {\pi}{8}$, $\gamma_1= \frac {\pi}{4}$. In both demonstrations, each data point is obtained from $N = 400$ independent runs, each based on $N_{\mathrm{shots}} = 2000$ measurement shots per measurement setting.}
  \label{experiment}
\end{figure}

The first demonstration aims to estimate two rotational parameters, $(\alpha,\theta)$, encoded in a qubit state given by 
\begin{equation}
    |\psi (\alpha, \theta)\rangle = R_z(-\alpha) R_y(-\theta) \sigma_x|0\rangle,
\end{equation}
where $R_i(\theta) = e^{- \frac{i}{2} \theta \sigma_i}$ are the rotation operators with $i\in\{x,y,z\}$, and $\sigma_{x,y,z}$ denote the Pauli X, Y, and Z gates, respectively. The corresponding SLDs for this parameterized qubit state are given by
\begin{equation}
\begin{aligned}
    L_{\alpha} &= \left(\begin{matrix}
    0 & i e^{i \alpha} \sin \theta \\ -i e^{-i \alpha} \sin \theta & 0
  \end{matrix}\right),\\
  L_{\theta} &= \left(\begin{matrix}
    \sin \theta & e^{i \alpha} \cos \theta \\ e^{-i \alpha} \cos \theta & -\sin \theta
  \end{matrix}\right).
\end{aligned}
\end{equation}
$F_Q$ and $F_{\operatorname{Im}}$ can then be obtained as
\begin{equation}
  F_Q = \left(\begin{matrix}
    \sin^2 \theta & 0 \\ 0 & 1
  \end{matrix}\right),  F_{\operatorname{Im}} = \left(\begin{matrix}
    0 & -\sin \theta \\ \sin \theta & 0
  \end{matrix}\right),
\end{equation}
yielding
\begin{equation}
    F_Q^{-\frac{1}{2}} F_{\operatorname{Im}} F_Q^{-\frac{1}{2}} = \left(\begin{matrix}
    0 & -1 \\ 1 & 0
  \end{matrix}\right)
\end{equation}
Accordingly, the fundamental tradeoff described by Eq.~(\ref{eq:mainbound}) becomes
\begin{equation}
  \operatorname{Tr}(F_Q^{-1} F_C) \leq 1.
\end{equation}
This coincides with the Gill-Massar bound, which is known to be tight for qubit state~\cite{GillM00}.

Our methodology yields infinitely many optimal measurements, providing flexibility in practical implementation. This versatility allows for the selection of measurement schemes that require fewer number of multi-qubit gates, thereby enhancing overall fidelity. Additionally, it enables the avoidance of circuit configurations that are especially vulnerable to platform noise.

To demonstrate this, we implement two distinct optimal measurements, $\mathcal{M}_1^{opt}$ and $\mathcal{M}_2^{opt}$, both of which saturate the tradeoff relation under ideal conditions.
Detailed circuit structures are provided in Appendix~\ref{sec:exp_theo_circuit}.
As shown in Fig.~\ref{experiment}, although both strategies are theoretically optimal, $\mathcal{M}_2^{opt}$ outperforms $\mathcal{M}_1^{opt}$ in practice due to its fewer CNOT gate requirements, making it less susceptible to platform noise. 

In another demonstration, we consider the estimation of three parameters $(\alpha_1, \theta_1, \gamma_1)$ encoded in a two-qubit state,
\begin{equation}
    |\psi(\alpha_1, \theta_1, \gamma_1 )\rangle = (I_2 \otimes R_z(\alpha_1)) \text{CNOT} (H R_y(\theta_1) \otimes R_y(\gamma_1))|00\rangle,
\end{equation}
where $H$ is the Hadamard gate, and the CNOT gate uses the first qubit as the control.
In this case, $F_Q$ and $F_{\operatorname{Im}}$ are given by
\begin{equation}
\begin{aligned}
    F_Q = \left(\begin{matrix}
    \frac{1}{4}(3 + \cos 2 \theta_1 - 2 \cos 2 \gamma_1 \sin^2 \theta_1) &  0 & 0  \\ 0 &1 & 0 \\ 0& 0& 1
  \end{matrix}\right), \\
  F_{\operatorname{Im}}= \left(\begin{matrix}
    0 &  -\cos \theta_1 \cos \gamma_1& \sin\theta_1 \sin \gamma_1  \\ \cos \theta_1 \cos \gamma_1 & 0 & 0 \\ -\sin\theta_1 \sin \gamma_1& 0& 0
  \end{matrix}\right),
\end{aligned}
\end{equation}
and the eigenvalues of $F_Q^{-\frac{1}{2}} F_{\operatorname{Im}} F_Q^{-\frac{1}{2}}$ are $\pm i \frac{\sqrt{2+ 2\cos 2\theta_1 \cos 2 \gamma_1}}{\sqrt{3+\cos 2\theta_1 - 2 \sin^2 \theta_1 \cos 2\gamma_1}}$ and 0.
Accordingly, the tradeoff relation in Eq.~(\ref{eq:mainbound}) becomes
\begin{equation}
    \operatorname{Tr}(F_Q^{-1} F_C) 
    \leq 2 + \frac{2\left|\cos \theta_1 \sin \gamma_1 \right|}{\sqrt{3+\cos 2\theta_1 - 2 \sin^2 \theta_1 \cos 2\gamma_1}}.
\end{equation}
The optimal measurement $\mathcal{M}^{opt}$, which saturates the tradeoff relation, is provided along with its corresponding quantum circuit in Appendix~\ref{sec:exp_theo_circuit}.
For comparison, we also implement a simpler, non-optimal measurement $\mathcal{M}$, realized as a projective measurement onto the eigenbasis of $\sigma_x\otimes\sigma_x$. This scheme involves only two Hadamard gates and has lower circuit complexity. As shown in Fig.~\ref{experiment}(b), the presence of noise in the circuit leads to a noticeable discrepancy between the measured performance and simulation results.
Nevertheless, due to the optimality of $\mathcal{M}^{opt}$, $\mathcal{M}^{opt}$ still outperforms $\mathcal{M}$ although $\mathcal{M}^{opt}$ requires more complicate circuits.

\section{Summary}
The main challenge in multiparameter quantum estimation lies in the incompatibility of optimal measurements for different parameters, leading to fundamental tradeoffs in achievable precision. In this work, we have presented a tight analytical tradeoff relation that quantifies these precision limits for an arbitrary number of parameters encoded in pure quantum states. Our approach not only establishes the ultimate bounds but also provides a systematic methodology for constructing optimal separable measurements that saturate these limits. To demonstrate the practical significance of our findings, we applied our framework to quantum radar, where we derived a refined Arthurs-Kelly relation that characterizes the ultimate precision for the simultaneous estimation of range and velocity. For separable photon sources, we explicitly constructed the optimal measurement achieving this bound, while for entangled biphoton sources, we quantified the advantage provided by entanglement. These optimal measurements are not unique—our framework enables the construction of infinitely many such measurements. This flexibility allows for the design of measurement schemes that are more robust against practical noise, a feature of practical relevance that we have demonstrated on a cloud-based superconducting quantum computing platform. These results highlight the transformative potential of our methodology in advancing quantum metrology applications, including sensing, imaging, and communication.  

Our work offers a deeper understanding of the interplay between incompatible parameters in quantum estimation and provides a structured approach to optimizing measurement strategies. Future research directions include extending these results to mixed states and exploring further applications in quantum-enhanced technologies.

\begin{acknowledgments}
This work is supported by the Quantum Science and Technology-National Science and Technology Major Project (2023ZD0300600), the Guangdong Provincial Quantum Science Strategic Initiative (GDZX2303007,GDZX2505003), the Research Grants Council of Hong Kong (14309223, 14309624, 14309022), 1+1+1 CUHK-CUHK(SZ)-GDST Joint Collaboration Fund (Grant No. GRDP2025-022). H. C. acknowledges the support from the National Natural Science Foundation of China (Grants No. 12505024, No. 92476201), Department of Science and Technology of Guangdong Province (Grant No. 2024QN11X234), Guangdong Basic and Applied Basic Research Foundation (Grant No. 2025A1515011441), Shenzhen Science and Technology Program (Grants No. ZDYJ20251211120900001, No. JCYJ20240813141350066).
\end{acknowledgments}

\appendix
\begin{widetext}

\section{Optimal approximation under a given measurement}\label{sec:f}
With a given projective measurement, $\{|m\rangle\langle m|\}$, we first identify the optimal $O_j=\sum_m f_j(m)|m\rangle\langle m|$ to approximate $L_j\otimes I$ such that the mean squared error 
 \begin{eqnarray}\label{eq:error_app}
 \aligned
 \epsilon_j^2&=\bra{\xi}\bra{\Psi_x}(O_j-L_j\otimes I)^2\ket{\Psi_x}\ket{\xi},
% &=\||o_j\rangle-|l_j\rangle\|_2,\\
 \endaligned
 \end{eqnarray}
is minimized. 

Since
% \begin{widetext}
  \begin{eqnarray}
 \aligned
 \epsilon_j^2
 %&=\||o_j\rangle-|l_j\rangle\|_2\\
=&\langle\xi|\langle \Psi_x|(\sum_m f_j(m)|m\rangle\langle m|-L_j\otimes I)^2|\Psi_x\rangle|\xi\rangle\\
=& \langle \Psi_x| L_j^2|\Psi_x\rangle+\sum_m f_j(m)^2p_m(x)-2f_j(m)\operatorname{Re}\{\langle\xi|\langle \Psi_x| m\rangle\langle m |L_j\otimes I|\Psi_x\rangle|\xi\rangle\}\\
% =&p_m(x)[f_j(m)-\frac{Re\{\langle\xi|\langle \Psi_x| m\rangle\langle m |L_j\otimes I|\Psi_x\rangle|\xi\rangle}{p_m(x)}]^2-\frac{Re\{\langle\xi|\langle \Psi_x| m\rangle\langle m |L_j\otimes I|\Psi_x\rangle|\xi\rangle\}^2}{p_m(x)}+\langle \Psi_x| L_j^2|\Psi_x\rangle
=&\langle \Psi_x| L_j^2|\Psi_x\rangle + \lim _{x^{\prime}\rightarrow x}\sum_m p_m(x^{\prime})\left[f_j(m)-\frac{\operatorname{Re}\{\langle\xi|\langle \Psi_{x^{\prime}}| m\rangle\langle m |L_j\otimes I|\Psi_{x^{\prime}}\rangle|\xi\rangle\}}{p_m(x^{\prime})}\right]^2\\
&-\lim _{x^{\prime}\rightarrow x}\sum_m \frac{\operatorname{Re}\{\langle\xi|\langle \Psi_{x^{\prime}}| m\rangle\langle m |L_j\otimes I|\Psi_{x^{\prime}}\rangle|\xi\rangle\}^2}{p_m(x^{\prime})},
\endaligned
 \end{eqnarray}
% \end{widetext}
here $p_m(x)=|\langle m|\Psi_x\rangle|\xi\rangle|^2$ is the probability of the measurement result $m$.
The limit $x^{\prime} \rightarrow x$ is nontrivial for the terms with $p_m(x)=0$, where we have the elements with the type 0/0.
The optimal $f_j(m)$ under a given measurement is then 
 \begin{eqnarray}
 \aligned
      f_j(m)&=\lim _{x^{\prime}\rightarrow x}\frac{\operatorname{Re}\{\langle\xi|\langle \Psi_{x^{\prime}}| m\rangle\langle m |L_j\otimes I|\Psi_{x^{\prime}}\rangle|\xi\rangle\}}{p_m(x^{\prime})}\\
&=\lim _{x^{\prime}\rightarrow x}\frac{\frac{1}{2}\operatorname{Tr}[|m\rangle\langle m|\{L_j\otimes I, |\Psi_{x^{\prime}}\rangle|\xi\rangle\langle \xi|\langle \Psi_{x^{\prime}}|\}]}{p_m(x^{\prime})}\\
&=\lim _{x^{\prime}\rightarrow x}\frac{\operatorname{Tr}[|m\rangle\langle m| \partial_{x^{\prime}_j}(|\Psi_{x^{\prime}}\rangle|\xi\rangle\langle \xi|\langle \Psi_{x^{\prime}}|)]}{p_m(x^{\prime})}\\
&=\lim _{x^{\prime}\rightarrow x}\frac{\partial_{x^{\prime}_j}\operatorname{Tr}[|m\rangle\langle m| |\Psi_{x^{\prime}}\rangle|\xi\rangle\langle \xi|\langle \Psi_{x^{\prime}}|]}{p_m(x^{\prime})}\\
&=\lim _{x^{\prime}\rightarrow x}\frac{\partial_{x^{\prime}_j}p_m(x^{\prime})}{p_m(x^{\prime})}.
 \endaligned    
 \end{eqnarray}
If $p_m(x^{\prime})|_{x^{\prime} = x+dx} =0$ up to any orders of $dx$, we can choose $f_j(m)$ arbitrarily, which for convenience will be taken as 0.

Now let 
\begin{eqnarray}\label{eq:lo}
|l_j\rangle=L_j\otimes I|\Psi_x\rangle|\xi\rangle,\\
%|l_2\rangle=L_2\otimes I|\Psi_x\rangle|\xi\rangle,\\
|o_j\rangle=O_j|\Psi_x\rangle|\xi\rangle,
%|o_2\rangle=O_2-\langle O_2\rangle|\Psi_x\rangle|\xi\rangle,
\end{eqnarray}
With the optimal choice of $f_j(m)$ under the given measurement, we have
\begin{eqnarray}
\aligned
\langle o_j|o_k\rangle&=\lim _{x'\rightarrow x}\sum_{m}\frac{\partial_{x'_j}p_m(x')\partial_{x'_k}p_m(x')}{p_m(x')}\\
&=(F_C)_{jk},    \\
\operatorname{Re}\langle o_j|l_j\rangle&=\lim _{x'\rightarrow x} \operatorname{Re}\sum_{m}\frac{\partial_{x'_j}p_m(x')}{p_m(x')}\langle \xi|\langle \Psi_{x'}|m\rangle\langle m|L_j\otimes I |\Psi_{x'}\rangle|\xi\rangle\\
&=\lim _{x'\rightarrow x}\sum_{m}\frac{\partial_{x'_j}p_m(x')}{p_m(x')}\operatorname{Re}[\langle m|L_j\otimes I|\Psi_{x'}\rangle|\xi\rangle\langle\xi|\langle \Psi_{x'}|m\rangle]\\
&=\lim _{x'\rightarrow x}\sum_{m}\frac{\partial_{x'_j}p_m(x')}{p_m(x')}\frac{1}{2}[\langle m|(L_j\otimes I|\Psi_{x'}\rangle|\xi\rangle\langle\xi|\langle \Psi_{x'}|+|\Psi_{x'}\rangle|\xi\rangle\langle\xi|\langle \Psi_{x'}|L_j\otimes I)|m\rangle]\\
&=\lim _{x'\rightarrow x}\sum_{m}\frac{\partial_{x'_j}p_m(x')}{p_m(x')} \langle m|\partial_{x'_j}(|\Psi_{x'}\rangle|\xi\rangle\langle\xi|\langle \Psi_{x'}|)|m\rangle\\
&=\lim _{x'\rightarrow x}\sum_{m}\frac{\partial_{x'_j}p_m(x')}{p_m(x')}\partial_{x'_j}\langle m|\Psi_{x'}\rangle|\xi\rangle\langle\xi|\langle \Psi_{x'}|m\rangle\\
&=\lim _{x'\rightarrow x}\sum_{m}\frac{\partial_{x'_j}p_m(x')}{p_m(x')}\partial_{x'_j}p_m(x')\\
&=(F_C)_{jj}.
\endaligned
\end{eqnarray}
Thus 
\begin{eqnarray}
    \aligned
    \sum_j\epsilon_j^2&=\sum_j \||o_j\rangle-|l_j\rangle\|^2\\
    &=\sum_j(\langle o_j|-\langle l_j|)(|o_j\rangle-|l_j\rangle)\\
    &=\sum_j\langle o_j|o_j\rangle-\langle o_j|l_j\rangle-\langle l_j|o_j\rangle+\langle l_j|l_j\rangle\\
    &=\sum_j\langle o_j|o_j\rangle-2\operatorname{Re}[\langle o_j|l_j\rangle]+\langle l_j|l_j\rangle\\
    &=\sum_j(F_C)_{jj}-2(F_C)_{jj}+(F_Q)_{jj}\\
    &=\operatorname{Tr}(F_Q-F_C).
    \endaligned
\end{eqnarray}

\section{Optimal $\{|o_j\rangle\}$}\label{sec:opt_oj}
In Appendix~\ref{sec:f}, the optimal $\{|o_j\rangle\}$ under a given POVM is obtained. In this section, we derive the optimal $\{|o_j\rangle\}$ over all POVMs. 
We first present a lemma, which is modified from a result obtained by Branciard~\cite{Branciard2013}.

\textbf{Lemma} Suppose $\vec{l}_1$ and $\vec{l}_2$ are two unit vectors in a Euclidean space $\mathcal{E}$, and $\vec{l}_1 \cdot \vec{l}_2 = \cos (\frac{\pi}{2} - \phi) = \sin \phi = \beta$, then for any two orthogonal vectors $\hat{o}_1$ and $\hat{o}_2$, we have
\begin{equation}
  \|\vec{l}_1 - \hat{o}_1\|^2 +\|\vec{l}_2 - \hat{o}_2\|^2 \geq 1-\sqrt{1-\beta^2}.
\end{equation}
Proof of Lemma: If $\|\hat{o}_1\| \neq 0$, define $\vec{o}_1 = \frac{\hat{o}_1}{\|\hat{o}_1\|}$. Otherwise, $\vec{o}_1$ is defined as any unit vector orthogonal to $\hat{o}_2$.
If $\|\hat{o}_2\| \neq 0$, define $\vec{o}_2 = \frac{\hat{o}_2}{\|\hat{o}_2\|}$. Otherwise, $\vec{o}_2$ is defined as any unit vector orthogonal to $\hat{o}_1$.
We use the notation $l_{1_\perp} = \sqrt{1-(\vec{l}_1 \cdot \vec{o}_1)^2}$, $l_{2_\perp} = \sqrt{1-(\vec{l}_2 \cdot \vec{o}_2)^2}$, then
\begin{equation}
  \begin{aligned}
      \|\vec{l}_1-\hat{o}_1\|^2 =& \|(\vec{l}_1-(\vec{l}_1 \cdot \vec{o}_1) \vec{o}_1)+( (\vec{l}_1 \cdot \vec{o}_1) \vec{o}_1-\hat{o}_1)\|^2 \\
      =& \|\vec{l}_1-(\vec{l}_1 \cdot \vec{o}_1) \vec{o}_1\|^2+\| (\vec{l}_1 \cdot \vec{o}_1) \vec{o}_1-\hat{o}_1\|^2\\
      \geq& \|\vec{l}_1-(\vec{l}_1 \cdot \vec{o}_1) \vec{o}_1\|^2= 1-(\vec{l}_1 \cdot \vec{o}_1)^2 =  l_{1_\perp} ^2.
  \end{aligned}
\end{equation}
The equality can be saturated if and only if $\hat{o}_1 = (\vec{l}_1 \cdot \vec{o}_1) \vec{o}_1$, which means $\hat{o}_1$ is the vector projection of the unit vector $\vec{l}_1$ onto the unit vector $\vec{o}_1$. This condition can be rewritten as $\vec{l}_{1} \cdot \hat{o}_1 = \|\hat{o}_1\|^2$.

Because $\vec{o}_1$ and $\vec{o}_2$ are orthogonal unit vectors, we have $(\vec{l}_1 \cdot \vec{o}_1)^2 +(\vec{l}_1 \cdot \vec{o}_2)^2 \leq \|\vec{l}_1\|^2 = 1$, then
\begin{equation}
  \begin{aligned}
    l_{1_\perp} ^2= 1-(\vec{l}_1 \cdot \vec{o}_1)^2
    \geq  (\vec{l}_1 \cdot \vec{o}_2)^2.
  \end{aligned}
\end{equation}
This equality can be saturated if $\vec{l}_1,\vec{l}_2, \vec{o}_1, \vec{o}_2$ are in the same plane.
i.e. $\vec{l}_1 \in \operatorname{Span}\{\vec{o}_1,\vec{o}_2\}$, $\vec{l}_2 \in \operatorname{Span}\{\vec{o}_1,\vec{o}_2\}$.
Similarly, we have
\begin{equation}
  \|\vec{l}_2-\hat{o}_2\|^2 \geq  l_{2_\perp} ^2  =  1-(\vec{l}_2 \cdot \vec{o}_2)^2 \geq  (\vec{l}_2 \cdot \vec{o}_1)^2.
\end{equation}
Then, we have
\begin{equation}
  \|\vec{l}_1-\hat{o}_1\|^2 + \|\vec{l}_2-\hat{o}_2\|^2 \geq (\vec{l}_1 \cdot \vec{o}_2)^2 + (\vec{l}_2 \cdot \vec{o}_1)^2.
\end{equation}
The inequality can be saturated if and only if
\begin{equation}
  \begin{aligned}
    &\vec{l}_{1} \cdot \hat{o}_1 = \|\hat{o}_1\|^2,\\
    &\vec{l}_{2} \cdot \hat{o}_2 = \|\hat{o}_2\|^2, \\
    &\vec{o}_1 \in \operatorname{Span}\{\vec{l}_1,\vec{l}_2\},\\
    &\vec{o}_2 \in \operatorname{Span}\{\vec{l}_1,\vec{l}_2\},\\
  \end{aligned}
\end{equation}
i.e. $\hat{o}_1$ is the projection of the unit vector $\vec{l}_1$ onto the unit vector $\vec{o}_1$, $\hat{o}_2$ is the projection of the unit vector $\vec{l}_2$ onto the unit vector $\vec{o}_2$, and $\vec{l}_1,\vec{l}_2, \vec{o}_1, \vec{o}_2$ are in the same plane.

We now provide the construction of the optimal $\hat{o}_1$ and $ \hat{o}_2$ that saturate the bound in the lemma. 

We first consider the case with $|\beta| = 1$, i.e. $\phi = \pm \frac{\pi}{2}$. In this case $\vec{l}_1$ and $\vec{l}_2$ are linearly dependent, $\vec{l}_1\parallel\vec{l}_2$. We introduce another unit vector $\vec{l}_{\perp}$ which is orthogonal to both $\vec{l}_1$ and $\vec{l}_2$, $\vec{l}_1 \cdot \vec{l}_{\perp} = \vec{l}_2 \cdot \vec{l}_{\perp} = 0$, as shown in Fig.1. Since $\vec{o}_1$ and $\vec{o}_2$ are orthogonal, it can be obtained by rotating $\vec{l}_{\perp}$ and $\vec{l}_1$. We thus rotate $\vec{l}_1$ and $\vec{l}_{\perp}$ clockwise with an angle $\varphi+\frac{\pi}{4}$ to get two orthogonal unit vectors $\vec{o}_1$ and $\vec{o}_{2}$, here we introduce $\varphi$ to simplify the calculations later. 
$\hat{o}_1$ and $\hat{o}_2$ can then be obtained by projecting $\vec{l}_1$ and  $\vec{l}_2$ onto $\vec{o}_{1}$ and $\vec{o}_{2}$ respectively. 
With simple triangle geometry, we have
\begin{equation}
  \begin{aligned}
  &\|\hat{o}_1\|^2= \cos^2(\varphi+\frac{\pi}{4}),\\
  &\|\hat{o}_2\|^2= \cos^2(\frac{\pi}{4}-\varphi),
  \end{aligned}
   \label{lemma1_eq_1_beta_1}
\end{equation}
\begin{equation}
  \begin{aligned}
    \vec{l}_1 \cdot \hat{o}_1= \cos^2(\varphi+\frac{\pi}{4}),\\
    \vec{l}_2 \cdot \hat{o}_2= \cos^2(\frac{\pi}{4}-\varphi),
  \end{aligned}
  \label{lemma1_eq_2_beta_1}
\end{equation}
%where $\varphi$ is an arbitrary state.
and
\begin{equation}
  \begin{aligned}
    \|\vec{l}_1 - \hat{o}_1\|^2 +\|\vec{l}_2 - \hat{o}_2\|^2
    &=\left(\vec{l}_1 \cdot \vec{o}_2\right)^2+\left(\vec{l}_2 \cdot \vec{o}_1\right)^2\\
    &= \sin^2(\varphi+\frac{\pi}{4})+\sin^2(\frac{\pi}{4}-\varphi) \\
    & =  \frac{1}{2}(1+\sin 2\varphi)+\frac{1}{2}(1-\sin 2\varphi) \\
    &= 1.
  \end{aligned}
\end{equation}
The inequality is saturated for any choice of $\varphi$ and the optimal $\hat{o}_1$ and $\hat{o}_2$ are given by
\begin{equation}
  \begin{aligned}
    &\hat{o}_1 = \frac{1}{2}(1-\sin 2 \varphi)\vec{l}_1 - \frac{1}{2} \beta \cos 2 \varphi \vec{l}_{\perp},\\
    &\hat{o}_2 = \frac{1}{2} \beta (1+\sin 2 \varphi)\vec{l}_1 + \frac{1}{2} \cos 2 \varphi \vec{l}_{\perp}.
  \end{aligned}
  \label{o1_o2_beta_eq_p1_1}
\end{equation}

\begin{figure}[htbp]
  \centering  %图片全局居中
  \includegraphics[width = 1\textwidth]{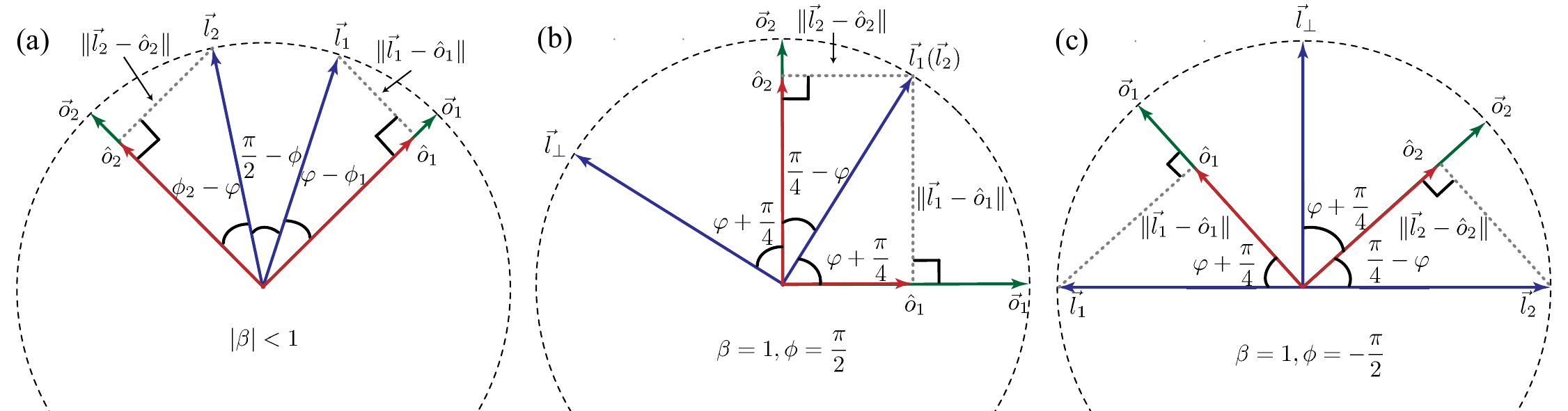}
  \caption{Geometric construction.}
  \label{geometric}
\end{figure}

When $|\beta|<1$, $\vec{l}_1$ and $\vec{l}_2$ are linearly independent, and they span a two-dimensional plane. As shown in Fig.~\ref{geometric}. We then rotate $\vec{l}_1$ clockwise with an angle $\varphi - \phi_1$ in the plane to get a unit vector $\vec{o}_1$ and $\vec{l}_2$ anti-clockwise with an angle $\phi_2 - \varphi$ in the plane to get a unit vector $\vec{o}_2$, here $\phi_1 = -\frac{\phi}{2}$, $\phi_2 = \frac{\phi}{2}$, $\phi=\arcsin\beta$ and $\varphi \in [ -\frac{|\phi|}{2}, \frac{|\phi|}{2}]$.  By projecting $\vec{l}_1$ onto $\vec{o}_1$ and projecting $\vec{l}_2$ onto $\vec{o}_2$, we get $\hat{o}_1$ and $\hat{o}_2$. With simple triangle geometries, we have
\begin{equation}
  \begin{aligned}
  &\|\hat{o}_1\|^2= \cos^2(\varphi-\phi_1),\\
  &\|\hat{o}_2\|^2= \cos^2(\phi_2-\varphi),
  \end{aligned}
  \label{lemma1_eq_1_beta}
\end{equation}
and
\begin{equation}
  \begin{aligned}
    \vec{l}_1 \cdot \hat{o}_1= \cos^2(\varphi-\phi_1),\\
    \vec{l}_2 \cdot \hat{o}_2= \cos^2(\phi_2-\varphi).
  \end{aligned}
   \label{lemma1_eq_2_beta}
\end{equation}

It is then straightforward to verify that 
\begin{equation}
  \begin{aligned}
    \|\vec{l}_1 - \hat{o}_1\|^2 +\|\vec{l}_2 - \hat{o}_2\|^2
    &=\left(\vec{l}_1 \cdot \vec{o}_2\right)^2+\left(\vec{l}_2 \cdot \vec{o}_1\right)^2\\
    &= \sin^2(\varphi-\phi_1) + \sin^2(\phi_2-\varphi)\\
    & = 1 - \cos \phi \cos 2\varphi\\
    & \geq 1 - \cos \phi = 1-\sqrt{1-\beta^2}.
  \end{aligned}
\end{equation}
The inequality is then saturated when we take $\varphi = 0$.
And the optimal $\hat{o}_1$ and $\hat{o}_2$ that saturate the bound are  
\begin{equation}
  \begin{aligned}
    &\hat{o}_1 = \frac{1+\cos \phi}{2 \cos \phi} \vec{l}_1 -\frac{\sin \phi}{2 \cos \phi} \vec{l}_2,\\
    &\hat{o}_2 = -\frac{\sin \phi}{2 \cos \phi}\vec{l}_1 + \frac{1+\cos \phi}{2 \cos \phi} \vec{l}_2.
  \end{aligned}
  \label{o1_o2_beta_less_1}
\end{equation}

We now apply the above constructions to quantum parameter estimation with a pure state $\ket{\Psi_x}$ that contains two parameters $x=(x_1,x_2)$. Let $L_1$ and $L_2$ be the SLDs corresponding to $x_1$ and $x_2$, respectively. As shown in the main text, for any POVM on $\ket{\Psi_x}$ we can replace it with a projective measurement, $\{|m\rangle\langle m|\}$, on the extended state $\ket{\Psi_x} \ket{\xi}$ with $\ket{\xi}$ as a state of the ancillary system. Two commuting observables $O_1$ and $O_2$ can then be constructed from the projective measurement to approximate $L_1\otimes I$ and $L_2\otimes I$. We then define 
%Given a state and an ancillary system, $\ket{\Psi_x} \ket{\xi}$, with $x = (x_1,x_2)$ as the parameters to be estimated. Let us define the ket vectors
\begin{equation}
  \begin{aligned}
    \ket{l_1} = L_1 \otimes I \ket{\Psi_x} \ket{\xi}, \\
    \ket{l_2} = L_2 \otimes I \ket{\Psi_x} \ket{\xi},  \\
    \ket{o_1} = O_1 \ket{\Psi_x} \ket{\xi},  \\
    \ket{o_2} = O_2  \ket{\Psi_x} \ket{\xi}.
  \end{aligned}
\end{equation}
We note that $\ket{o_1}$ and $\ket{o_2}$ are not necessarily quantum states since they may not be normalized, we write them with the ket notation just for convenience. And without loss of generality, we assume $F_Q = I$(if not, we can first make a reparametrization to make $F_Q = I$ as $\operatorname{Tr}(F_Q^{-1}F_C)$ is invariant under reparametrization). 

We then construct two real unit vectors from $\ket{l_1}$ and $\ket{l_2}$ as
\begin{equation}
  \begin{aligned}
    \vec{l}_1 = \left(\begin{matrix}
      \operatorname{Re} \ket{l_1} \\ \operatorname{Im} \ket{l_1}
    \end{matrix}\right), \quad
    \vec{l}_2 = \left(\begin{matrix}
      \operatorname{Im} \ket{l_2} \\ -\operatorname{Re} \ket{l_2}
    \end{matrix}\right).
  \end{aligned}
\end{equation}
The inner product of these two real vectors is
\begin{equation}
  \vec{l}_1 \cdot \vec{l}_2 = \operatorname{Im} \inp{l_1}{l_2} = \beta.
\end{equation}
Similarly we can obtain two real vectors from $\ket{o_1}$ and $\ket{o_2}$ as
\begin{equation}
  \begin{aligned}
    \hat{o}_1 = \left(\begin{matrix}
      \operatorname{Re} \ket{o_1} \\ \operatorname{Im} \ket{o_1}
    \end{matrix}\right), \quad
    \hat{o}_2 = \left(\begin{matrix}
      \operatorname{Im} \ket{o_2} \\ -\operatorname{Re} \ket{o_2}
    \end{matrix}\right).
  \end{aligned}
\end{equation}
%where $\vec{l}_1$ and $\vec{l}_2$ are unit vectors, with
From $[O_1,O_2]=0$, we have $\operatorname{Im}\langle o_1|o_2\rangle=0$, thus $\hat{o}_1\cdot \hat{o}_2=\operatorname{Im}\langle o_1|o_2\rangle=0$, i.e., $\hat{o}_1$ and $\hat{o}_2$ are orthogonal to each other. It is also straightforward to see
\begin{equation}
   \|\vec{l}_j-\hat{o}_j\|^2 = \||l_j\rangle-|o_j\rangle\|^2= \epsilon_j^2.
\end{equation}
$\vec{l}_1, \vec{l}_2, \hat{o}_1$ and $\hat{o}_2$ thus satisfy the assumptions of Lemma1, we then have
\begin{equation}
   \epsilon_1^2 + \epsilon_2^2  \geq 1- \sqrt{1-\beta^2}.
   \label{eq:minimum_error}
\end{equation}
To construct the optimal $\ket{o_1}$ and $\ket{o_2}$, we first consider the cases with $\beta = \pm 1$. Similarly, in this case, as $\vec{l}_1$ and  $\vec{l}_2$ are linearly dependent, we can introduce another unit vector, $\vec{l}_{\perp}$, that is orthogonal to $\vec{l}_1$ and  $\vec{l}_2$. We can write $\vec{l}_{\perp}$ as 
\begin{equation}
  \vec{l}_{\perp} = \left(\begin{array}{l}
\operatorname{Im}|l_{\perp}\rangle \\
-\operatorname{Re}|l_{\perp}\rangle
\end{array}\right),
\end{equation}
where $|l_{\perp}\rangle$ satisfies
\begin{equation}
    \operatorname{Im} \inp{l_{\perp}}{l_1} = 0; \quad \operatorname{Re} \inp{l_{\perp}}{l_2} = 0; \quad
    \inp{l_{\perp}}{l_{\perp}} = 1,
\end{equation}
which are just the conditions of $\vec{l}_1 \cdot \vec{l}_{\perp}=\vec{l}_2 \cdot \vec{l}_{\perp} =0$ and $\|\vec{l}_{\perp}\|=1$.
The optimal $\{\ket{o_j}\}$ that saturate the bound can be obtained similarly as in Eq.~(\ref{o1_o2_beta_eq_p1_1}) with
\begin{equation}
  \begin{aligned}
    \ket{o_1} &= \frac{1}{2}(1-\sin 2\varphi) |l_1\rangle + \frac{i}{2} \beta \cos 2\varphi \ket{l_{\perp}}, \\
    \ket{o_2} &= \frac{i}{2}\beta (1+\sin 2\varphi) |l_1\rangle + \frac{1}{2}\cos 2\varphi \ket{l_{\perp}},
  \end{aligned}
  \label{construct_beta1}
\end{equation}
from which we can get the classical Fisher information matrix with $
    (F_C)_{jk} = \inp{o_j}{o_k}$, which gives $F_C = \left(\begin{matrix}
    \frac{1}{2}(1-\sin 2\varphi) & \frac{1}{2} \cos 2\varphi \operatorname{Re} \inp{l_{\perp}}{l_1} \\
    \frac{1}{2} \cos 2\varphi \operatorname{Re} \inp{l_{\perp}}{l_1} & \frac{1}{2}(1+\sin 2\varphi)
\end{matrix}\right).$

We note that $\vec{l}_{\perp}$ only needs to be orthogonal to $\vec{l}_1$ and  $\vec{l}_2$, which is not unique, so is $\ket{l_{\perp}}$. Some choices of $\ket{l_{\perp}}$ can lead to a singular classical Fisher information matrix, which is the case if we take $\ket{l_{\perp}} = \ket{l_1}$. And some choices of $\ket{l_{\perp}}$ can lead to a diagonal classical Fisher information matrix, which is the case when $\ket{l_{\perp}}$ satisfies $\inp{l_{\perp}}{\Psi_x}\ket{\xi} = \inp{l_{\perp}}{l_1}=\inp{l_{\perp}}{l_2} = 0$. For example, if $\ket{l_{\perp}}$ is taken as $|\Phi\rangle|\xi^{\perp}\rangle$ where $|\xi^{\perp}\rangle$ is orthogonal to $|\xi\rangle$ and $|\Phi\rangle$ is an arbitrary state, the classical Fisher information matrix is then diagonal.

When $|\beta|<1$, $\beta =\sin \phi$ with $\phi \in \left(-\frac{\pi}{2}, \frac{\pi}{2}\right)$.
The optimal $\{\ket{o_j}\}$ can be obtained similar as in Eq.~(\ref{o1_o2_beta_less_1}) with
\begin{equation}
  \begin{aligned}
    \ket{o_1} = a|l_1\rangle -i b|l_2\rangle, \\
    \ket{o_2} = i b|l_1\rangle +a |l_2\rangle,
  \end{aligned}
  \label{construct_beta_less_1}
\end{equation}
where $a= \frac{1 +\cos\phi}{2\cos \phi}$, $b = -\frac{\sin \phi}{2\cos \phi}$. This gives the classical Fisher information matrix as $F_C=\left(\begin{matrix}
    \frac{1+\cos \phi}{2} & 0 \\
    0 & \frac{1+\cos \phi}{2}
\end{matrix}\right).$

\section{Verify the optimal measurement by directly computing  $F_C$}\label{sec:appendixverify}
In this section, we verify the optimality of the measurement constructed in the main text by directly computing the classical Fisher information matrix $F_C$ from the measurement and showing that it saturates the tradeoff relation.

Without loss of generality, we assume $F_Q = I$ and  $F_{\operatorname{Im}}$ takes the block diagonal form \begin{eqnarray}
F_{\operatorname{Im}}=\left[\begin{array}{ccccccccccc}
0 & \beta_1 &  0 &  \ldots & &   & & & \\
-\beta_1 & 0 &  & & & & & & \\
 0 & \cdots & 0 & \beta_2 & & & & & & \\
& & -\beta_2 & 0 & & & & & \\
 \vdots &  & & & \ddots & \vdots &  & & & \\
 0 & &\cdots & 0 &  \cdots & 0 & \beta_r & & & \\
& & & & &  -\beta_r & 0 & & & \\
& & & & & & & 0 & & \\
& & & & & & & & \ddots & \\
& & & & & & & & & 0
\end{array}\right].
\end{eqnarray}  %In this case the different $\{|l_j\rangle = L_j\otimes I |\Psi_x\rangle |\xi\rangle\}$ from different blocks are orthogonal to each other. 

We first recall the construction of the optimal measurements from the state $|\Psi_x\rangle |\xi\rangle$ and the optimal $\{|o_1\rangle, \cdots, |o_n\rangle\}$. 

The optimal $\{|o_1\rangle, \cdots, |o_n\rangle\}$ are obtained from $\{|l_j\rangle = L_j\otimes I |\Psi_x\rangle |\xi\rangle\}$ as following: 
\begin{itemize}
       \item for $1\leq j\leq r$ and $|\beta_j|<1$, 
    \begin{equation}
   \begin{aligned}
& \left|o_{2j-1}\right\rangle=a_j\left|l_{2j-1}\right\rangle-i b_j\left|l_{2j}\right\rangle, \\
& \left|o_{2j}\right\rangle=i b_j\left|l_{2j-1}\right\rangle+a_j\left|l_{2j}\right\rangle,
\end{aligned}
\end{equation}
where $a_j= \frac{1 +\cos\phi_j}{2\cos \phi_j}$, $b_j = -\frac{\sin \phi_j}{2\cos \phi_j}$, $\beta_j=\sin \phi_j$ with $\phi_j \in (-\frac{\pi}{2},\frac{\pi}{2})$.
   \item  for $1\leq j\leq r$ and $|\beta_j|=1$, 
   \begin{equation}
 \begin{aligned}
   \ket{o_{2j-1}} &= \frac{1}{2} (1-\sin 2\varphi_j) |l_{2j-1}\rangle + \frac{i}{2} \beta_j \cos 2\varphi_j \ket{l_{j\perp}}, \\
   \ket{o_{2j}} &= \frac{i}{2}\beta_j(1+\sin 2\varphi_j) |l_{2j-1}\rangle + \frac{1}{2}\cos 2\varphi_j \ket{l_{j\perp}},
 \end{aligned}
\end{equation}
where $\varphi_j$ can take any real value, $|l_{j\perp}\rangle$ is an arbitrary state orthogonal to $|\Psi_x\rangle|\xi\rangle$ and all $\{|l_{k}\rangle\}$. Moreover, $\langle l_{j\perp}|l_{k\perp}\rangle = \delta_{jk}$. One choice for $|l_{j\perp}\rangle=|\Phi\rangle|\xi_j^{\perp}\rangle$, where $|\xi_j^{\perp}\rangle$ is orthogonal to $|\xi\rangle$ and $|\Phi\rangle$ is an arbitrary state.  
\item for $j>2r$, $|o_j\rangle=|l_j\rangle$.
\end{itemize}

The Gram-Schmidt orthonormalization on $|\Psi_x\rangle |\xi\rangle$ and $\{|o_1\rangle, \cdots, |o_n\rangle\}$ then leads to an orthonormal set of states $\{|a_0\rangle, |a_1\rangle,\cdots, |a_n\rangle\}$ as
\begin{itemize}
      \item for $j=0$,
    %\begin{equation}
    %\begin{aligned}
       $ |a_0\rangle = |\Psi_x\rangle |\xi\rangle;$
    %\end{aligned}
%\end{equation}
\item for $1\leq j \leq r$ and $|\beta_j|<1$,
\begin{equation}
    \begin{aligned}
        |a_{2j-1}\rangle & = \frac{\sqrt{1+\sqrt{1-\beta_j^2}}}{\sqrt{2}\sqrt{1-\beta_j^2}} \left|l_{2j-1}\right\rangle+i \frac{\beta_j}{\sqrt{2}\sqrt{1-\beta_j^2}\sqrt{1+\sqrt{1-\beta_j^2}}}\left|l_{2j}\right\rangle; \\
        |a_{2j}\rangle &= -i \frac{\beta_j}{\sqrt{2}\sqrt{1-\beta_j^2}\sqrt{1+\sqrt{1-\beta_j^2}}}\left|l_{2j-1}\right\rangle+ \frac{\sqrt{1+\sqrt{1-\beta_j^2}}}{\sqrt{2}\sqrt{1-\beta_j^2}} \left|l_{2j}\right\rangle;
    \end{aligned}
    \label{eq:aj_less1}
\end{equation}
\item for $1\leq j \leq r$ and $|\beta_j|=1$,
\begin{equation}
    \begin{aligned}
        |a_{2j-1}\rangle &= \frac{\sqrt{2}}{2} \sqrt{1-\sin 2\varphi_j} |l_{2j-1}\rangle + i\frac{\sqrt{2}}{2} \beta_j \frac{\cos 2\varphi_j}{\sqrt{1-\sin 2\varphi_j}} \ket{l_{j\perp}}; \\
        |a_{2j}\rangle &= i\frac{\sqrt{2}}{2}\beta_j\sqrt{1+\sin 2\varphi_j} |l_{2j-1}\rangle + \frac{\sqrt{2}}{2}\frac{\cos 2\varphi_j}{\sqrt{1+\sin 2\varphi_j}}\ket{l_{j\perp}} ;
    \end{aligned}
    \label{eq:aj_1}
\end{equation}
\item for $j>2r$, 
%\begin{equation}
    $|a_j\rangle = |l_j\rangle.$
%\end{equation}
\end{itemize}
$\{|a_0\rangle, |a_1\rangle, \cdots, |a_n\rangle\}$ are then expanded into a complete basis by adding additional orthonormal vectors $\{|a_k\rangle| n+1 \leq k \leq d-1\}$, with $d$ is the dimension of the system+ancilla. The optimal measurement basis then corresponds to the rows of the unitary $U=BA^{-1}$ where $A$ is the unitary matrix with $\{|a_0\rangle, |a_1\rangle, \cdots, |a_{d-1}\rangle\}$ as the columns and $B$ is an orthogonal matrix with $\{|b_0\rangle, |b_1\rangle, \cdots, |b_{d-1}\rangle\}$, a set of arbitrary chosen real orthonormal vectors, as the columns. Here the only constraint we put on $B$ is that the first column of $B$ (i.e., $|b_0\rangle$) contains no zero entries. Different choices of $B$ lead to different optimal measurements. 

Since the optimal measurement corresponds to the rows of $U$, i.e., $U=\left(\begin{matrix}
    \langle 1|\\
    \vdots\\
    \langle m|\\
    \vdots\\
    \langle d|
\end{matrix}\right)$, and $UA=B$, we then have $|b_0\rangle=(\langle 1|\Psi_x\rangle|\xi\rangle,\langle 2|\Psi_x\rangle|\xi\rangle,\cdots, \langle m|\Psi_x\rangle|\xi\rangle,\cdots,\langle d|\Psi_x\rangle|\xi\rangle)^T$, which determines the probabilities of the measurement outcome $p_m=|\langle m|\Psi_x\rangle|\xi\rangle|^2=\operatorname{Tr}(|m\rangle\langle m|\Psi_x\rangle|\xi\rangle\langle\xi|\langle \Psi_x|)$. 
The derivative of the probability with respect to $x_j$ is 
\begin{equation}
    \begin{aligned}
        \partial_{x_j} p_m&=\operatorname{Tr}[|m\rangle\langle m| \frac{1}{2}(L_j\otimes I|\Psi_x\rangle|\xi\rangle\langle\xi|\langle \Psi_x|+|\Psi_x\rangle|\xi\rangle\langle\xi|\langle \Psi_x|L_j\otimes I)]\\
        &=\frac{1}{2}[\operatorname{Tr}(|m\rangle\langle m |l_j\rangle\langle\xi|\langle \Psi_x|)+\operatorname{Tr}(|m\rangle\langle m |\Psi_x\rangle|\xi\rangle\langle l_j|)]\\
               % \partial_{x_i} p_m &= \langle \partial_{x_i} \Psi_x| \langle  \xi |m\rangle \langle m | \Psi_x\rangle |\xi\rangle + \langle \Psi_x| \langle  \xi |m\rangle \langle m | \partial_{x_i} \Psi_x\rangle |\xi\rangle \\
        %&= 2 \operatorname{Re}\{ \frac{1}{2}\left\langle\Psi_x\right|\langle\xi | m\rangle\langle m|L_i \otimes I |\Psi_x\rangle |\xi\rangle
        % - \langle \partial_{x_i} \Psi_x| \Psi_x\rangle \left|\langle m|\Psi_x\rangle |\xi\rangle\right|^2\} \\
         &=\operatorname{Re}\{\left\langle\Psi_x\right|\langle\xi | m\rangle\langle m|l_j\rangle \}.
    \end{aligned}
    \label{eq:partial}
\end{equation}
The entries of the classical Fisher information matrix (CFIM) are then given by(note that $\{\langle m|\Psi_x\rangle|\xi\rangle\}$ are all real and nonzero due to the choice of $|b_0\rangle$)
\begin{equation}
    \begin{aligned}
        (F_C)_{jk} &= \sum_m \frac{\partial_{x_{j}} p_m \partial_{x_{k}} p_m}{p_m} = \sum_m \frac{\operatorname{Re}\{\left\langle\Psi_x\right|\langle\xi | m\rangle\langle m|l_j\rangle \operatorname{Re}\{\left\langle\Psi_x\right|\langle\xi | m\rangle\langle m|l_k\rangle}{|\langle m | \Psi_x\rangle |\xi\rangle|^2}= \sum_m \operatorname{Re}\{\langle m| l_j\rangle\} \operatorname{Re}\{\langle m|l_k \rangle \}.
    \end{aligned}
\end{equation}
Since we know $\langle m|a_j\rangle=B_{mj}$, we can write ${|l_j\rangle}$ in terms of ${|a_j\rangle}$ to compute $\langle m| l_j\rangle$. 

From Eq.~(\ref{eq:aj_less1}) and Eq.~(\ref{eq:aj_1}), we can directly obtain
\begin{itemize}
    \item for $1\leq j\leq r$ and $|\beta_j|<1$,
\begin{equation}
    \begin{aligned}
        |l_{2j-1}\rangle = \frac{\sqrt{1+\sqrt{1-\beta_j^2}}}{\sqrt{2}}|a_{2j-1}\rangle - \frac{i\beta_j}{\sqrt{2}\sqrt{1+\sqrt{1-\beta_j^2}}}|a_{2j}\rangle, \\
        |l_{2j}\rangle = \frac{i\beta_j}{\sqrt{2}\sqrt{1+\sqrt{1-\beta_j^2}}}|a_{2j-1}\rangle +\frac{\sqrt{1+\sqrt{1-\beta_j^2}}}{\sqrt{2}}|a_{2j}\rangle.
    \end{aligned}
\end{equation}
\item for $1\leq j\leq r$ and $|\beta_j|=1$, 
\begin{equation}
    \begin{aligned}
        |l_{2j-1}\rangle = \frac{\sqrt{1-\sin2\varphi_j}}{\sqrt{2}}|a_{2j-1}\rangle - i \beta_j\frac{\sqrt{1+\sin 2\varphi_j}}{\sqrt{2}} |a_{2j}\rangle, \\
        |l_{2j}\rangle = i \beta_j \frac{\sqrt{1-\sin2\varphi_j}}{\sqrt{2}}|a_{2j-1}\rangle + \frac{\sqrt{1+\sin 2\varphi_j}}{\sqrt{2}} |a_{2j}\rangle.
    \end{aligned}
\end{equation}
\item for $j>2r$, $|l_j\rangle=|a_j\rangle$.
\end{itemize}
Under the parametrization $F_Q=I$, only the diagonal entries of $F_C$ come into $\operatorname{Tr}(F_Q^{-1}F_C)$, we will thus focus on the computation of the diagonal entries.

For $1\leq j\leq r$ and $|\beta_j|<1$, we have $\langle m|l_{2j-1}\rangle=\frac{\sqrt{1+\sqrt{1-\beta_j^2}}}{\sqrt{2}}B_{m,2j-1} - i\frac{\beta_j}{\sqrt{2}\sqrt{1+\sqrt{1-\beta_j^2}}}B_{m,2j}$. Since $B$ is real orthogonal, all entries of $B$ are real, we then have $\operatorname{Re}\{\langle m| l_{2j-1} \rangle\}=\frac{\sqrt{1+\sqrt{1-\beta_j^2}}}{\sqrt{2}}B_{m,2j-1}$. Thus
\begin{equation}
    \begin{aligned}
        (F_C)_{2j-1, 2j-1} &= \sum_m \operatorname{Re}\{\langle m| l_{2j-1} \rangle \}^2  = \frac{1+\sqrt{1-\beta_j^2}}{2} \sum_m B_{m,2j-1}^2 = \frac{1+\sqrt{1-\beta_j^2}}{2}.
               % (F_C)_{2j-1, 2j} &= \sum_m \operatorname{Re}\{\langle m| l_{2j-1} \rangle \} \operatorname{Re}\{\langle m| l_{2j} \rangle \}  = \frac{1+\sqrt{1-\beta_j^2}}{2} \sum_m \langle m| a_{2j-1} \rangle \langle m| a_{2j} \rangle= 0
    \end{aligned}
\end{equation}
Similarly, we have $(F_C)_{2j, 2j} = \sum_m \operatorname{Re}\{\langle m| l_{2j} \rangle \}^2 = \frac{1+\sqrt{1-\beta_j^2}}{2}$. Thus $(F_C)_{2j-1, 2j-1}+(F_C)_{2j, 2j}=1+\sqrt{1-\beta_j^2}$.

For $1\leq j\leq r$ and $|\beta_j|=1$, we can similarly obtain
\begin{equation}
    \begin{aligned}
        (F_C)_{2j-1, 2j-1} &= \sum_m \operatorname{Re}\{\langle m| l_{2j-1} \rangle \}^2  = \frac{1-\sin 2\varphi_j}{2} \sum_m B_{m,2j-1}^2 = \frac{1-\sin 2\varphi_j}{2}, \\
        (F_C)_{2j, 2j} &= \sum_m \operatorname{Re}\{\langle m| l_{2j} \rangle \}^2  = \frac{1+\sin 2\varphi_j}{2} \sum_m B_{m,2j}^2 = \frac{1+\sin 2\varphi_j}{2}.
    \end{aligned}
\end{equation}
In this case, $(F_C)_{2j-1, 2j-1}+(F_C)_{2j, 2j}=1+\sqrt{1-\beta_j^2}=1$, which can also be written as $1+\sqrt{1-\beta_j^2}$ since $|\beta_j|=1$.

For $2r<j\leq n$, we have $|l_j\rangle=|a_j\rangle$ and 
\begin{equation}
    \begin{aligned}
        (F_C)_{j, j} &= \sum_m \operatorname{Re}\{\langle m| l_{j} \rangle \}^2  =  \sum_m B_{m,j}^2 = 1.
    \end{aligned}
\end{equation}
Summing all the diagonal entries, we obtain
\begin{equation}
    \begin{aligned}
        \operatorname{Tr}(F_Q^{-1}F_C) &= n-2r +\sum_{j=1}^r \left(1+ \sqrt{1-\beta_j^2}\right)\\
        & = n - \sum_{j=1}^r \left(1- \sqrt{1-\beta_j^2}\right).
    \end{aligned}
\end{equation}
This directly verifies that the constructed measurements saturate the tradeoff relations, and are thus optimal.

\section{Recover the conditions for the optimal measurement when the weak commutativity condition holds}
Here, we demonstrate how the conditions derived in~\cite{Luca2017} for optimal measurements in the special case where the weak commutativity condition \(\operatorname{Im}\langle l_j | l_k \rangle = 0\) holds for all \(j, k \in \{1, \dots, n\}\) can be recovered within our framework.

Without loss of generality, we assume we are working under the parametrization that $F_Q=I$. Let $\{L_j|j=1,\cdots, n\}$ be the SLDs for $|\Psi_x\rangle$ with $x=(x_1,\cdots,x_n)$ and $|l_j\rangle=L_j\otimes I |\Psi_x\rangle|\xi\rangle$. Given a measurement on the system+ancilla, denoted as $\{|m\rangle\langle m|\}$,
we have $O_j=\sum_m f_j(m)|m\rangle\langle m|$, and $|o_j\rangle=O_j|\Psi_x\rangle|\xi\rangle$. For the optimal choice of $f_j(m)$, $\langle O_j\rangle=0$. 
When the weak commutativity condition $\operatorname{Im}\langle l_j|l_k\rangle=0$, $\forall j, k\in \{1,\cdots, n\}$, 
$|o_j\rangle$ can be just taken as $|l_j\rangle$ and $F_C$ equals to $F_Q$. We thus have 
\begin{equation}\label{eq:ojlj}
|l_j\rangle=O_j|\Psi_x\rangle|\xi\rangle=\sum_m f_j(m)|m\rangle\langle m|\Psi_x\rangle|\xi\rangle.
\end{equation}
%where $O_j=\sum_m f_j(m)|m\rangle\langle m|$. 
When $p_m(x)=|\langle m|\Psi_x\rangle|\xi\rangle|^2\neq 0$, we have $f_j(m)=\frac{\partial_{x_j} p_m(x)}{p_m(x)}$, in this case
\begin{equation}\label{eq:mlj}
    \langle m|l_j\rangle=\frac{\partial_{x_j} p_m(x)}{p_m(x)}\langle m| \Psi_x\rangle|\xi\rangle.
\end{equation}
%let $(b_1)_m=\langle m|\Psi_x\rangle|\xi\rangle$, then $p_m(x)=(b_1)_m (b_1)_m^*$, 
Since $\frac{1}{2}L_j|\Psi_x\rangle=|\partial_{x_j}\Psi_x\rangle +\langle\partial_{x_j}\Psi_x|\Psi_x\rangle|\Psi_x\rangle$, we have $|l_j\rangle=2|\partial_{x_j}\Psi_x\rangle |\xi\rangle+2\langle\partial_{x_j}\Psi_x|\Psi_x\rangle|\Psi_x\rangle|\xi\rangle$, and $\frac{\partial_{ x_j}p_m(x)}{p_m(x)}=\frac{\langle m|\partial_{x_j}\Psi_x\rangle|\xi\rangle \langle \xi|\langle \Psi_x|m\rangle+\langle m|\Psi_x\rangle|\xi\rangle\bra{\xi}\bra{\partial_{x_j}\Psi_x}m\rangle}{\langle \xi|\langle \Psi_x| m \rangle\langle m|\Psi_x\rangle|\xi\rangle}$,  Eq.~(\ref{eq:mlj}) then becomes 
\begin{eqnarray}
 \aligned
2\langle m|\partial_{x_j}\Psi_x\rangle|\xi\rangle+2\inp{\partial_{x_j} \Psi_x}{\Psi_x}\langle m|\Psi_x\rangle|\xi\rangle=\frac{\langle m|\partial_{x_j}\Psi_x\rangle|\xi\rangle \langle \xi|\langle \Psi_x|m\rangle+\langle m|\Psi_x\rangle|\xi\rangle\bra{\xi}\bra{\partial_{x_j}\Psi_x}m\rangle}{\langle \xi|\langle \Psi_x| m \rangle},
\endaligned
 \end{eqnarray}
% as $b_m=\langle m|\Psi_x\rangle|\xi\rangle$, thus
which is equivalent to
 \begin{eqnarray}
 \aligned
\langle\xi|\langle\Psi_x|m\rangle\langle m|\partial_{x_j}\Psi_x\rangle|\xi\rangle -\langle m|\Psi_x\rangle|\xi\rangle\bra{\xi}\bra{\partial_{x_j}\Psi_x}m\rangle=-2\langle\partial_{x_j}\Psi_x|\Psi_x\rangle\langle m|\Psi_x\rangle|\xi\rangle\langle\xi|\langle\Psi_x|m\rangle .
\endaligned
 \end{eqnarray}
This can be written as
\begin{eqnarray}\label{eq:weakcomm1}
 \aligned
-2i \operatorname{Im} [ \bra{\xi}\bra{\partial_{x_j}\Psi_x}m\rangle\langle m|\Psi_x\rangle|\xi\rangle ]=-2i|\langle\xi|\langle\Psi_x|m\rangle|^2 \operatorname{Im} [\langle\partial_{x_j}\Psi_x|\Psi_x\rangle].
\endaligned
 \end{eqnarray}
From this we obtain 
\begin{equation}
\operatorname{Im} [ \bra{\xi}\bra{\partial_{x_j}\Psi_x}m\rangle\langle m|\Psi_x\rangle|\xi\rangle ]=|\langle\xi|\langle\Psi_x|m\rangle|^2 \operatorname{Im} [\langle\partial_{x_j}\Psi_x|\Psi_x\rangle],    
\end{equation}
which recovers Theorem 2 in~\cite{Luca2017}.

When $p_m(x) = 0$, we have $f_j(m) = \lim _{x^{\prime}\rightarrow x}\frac{\partial_{x^{\prime}_j}p_m(x^{\prime})}{p_m(x^{\prime})}$. In this case, for $\ket{m} \neq \ket{\Psi_x}\ket{\xi}$,
\begin{equation}
    \label{eq:mljp0}
    \inp{m}{l_j} = \lim_{x^{\prime} \rightarrow x} \frac{\partial_{x^{\prime}_j}p_m(x^{\prime})}{p_m(x^{\prime})} \inp{m}{\Psi_{x^{\prime}}} \ket{\xi}.
\end{equation}
Substituting
$|l_j\rangle=2|\partial_{x_j}\Psi_x\rangle |\xi\rangle+2\langle\partial_{x_j}\Psi_x|\Psi_x\rangle|\Psi_x\rangle|\xi\rangle$, Eq.~(\ref{eq:mljp0}) becomes 
\begin{eqnarray}
 \aligned
     \label{eq:mljp0lim}
2\langle m|\partial_{x_j}\Psi_{x}\rangle|\xi\rangle+2\inp{\partial_{x_j} \Psi_{x}}{\Psi_{x}}\langle m|\Psi_{x}\rangle|\xi\rangle= \lim_{x^{\prime} \rightarrow x} \frac{2 \operatorname{Re}\{\bra{\xi}\bra{\partial_{x^{\prime}_j}\Psi_{x^{\prime}}}m\rangle\langle m|\Psi_{x^{\prime}}\rangle|\xi\rangle\}}{\langle \xi|\langle \Psi_{x^{\prime}}| m \rangle},
\endaligned
 \end{eqnarray}
Excluding the case that $\inp{m}{\partial_{x_j}\Psi_x}\ket{\xi} = 0$ for all $j$, we expand both the denominator and the numerator on the right-hand side with respect to $x'=x+\delta x$.
%For the $x_j$ such that $\inp{m}{\partial_{x_j}\Psi_x}\ket{\xi} \neq 0$, 
By replacing $\ket{\Psi_{x'}}$ with $\ket{\Psi_x}+\sum_k\ket{\partial_{x_k}\Psi_x}\delta x_k$, we have
\begin{equation}
    %\begin{aligned}
        2\langle m|\partial_{x_j}\Psi_x\rangle|\xi\rangle %+O(\delta x) +\sum_{k=1}^n 2\inp{\partial_{x_j} \Psi_x}{\Psi_x}\langle m|\partial_{x_k} \Psi_x\rangle|\xi\rangle \delta x_k + O(\delta x^2)\\
        =\frac{\sum_{k=1}^n 2 \operatorname{Re}\{\bra{\xi}\bra{\partial_{x_j}\Psi_x}m\rangle\langle m|\partial_{x_k}\Psi_x\rangle|\xi\rangle\}\delta x_k + O(\delta x^2)}{\sum_{k=1}^n \langle \xi|\langle \partial_{x_k}\Psi_x| m \rangle \delta x_k + O(\delta x^2)}
    %\end{aligned}
\end{equation}
which is equivalent to
\begin{equation}
   \langle \xi|\langle \partial_{x_k}\Psi_x| m \rangle \langle m|\partial_{x_j}\Psi_x\rangle|\xi\rangle = \operatorname{Re}\{\bra{\xi}\bra{\partial_{x_j}\Psi_x}m\rangle\langle m|\partial_{x_k}\Psi_x\rangle|\xi\rangle\}
\end{equation}
for all $k$.
Thus we have $\operatorname{Im}\{\bra{\xi}\bra{\partial_{x_k}\Psi_x}m\rangle\langle m|\partial_{x_j}\Psi_x\rangle|\xi\rangle\} =0$ for all $j,k$, which is exactly Eq.~(7) in~\cite{Luca2017}.

\section{Connections to previous bounds}\label{sec:appendixPre}
A widely used bound in multiparameter quantum estimation is the Gill-Massar bound $\operatorname{Tr}(F_Q^{-1} F_C)\leq d-1$~\cite{GillM00}. The Gill-Massar bound is in general not tight, but for the special case with $2d-2$ parameters encoded in a $d$-dimensional pure quantum state, the bound becomes tight~\cite{GillM00}. We now show that the special case of the Gill-Massar bound can be recovered from our bound.  

Consider a $d$-dimensional pure state $|\Psi_x\rangle$ encoding $2d-2$ independent parameters $x=(x_1,\cdots, x_{2d-2})$. Again without loss of generality (recall $\operatorname{Tr}(F_Q^{-1}F_C)$ is invariant under reparametrization), we assume we are working under the parametrization with $F_Q = I$ and $F_{\operatorname{Im}}$ takes the block diagonal form as,
\begin{eqnarray}\label{eq:block}
{F}_{\operatorname{Im}}=\left[\begin{array}{ccccccccccc}
0 & \beta_1 &  0 &  \ldots & &   & & & \\
-\beta_1 & 0 &  & & & & & & \\
 0 & \cdots & 0 & \beta_2 & & & & & & \\
& & -\beta_2 & 0 & & & & & \\
 \vdots &  & & & \ddots & \vdots &  & & & \\
 0 & &\cdots & 0 &  \cdots & 0 & \beta_r & & & \\
& & & & &  -\beta_r & 0 & & & \\
& & & & & & & 0 & & \\
& & & & & & & & \ddots & \\
& & & & & & & & & 0
\end{array}\right],
\end{eqnarray}
where $r\leq d-1$.

We first show that $r=d-1$, i.e., all blocks are $2\times 2$. 
Let $|l_j\rangle = L_j\otimes I|\Psi_x\rangle|\xi\rangle$, which are all orthogonal to the state since $\langle \xi|\langle \Psi_x|l_j\rangle=0$ for $1\leq j\leq 2d-2$. Since the $jk$-th entry of $F=F_Q+i{F}_{\operatorname{Im}}$ equals to $\langle l_j|l_k\rangle$ and $F_{2j-1,k}=0$ for $1\leq j\leq r$, $\forall k\notin \{2j-1,2j\}$, we have
$\langle l_{2j-1}|l_{k}\rangle=0$, for $1\leq j\leq r$, $\forall k\notin \{2j-1,2j\}$, i.e., $|l_1\rangle$ is orthogonal to all $|l_k\rangle$ except $|l_1\rangle$ and $|l_2\rangle$; $|l_3\rangle$ is orthogonal to all $|l_k\rangle$ except $|l_3\rangle$ and $|l_4\rangle$, etc. In particular, $\{|l_1\rangle, |l_3\rangle,\cdots, |l_{2r-1}\rangle\}$ are orthogonal to each other. 

If $r<d-1$, then $\forall j>2r$, we have $\langle l_j|l_k\rangle=0$ when $k\neq j$, since $F_{jk}=0$ in this case. This implies that 
 $\{|l_1\rangle, |l_3\rangle,\cdots, |l_{2r-1}\rangle, |l_{2r+1}\rangle, |l_{2r+2}\rangle, |l_{2r+3}\rangle\cdots, |l_{2d-2}\rangle\}$ are all orthogonal to each other. Since $|l_j\rangle = L_j\otimes I|\Psi_x\rangle|\xi\rangle=|\tilde{l}_j\rangle|\xi\rangle$ with $|\tilde{l}_j\rangle=L_j|\Psi_x\rangle$, we also have $\{|\tilde{l}_1\rangle, |\tilde{l}_3\rangle,\cdots, |\tilde{l}_{2r-1}\rangle, |\tilde{l}_{2r+1}\rangle, |\tilde{l}_{2r+2}\rangle, |\tilde{l}_{2r+3}\rangle\cdots, |\tilde{l}_{2d-2}\rangle\}$ are all orthogonal to each other, 
 which are totally $r+(2d-2-2r)=2d-2-r$ number of orthogonal vectors in the d-dimensional Hilbert space. Furthermore, since all $|\tilde{l}_j\rangle$ are orthogonal to $|\Psi_x\rangle$, we should have 
 \begin{equation}
     2d-2-r\leq d-1,
 \end{equation}
which implies $r\geq d-1$. Since $F$ is $(2d-2)\times (2d-2)$, we also have $r\leq d-1$, thus $r=d-1$ and 
\begin{eqnarray}
F_{\operatorname{Im}}=\left[\begin{array}{ccccccc}
0 & \beta_1 &  0 &  0  && \ldots &0  \\
-\beta_1 & 0 & 0 &  0  && \ldots &0\\
 0 & 0 & 0 & \beta_2 & & & \vdots  \\
0& 0& -\beta_2 & 0 & & & \\
 \vdots & \vdots  & & & \ddots&   & \vdots  \\
 \vdots & \vdots &\ldots  &  \ldots & & 0 & \beta_{d-1}\\
 0& 0 & \ldots & \ldots & &  -\beta_{d-1} & 0 \\
\end{array}\right].
\end{eqnarray}

We next show $|\beta_j|=1$ for $1\leq j\leq d-1$. First note that $\{|\Psi_x\rangle, |\tilde{l}_1\rangle, |\tilde{l}_3\rangle,\cdots, |\tilde{l}_{2d-3}\rangle\}$ form a complete basis for the $d$-dimensional system space since they are orthonormal. As $\langle \tilde{l}_2|\tilde{l}_k\rangle=0$, $\forall k\notin {1,2}$, and $\langle \tilde{l}_2|\Psi_x\rangle=0$, $|\tilde{l}_2\rangle$ is thus orthogonal to all the vectors in the basis except $|\tilde{l}_1\rangle$, thus $|\tilde{l}_2\rangle$ must be linearly dependent with $|\tilde{l}_1\rangle$, $|\tilde{l}_2\rangle=\alpha |\tilde{l}_1\rangle$. Since 
$\langle\tilde{l}_1|\tilde{l}_1\rangle=\langle l_1|l_1\rangle=F_{11}=1$ and
$\langle\tilde{l}_2|\tilde{l}_2\rangle=\langle l_2|l_2\rangle=F_{22}=1$, we then have $|\alpha|=1$. From which we then have  $|\beta_1|=|F_{12}|=|\langle l_1|l_2\rangle|=|\langle \tilde{l}_1|\tilde{l}_2\rangle|=|\alpha|$. The proof is similar for $j=2,\cdots, d-1$.

The eigenvalues of $F_{\operatorname{Im}}$ are then $\{\lambda_1,\cdots, \lambda_{2d-2}\}=\{\pm i, \cdots, \pm i\}$, our tradeoff relation then reduces
\begin{eqnarray}
    \aligned
    \operatorname{Tr}(F_Q^{-1}F_C)&\leq 2d -2 -\frac{1}{2}\sum_{q=1}^{2d-2}(1-\sqrt{1-|\lambda_q|^2})\\
    & = 2d-2 - \frac{1}{2}(2d-2)\\
    & = d-1.
    \endaligned
    \end{eqnarray}
This recovers the Gill-Massar bound. We note that the number of independent real parameters encoded in a $d$-dimensional pure state is at most $2d-2$ ($-2$ corresponds to the normalization and global phase). When the number of parameters encoded in the $d$-dimensional pure state is less than $2d-2$, the Gill-Massar bound is in general not saturable, thus less tight than our bound.

Matsumoto obtained a bound in terms of $\operatorname{Tr}(F_QF_C^{-1})$ through a direct optimization using the Lagrange multiplier, which is $\operatorname{Tr}(F_QF_C^{-1})\geq \sum_{q=1}^n\frac{2}{1+\sqrt{1-|\lambda_q|^2}}$~\cite{MatsumotoThesis}. %where $\{\lambda_q\}$ are eigenvalues of $F_Q^{-\frac{1}{2}}F_{\operatorname{Im}}F_Q^{-\frac{1}{2}}$. Here 
We show that Matsumoto's bound can be obtained from our bound via the Cauchy–Schwarz inequality. On the other hand, our bound can not be obtained from Matsumoto's bound.

Without loss of generality, we assume $F_Q=I$, and
\begin{eqnarray}\label{eq:suppblock}
{F}_{\operatorname{Im}}=\left[\begin{array}{ccccccccccc}
0 & \beta_1 &  0 &  \ldots & &   & & & \\
-\beta_1 & 0 &  & & & & & & \\
 0 & \cdots & 0 & \beta_2 & & & & & & \\
& & -\beta_2 & 0 & & & & & \\
 \vdots &  & & & \ddots & \vdots &  & & & \\
 0 & &\cdots & 0 &  \cdots & 0 & \beta_r & & & \\
& & & & &  -\beta_r & 0 & & & \\
& & & & & & & 0 & & \\
& & & & & & & & \ddots & \\
& & & & & & & & & 0
\end{array}\right].
\end{eqnarray}
Denote the $j$-th $2\times 2$ block of $F_C$  and $F_C^{-1}$ as $F_{Cj}$ and $Q_{j}$ respectively, here $1\leq j\leq r$, i.e.,
\begin{eqnarray}\label{eq:FCblock1}
{F}_{C}=\left[\begin{array}{ccccccccc}
F_{C1} & * &    &  \ldots & &     \\
* & F_{C2} &  & & &  \\
% * & * & F_{C3} &  & & & \\
%\vdots & & \vdots & \vdots & &  \\
 \vdots &  &\ddots &  &  & \vdots &  \\
 * & & & F_{Cr} &   &  &  &   \\
\vdots &  & & & \ddots & \vdots &  \\
 %& & & & & & & 0 & & \\
%& & & & & & & & \ddots & \\
%& & & & & & & & & 0
\end{array}\right],
\end{eqnarray}
\begin{eqnarray}\label{eq:FCblock2}
{F}_{C}^{-1}=\left[\begin{array}{ccccccccc}
Q_1 & * &    &  \ldots & &     \\
* & Q_2 &  & & &  \\
% * & * & F_{C3} &  & & & \\
%\vdots & & \vdots & \vdots & &  \\
 \vdots &  &\ddots &  &  & \vdots &  \\
 * & & & Q_r &   &  &  &   \\
\vdots &  & & & \ddots & \vdots &  \\
 %& & & & & & & 0 & & \\
%& & & & & & & & \ddots & \\
%& & & & & & & & & 0
\end{array}\right],
\end{eqnarray}
we then have  $\operatorname{Tr}(Q_{j})\geq \operatorname{Tr}(F_{Cj}^{-1})$ where the equality is achieved when $F_C$ is a block diagonal matrix. Note that $\operatorname{Tr}(F_{Cj})\leq 1+\sqrt{1-\beta_j^2}$, by using the Cauchy–Schwarz inequality  $\operatorname{Tr}(F_{Cj}^{-1})\geq \frac{4}{\operatorname{Tr}(F_{Cj})}$, we then have 
\begin{eqnarray}\label{eq:cauchy}
\aligned
\operatorname{Tr}(Q_{j})&\geq \operatorname{Tr}(F_{Cj}^{-1})\\
&\geq \frac{4}{1+\sqrt{1-\beta_j^2}}.
\endaligned
\end{eqnarray}
For $j > 2r$, we have $(F_C)_{jj}\leq (F_Q)_{jj}=1$ and $(F_C^{-1})_{jj}\geq \frac{1}{(F_C)_{jj}}\geq 1$. Thus
\begin{eqnarray}
\aligned
    \operatorname{Tr}(F_C^{-1})&=\sum_{j=1}^r \operatorname{Tr}(Q_j)+\sum_{j=2r+1}^n(F_C^{-1})_{jj}\\
    &\geq \sum_{j=1}^r\frac{4}{1+\sqrt{1-\beta_j^2}}+n-2r\\
    &=\sum_{q=1}^n\frac{2}{1+\sqrt{1-|\lambda_q|^2}}.
  \endaligned  
\end{eqnarray}
The last equality holds as the eigenvalues of $F_{\operatorname{Im}}$ are $\{\pm i\beta_1,\cdots, \pm i\beta_r, 0,\cdots, 0\}$.

When $F_Q\neq I$, the bound can be written as $\operatorname{Tr}(F_QF_C^{-1})\geq \sum_{q=1}^n\frac{2}{1+\sqrt{1-|\lambda_q|^2}}$, where $\{\lambda_q\}$ are eigenvalues of $F_Q^{-\frac{1}{2}}F_{\operatorname{Im}}F_Q^{-\frac{1}{2}}$, which is just the Matsumoto's bound. The inequality is saturated when $F_C$ is a diagonal matrix with $F_{Cj}$ proportional to the Identity matrix for $1\leq j\leq r$, and $(F_C)_{kk}=1$ for $2r+1\leq k\leq n$, which can be satisfied by the optimal choices of $\{|o_j\rangle\}$ in the main text. 

Our bound is stronger than Matsumoto's bound since Matsumoto's bound can be obtained from our bound, but not vice-versa. We use the example of quantum radar to illustrate the difference. Consider using the separable photons for simultaneous estimation of the range and velocity, we have shown in the main text that the Arthurs-Kelly relation, $\hat{\sigma}_t\hat{\sigma}_\omega\geq 1$, can be directly obtained from our bound. On the other hand, from the  Matsumoto's bound, we have(note $\beta=-1$ in this case) 
\begin{equation}
    \operatorname{Tr}(F_QF_C^{-1})\geq 4.
\end{equation}
As 
%\begin{equation}
 $ F_Q = \left(\begin{matrix}
    4 \sigma^2 & 0 \\ 0 & \frac{1}{\sigma^2}
  \end{matrix}\right),$
%\end{equation}
and $\hat{\sigma}_t\geq (F_C^{-1})_{11}$, $\hat{\sigma}_\omega\geq (F_C^{-1})_{22}$, the Matsumoto's bound then gives
\begin{equation}\label{eq:suppmat}
    4 \sigma^2 \hat{\sigma}_t+\frac{1}{\sigma^2} \hat{\sigma}_\omega\geq 4.
\end{equation}
This is weaker than the Arthurs-Kelly relation since from the Arthurs-Kelly relation we can get the above bound as $4 \sigma^2 \hat{\sigma}_t+\frac{1}{\sigma^2} \hat{\sigma}_\omega\geq 4\sqrt{\hat{\sigma}_t\hat{\sigma}_\omega}\geq 4$. However, we can not get the Arthurs-Kelly relation from the above bound, for example,  $\hat{\sigma}_t=\frac{1}{\sigma^2}$ and $\hat{\sigma}_\omega = 0$ satisfy the above bound but violate the Arthurs-Kelly relation, showing the bound is strictly looser than the Arthurs-Kelly relation.

Chen et al.~\cite{HongzhenPra,HongzhenPRL} obtained an analytical bound on $\operatorname{Tr}(F_Q^{-1}F_C)$ for pure states as
    $\operatorname{Tr}(F_Q^{-1}F_C)\leq n-f(n)\|F_Q^{-\frac{1}{2}}F_{\operatorname{Im}}F_Q^{-\frac{1}{2}}\|_F^2$ with $f(n) \in\left\{\frac{1}{4(n-1)}, \frac{n-2}{(n-1)^2}, \frac{1}{5}\right\}$. Since larger $f(n)$ leads to a tighter bound, one may take $f(n)= \max\left\{\frac{1}{4(n-1)}, \frac{n-2}{(n-1)^2}, \frac{1}{5}\right\} \leq \frac{1}{4}$ for all $n$. This bound can be rewritten as $\operatorname{Tr}\left(F_Q^{-1} F_C\right) \leq n- f(n) \sum_{q=1}^n\left|\lambda_q\right|^2$, where $\{\lambda_q\}$ are the eigenvalues of $F_Q^{-1 / 2} F_{\operatorname{Im}} F_Q^{-1 / 2}$. Using the fact that $1-\sqrt{1-t} \geq \frac{t}{2}$ for $t \in [0,1]$ with $t=|\lambda_q|^2$, we have
\begin{equation} \label{eq:compare1}
    \begin{aligned}
        \operatorname{Tr}(F_Q^{-1}F_C)&\leq n-\frac{1}{2}\sum_{q=1}^n(1-\sqrt{1-|\lambda_q|^2})\\
    &\leq n-\frac{1}{2}\sum_{q=1}^n\left[1-\left(1-\frac{1}{2}|\lambda_q|^2\right)\right]\\
    &=n-\frac{1}{4}\sum_{q=1}^n|\lambda_q|^2\\
    &\leq n-f(n)\sum_{q=1}^n|\lambda_q|^2.
    \end{aligned}
\end{equation}
Eq.~(\ref{eq:mainbound}) thus yields a strictly tighter bound than Refs.~\cite{HongzhenPra,HongzhenPRL}.

\section{Connections to other metrics}\label{sec:appendix_measures}
The upper bound on $\operatorname{Tr}(F_Q^{-1}F_C)$ can also be converted into lower bounds on $\operatorname{Tr}(G \operatorname{Cov}(\hat{x}))$ with any positive-semidefinite weight matrix $G\geq 0$. Specifically, via the Cauchy-Schwarz inequality, we have
\begin{equation}
    \begin{aligned}
        \operatorname{Tr}(GF_C^{-1})\operatorname{Tr}(F_Q^{-1}F_C)&=\operatorname{Tr}(G^{\frac{1}{2}}F_C^{-\frac{1}{2}}F_C^{-\frac{1}{2}}G^{\frac{1}{2}})\operatorname{Tr}(F_Q^{-\frac{1}{2}}F_C^{\frac{1}{2}}F_C^{\frac{1}{2}}F_Q^{-\frac{1}{2}}) \\
        &\geq \left(\operatorname{Tr}\left|G^{\frac{1}{2}}F_Q^{-\frac{1}{2}}\right|\right)^2 =\left(\operatorname{Tr} \sqrt{F_Q^{-\frac{1}{2}} G F_Q^{-\frac{1}{2}}}\right)^2,
    \end{aligned}
    \label{eq:CS_trace_ratio}
\end{equation}
where $|A|:=\sqrt{A^\dagger A}$. Combining with the classical Cram\'er-Rao inequality
$N \operatorname{Cov}(\hat{x})\geq F_C^{-1}$ (here we take $N$ as the number of copies), we then have
\begin{equation}
     N\operatorname{Tr}(G\operatorname{Cov}(\hat{x}))\geq\operatorname{Tr}(GF_C^{-1})\geq \frac{\left(\operatorname{Tr} \sqrt{F_Q^{-\frac{1}{2}} G F_Q^{-\frac{1}{2}}}\right)^2}{\operatorname{Tr}(F_Q^{-1}F_C)}.
\end{equation}
An upper bound $\operatorname{Tr}(F_Q^{-1}F_C)\leq C$ can then be directly transformed into a lower bound on $N\operatorname{Tr}(G\operatorname{Cov}(\hat{x}))$ as
\begin{equation}\label{eq:boundcov}
   N\operatorname{Tr}(G\operatorname{Cov}(\hat{x}))\geq \frac{\left(\operatorname{Tr} \sqrt{F_Q^{-\frac{1}{2}} G F_Q^{-\frac{1}{2}}}\right)^2}{C}. 
\end{equation} 
For the Gill-Massar bound, $ C=d-1$ while for our bound $C=n-\frac{1}{2}\sum_{q=1}^n(1-\sqrt{1-|\lambda_q|^2})$.

In Ref.~\cite{Federico2021}, the incompatibility measure is defined as the ratio between the achievable precision and the quantum Cram\'er-Rao bound over an optimized weight matrix $G$:
\begin{equation}\label{eq:incomratio}
 r_N^s(x):=\inf _{M_N \in \mathcal{M}_N^{(\mathrm{LU\mbox{-}S})}} \sup _{G \geq 0}  \frac{N\operatorname{Tr}\left(G  \operatorname{Cov}(\hat{x})\right)}{\operatorname{Tr}\left(G F_Q^{-1}(x)\right)},
 \end{equation}
$\mathcal{M}_N^{(\mathrm{LU\mbox{-}S})}$ denotes the set of locally unbiased separable measurements. Since $N\operatorname{Tr}\left(G  \operatorname{Cov}(\hat{x})\right)\geq \operatorname{Tr} \left(G F_Q^{-1}(x)\right)$, $r_N^s(x)\geq 1$. The gap between $r_N^s(x)$ and 1 quantifies the incompatibility. 
Since the inequality in Eq.(\ref{eq:boundcov}) holds for any locally unbiased separable measurement, we then have 
\[ r_N^s(x)\geq \sup _{G \geq 0}  \frac{\left(\operatorname{Tr} \sqrt{F_Q^{-\frac{1}{2}} G F_Q^{-\frac{1}{2}}}\right)^2}{C\operatorname{Tr}\left(G F_Q^{-1}\right)}.
\]
Furthermore, by the Cauchy–Schwarz inequality   $\operatorname{Tr}\left(G F_Q^{-1}\right) \operatorname{Tr}(I)\geq \left(\operatorname{Tr} \sqrt{F_Q^{-\frac{1}{2}} G F_Q^{-\frac{1}{2}}}\right)^2$, we have $\frac{\left(\operatorname{Tr} \sqrt{F_Q^{-\frac{1}{2}} G F_Q^{-\frac{1}{2}}}\right)^2}{\operatorname{Tr}\left(G F_Q^{-1}\right)}\leq \operatorname{Tr}(I)=n$, where $n$ is the number of parameters. We thus have $\sup _{G \geq 0}  \frac{\left(\operatorname{Tr} \sqrt{F_Q^{-\frac{1}{2}} G F_Q^{-\frac{1}{2}}}\right)^2}{\operatorname{Tr}\left(G  F_Q^{-1}\right)}=n$ where the equality can be achieved by taking $G=F_Q$. This then leads to an analytical bound 
\begin{equation}\label{eq:incompatibility_GM}
    r_N^s(x) \geqslant \frac{n}{C}.
\end{equation}

In Ref.~\cite{Federico2021}, an analytical bound is derived from the Gill-Massar inequality $\operatorname{Tr}(F_Q^{-1}F_C)\leq C=d-1$, where $d$ is the dimension of a single probe, resulting in $r_N^s(x)\geq \frac{n}{d-1}$. Since this holds for any $N$, it gives 
\begin{equation}
    \underline{r}^s(x):=\liminf _{N \rightarrow \infty} r_N^s(x) \geq \frac{n}{d-1},
\end{equation}
which reproduces Eq.~(37) in Ref.~\cite{Federico2021} (in Ref.~\cite{Federico2021} the number of parameters is denoted as $d$ and the dimension of the system is denoted as $D$). This bound is only nontrivial when $n\geq d-1$, a restriction inherited from the Gill-Massar bound. 

Our improved bound, $\operatorname{Tr}(F_Q^{-1}F_C)\leq C=n-\frac{1}{2} \sum_{q=1}^n\left(1-\sqrt{1-\left|\lambda_q\right|^2}\right)$ leads to a tighter result:
\begin{equation}\label{eq:ours_r}
    \underline{r}^s(x) \geqslant \frac{n}{n-\frac{1}{2} \sum_{q=1}^n\left(1-\sqrt{1-\left|\lambda_q\right|^2}\right)},
\end{equation}
which is nontrivial for any $n$. The gap between $\underline{r}^s(x)$ and 1, 
\begin{equation}
    \underline{r}^s(x)-1\geq \frac{\frac{1}{2} \sum_{q=1}^n\left(1-\sqrt{1-\left|\lambda_q\right|^2}\right)}{n-\frac{1}{2} \sum_{q=1}^n\left(1-\sqrt{1-\left|\lambda_q\right|^2}\right)},
\end{equation}
quantifies the incompatibility. For mixed states, our analytical bound on $\operatorname{Tr}(F_Q^{-1}F_C)$ remains valid, though not universally tight. Consequently, Eq.(\ref{eq:ours_r}) remains valid for mixed states, and yields nontrivial bounds for any $n$, thereby improving over the existing bound.

Our bound also holds for collective measurements performed on $k$ copies of the state, $\rho_x^{\otimes k}$. This can be seen easily by treating $\rho_x^{\otimes k}$ as a single state, whose quantum geometric tensor satisfies $F^{(k)}=F_Q^{(k)}+iF_{\operatorname{Im}}^{(k)}=k(F_Q+iF_{\operatorname{Im}})$, so that $F_Q^{(k)}=kF_Q$ and $F_{\operatorname{Im}}^{(k)}=kF_{\operatorname{Im}}$. Consequently, $(F_Q^{(k)})^{-\frac{1}{2}} F_{\operatorname{Im}}^{(k)}(F_Q^{(k)})^{-\frac{1}{2}}=F_Q^{-\frac{1}{2}} F_{\operatorname{Im}} F_Q^{-\frac{1}{2}}$. Since the bound depends only on the eigenvalues of this matrix, it remains unchanged, yielding \begin{equation}\label{eq:boundN}
\operatorname{Tr}\left(\left(F_Q^{(k)}\right)^{-1}F_C^{(k)}\right)\leq n-\frac{1}{2} \sum_{q=1}^n\left(1-\sqrt{1-\left|\lambda_q\right|^2}\right),  
\end{equation}
where $\{\lambda_q\}$ are eigenvalues of $(F_Q^{(k)})^{-\frac{1}{2}} F_{\operatorname{Im}}^{(k)}(F_Q^{(k)})^{-\frac{1}{2}}=F_Q^{-\frac{1}{2}} F_{\operatorname{Im}}F_Q^{-\frac{1}{2}}$. We note that since $\rho_x$ can encode at most $d^2-1$ independent parameters, hence $n\leq d^2-1$.  The Gill-Massar bound applied to collective measurement on $\rho_x^{\otimes k}$ leads to trivial bounds  $\operatorname{Tr}\left(\left(F_Q^{(k)}\right)^{-1}F_C^{(k)}\right)\leq n\leq d^k-1$ for $k\geq 2$. 

Applying the Cauchy–Schwarz inequality similarly gives
\begin{equation}
    \begin{aligned}
\operatorname{Tr}\left(G\left(F_C^{(k)}\right)^{-1}\right)\operatorname{Tr}\left((F_Q^{(k)})^{-1}F_C^{(k)}\right)
        &\geq \left(\operatorname{Tr}\left|G^{\frac{1}{2}}\left(F_Q^{(k)}\right)^{-\frac{1}{2}}\right|\right)^2 =\left(\operatorname{Tr} \sqrt{\left(F_Q^{(k)}\right)^{-\frac{1}{2}} G \left(F_Q^{(k)}\right)^{-\frac{1}{2}}}\right)^2.
    \end{aligned}
    \label{eq:CS_trace_ratiok}
\end{equation}
If the collective measurement is repeated $\frac{N}{k}$ times with a total $N$ copies of state, we obtain 
\begin{equation}
\frac{N}{k}\operatorname{Tr}(G\operatorname{Cov}(\hat{x}))\geq \operatorname{Tr}\left(G\left(F_C^{(k)}\right)^{-1}\right)\geq \frac{\left(\operatorname{Tr} \sqrt{\left(F_Q^{(k)}\right)^{-\frac{1}{2}} G \left(F_Q^{(k)}\right)^{-\frac{1}{2}}}\right)^2}{\operatorname{Tr}\left(\left(F_Q^{(k)}\right)^{-1}F_C^{(k)}\right)}=\frac{\left(\operatorname{Tr} \sqrt{F_Q^{-\frac{1}{2}} G F_Q^{-\frac{1}{2}}}\right)^2}{k\operatorname{Tr}\left(\left(F_Q^{(k)}\right)^{-1}F_C^{(k)}\right)}.
\end{equation} 
Combining with the bound in Eq.(\ref{eq:boundN}) leads to 
\begin{equation}
    r^{(k)}_N(x):=\inf _{M_N \in \mathcal{M}_N^{(\mathrm{k})}} \sup _{G \geq 0}  \frac{N \operatorname{Tr}\left(G  \operatorname{Cov}(\hat{x})\right)}{\operatorname{Tr}\left(G F_Q^{-1}(x)\right)}\geq \sup _{G \geq 0}  \frac{\left(\operatorname{Tr} \sqrt{F_Q^{-\frac{1}{2}} G F_Q^{-\frac{1}{2}}}\right)^2}{\operatorname{Tr}\left(G F_Q^{-1}(x)\right)C}=\frac{n}{C},
    \label{eq:rNs_def}
\end{equation}
where $C=n-\frac{1}{2} \sum_{q=1}^n\left(1-\sqrt{1-\left|\lambda_q\right|^2}\right)$, $\mathcal{M}_N^{(\mathrm{k})}$ denotes the set of collective measurements on at most $k$ copies of states. This leads to 
\begin{equation}
    \underline{r}^{(k)}(x):=\liminf _{N \rightarrow \infty} r_N^{(k)}(x) \geqslant \frac{n}{n-\frac{1}{2} \sum_{q=1}^n\left(1-\sqrt{1-\left|\lambda_q\right|^2}\right)}.
\end{equation}
These incompatibility measures for collective measurement $\{\underline{r}^{(k)}\}$, defined in Eq.~(44) of Ref.~\cite{Federico2021}, previously lacked explicit lower bounds. We not only supply a closed-form expression that applies to all 
$k$, but also guarantee its non-triviality for an arbitrary number of parameters $n$. Our results thus significantly extend the applicability of the incompatibility measure defined in ~\cite{Federico2021}.

\section{Example: squeezed coherent state}\label{sec:example}
A squeezed coherent state can be written as
 $ |\eta, r, 0 \rangle = D(\eta) S(r) \ket{0},$
where $\ket{0}$ is the vacuum state, $D(\eta)$ is the displacement operator,
\begin{equation}
  D(\eta) = \exp\left(\eta a^{\dagger} - \eta^{*} a \right),
\end{equation}
and $S(r)$ is the squeezing operator with
\begin{equation}
  S(r) = \exp\left[\frac{r}{2}(a^2 - a^{\dagger 2})\right],
\end{equation}
where $r$ is the squeezing parameter. We consider the estimation of $\eta$ and $r$. Note that $\eta$ is generally a complex number, the parameters are thus $x_1 = \operatorname{Re}\eta$, $x_2 = \operatorname{Im}\eta$, $x_3 = r$. 

We first calculate the SLDs for which the following properties will be used, 
\begin{equation}
  \begin{aligned}
    &\frac{\partial D(\eta)}{\partial x_1} = \left(i x_2 + (a^{\dagger}-a)\right)D(\eta),\\
    &\frac{\partial D(\eta)}{\partial x_2} = \left(-i x_1 + i(a+a^{\dagger})\right)D(\eta),\\
    &\frac{\partial S(r)}{\partial x_3} = \frac{1}{2}(a^2 - a^{\dagger 2})S(r),
  \end{aligned}
\end{equation}
%and
\begin{equation}
  \begin{aligned}
    & D(\eta) a^2 D^{\dagger}(\eta) = a^2 - 2 \eta a + \eta^2, \\
    & D(\eta) a^{\dagger 2} D^{\dagger}(\eta) = a^{\dagger 2} - 2 \eta^{*} a^{\dagger} + \eta^{*2},\\
     & D^{\dagger}(\eta) a D(\eta) = a + \eta, \\
    & S^{\dagger}(r) a S(r) = \cosh r a - \sinh r a^{\dagger}.    
  \end{aligned}
\end{equation}
With these properties, we can get
\begin{equation}
  \begin{aligned}
    &\frac{\partial |\eta, r, 0 \rangle}{\partial x_1} = \frac{\partial D(\eta)}{\partial x_1} S(r) \ket{0} = \left(i x_2 + (a^{\dagger}-a)\right)|\eta, r, 0 \rangle, \\
    &\frac{\partial |\eta, r, 0 \rangle}{\partial x_2} = \frac{\partial D(\eta)}{\partial x_2} S(r) \ket{0}= \left(-i x_1 + i (a+a^{\dagger})\right)|\eta, r, 0 \rangle,\\
    &\frac{\partial |\eta, r, 0 \rangle}{\partial x_3} =  D(\eta) \frac{\partial S(r)}{\partial x_3} \ket{0} =\left(\frac{1}{2}(a^2 - a^{\dagger 2})+ x_1(a^{\dagger}-a) - i x_2 (a+a^{\dagger}) + 2i x_1 x_2\right)|\eta, r, 0 \rangle.
    \label{derivatives_1}
  \end{aligned}
\end{equation}
Note that
\begin{equation}
  \begin{aligned}
    a \ket{\eta, r,0} &= a D(\eta) S(r) \ket{0} \\
    & = \left(D(\eta) a + \eta D(\eta) \right) S(r)\ket{0} \\
    & = D(\eta) a S(r) \ket{0} + \eta \ket{\eta, r,0} \\
    & = D(\eta)\left(\cosh r S(r) a - \sinh r S(r) a^{\dagger}\right)\ket{0}+ \eta \ket{\eta, r,0} \\
    & = -\sinh r \ket{\eta, r,1}+ \eta \ket{\eta, r,0},
  \end{aligned}
\end{equation}
here $|\eta, r, n \rangle = D(\eta) S(r) \ket{n}$, and similarly we have 
\begin{equation}
  \begin{aligned}
    &a^{\dagger} \ket{\eta, r,0} = \cosh r \ket{\eta, r,1}+ \eta^{*} \ket{\eta, r,0},\\
    &a^{2} \ket{\eta, r,0}  = (\eta^2 - \sinh r \cosh r) \ket{\eta, r,0} - 2\eta \sinh r \ket{\eta, r,1} + \sqrt{2} \sinh^2 r \ket{\eta, r,2},\\
    &a^{\dagger 2} \ket{\eta, r,0}  = (\eta^{*2} - \sinh r \cosh r) \ket{\eta, r,0} + 2\eta^{*} \cosh r \ket{\eta, r,1} + \sqrt{2} \cosh^2 r \ket{\eta, r,2}.
  \end{aligned}
\end{equation}
Eq.~(\ref{derivatives_1}) can then be written as 
\begin{equation}
  \begin{aligned}
    &\frac{\partial |\eta, r, 0 \rangle}{\partial x_1} = -i x_2|\eta, r, 0 \rangle + e^{x_3} |\eta, r, 1 \rangle,\\
    &\frac{\partial |\eta, r, 0 \rangle}{\partial x_2} = i x_1 |\eta, r, 0 \rangle + i e^{-x_3} |\eta, r, 1 \rangle,\\
    &\frac{\partial |\eta, r, 0 \rangle}{\partial x_3} = -\frac{\sqrt{2}}{2} |\eta, r, 2 \rangle.
  \end{aligned}
\end{equation}

The SLDs,
%\begin{equation}
 $ L_i = 2\left(\frac{\partial |\eta, r, 0 \rangle}{\partial x_i} \langle \eta, r, 0| + |\eta, r, 0 \rangle \frac{\partial \langle\eta, r, 0|}{\partial x_i}\right)$, can then be obtained as
%\end{equation}
\begin{equation}
  \begin{aligned}
    &L_1 = 2 e^{x_3} \ket{\eta, r, 1} \bra{\eta, r, 0} +  2 e^{x_3} \ket{\eta, r, 0} \bra{\eta, r, 1},\\
    &L_2 = 2 i e^{-x_3} \ket{\eta, r, 1} \bra{\eta, r, 0} -  2 i  e^{- x_3} \ket{\eta, r, 0} \bra{\eta, r, 1},\\
    &L_3 = -\sqrt{2} \ket{\eta, r, 2} \bra{\eta, r, 0} -\sqrt{2} \ket{\eta, r, 0} \bra{\eta, r, 2}.
  \end{aligned}
\end{equation}
From these we obtain $F_Q$ and $F_{\operatorname{Im}}$ as
\begin{equation}
  F_Q = \left(\begin{matrix}
    4 e^{2 x_3} & 0 & 0 \\ 0 & 4 e^{-2 x_3} & 0 \\ 0 & 0& 2
  \end{matrix}\right), \quad F_{\operatorname{Im}} = \left(\begin{matrix}
  0 & 4   & 0 \\ -4 & 0 & 0 \\ 0 & 0& 0
  \end{matrix}\right).
\end{equation}
Since $F_Q\neq I$, we first make a reparametrization with $\left(\begin{matrix}
    x_1^{\prime}\\ x_2^{\prime}\\ x_3^{\prime}
  \end{matrix}\right) = F_Q^{\frac{1}{2}}  \left(\begin{matrix}
    x_1\\ x_2\\ x_3
  \end{matrix}\right)$ under which $\tilde{F}_Q = I$, and the SLDs for the new parameters are given by
  \begin{equation}
      L_{1}^{\prime} =  \frac{L_1}{2 e^{x_3}}, \quad
      L_{2}^{\prime} = \frac{L_2}{2 e^{-x_3}}, \quad
      L_{3}^{\prime} = \frac{L_3}{\sqrt{2}}.
  \end{equation}
Under this reparametrization,
  \begin{equation}
    \tilde{F}_{\operatorname{Im}} = \left(\begin{matrix}
       0 & 1 & 0 \\ -1 & 0 & 0 \\ 0 & 0 & 0
    \end{matrix}\right),
  \end{equation}
whose eigenvalues are $\pm i$ and $0$.
The tradeoff relation can then be obtained as
  \begin{equation}
     \operatorname{Tr}(F_Q^{-1} F_C) \leq \frac{1}{2} \sum_j(1+\sqrt{1-|\lambda_j|^2})= 2.
     \label{bound_example_3para}
  \end{equation}
To construct the optimal measurement, we first let
  \begin{equation}
    \begin{aligned}
      &|l_{1}\rangle = L_{1}^{\prime} \otimes I \ket{\eta, r, 0} \ket{0} =  \ket{\eta, r, 1} \ket{0},\\
      &|l_{2}\rangle = L_{2}^{\prime} \otimes I \ket{\eta, r, 0} \ket{0} = i \ket{\eta, r, 1} \ket{0},\\
      &|l_{3}\rangle = L_{3}^{\prime} \otimes I \ket{\eta, r, 0} \ket{0} = - \ket{\eta, r, 2} \ket{0},\\
      &|l_\perp \rangle %=i I \otimes a^{\prime \dagger} \ket{\eta, r, 0} \ket{0} 
      = i \ket{\eta, r, 0} \ket{1},
    \end{aligned}
  \end{equation}
  where an ancillary mode is used. $|l_\perp \rangle$ is introduced as $|l_1\rangle$ and $|l_2\rangle$ are linearly dependent in this case, here $|l_\perp \rangle$ satisfies $\inp{l_1}{l_\perp} = \inp{l_2}{l_\perp} =0 $ and $\inp{l_\perp}{l_\perp} = 1$.
  We can then construct the optimal $\{\ket{o_1}, \ket{o_2}, \ket{o_3}\}$ as in Eq.~(\ref{construct_beta1}) with $\varphi=0$,
  \begin{equation}
    \begin{aligned}
      &\ket{o_1}  = \frac{1}{2} \ket{l_1} + \frac{i}{2} \ket{l_\perp} = \frac{1}{2} \ket{\eta, r, 1} \ket{0} - \frac{1}{2}  \ket{\eta, r, 0} \ket{1}, \\
      &\ket{o_2}  = \frac{i}{2} \ket{l_1} + \frac{1}{2} \ket{l_\perp} = \frac{i}{2} \ket{\eta, r, 1} \ket{0} + \frac{i}{2}  \ket{\eta, r, 0} \ket{1},   \\
      &\ket{o_3}  = \ket{l_3} =- \ket{\eta, r, 2} \ket{0}.
    \end{aligned}
  \end{equation}
  It is easy to compute that $\inp{o_1}{o_1} = \inp{o_2}{o_2} = \frac{1}{2}, \inp{o_3}{o_3}=1, \inp{o_1}{o_2} = \inp{o_1}{o_3} = \inp{o_2}{o_3} = 0$.
 We then perform the Gram-Schmidt orthonormalization to get
  \begin{equation}
    \begin{aligned}
      &\ket{a_0} = \ket{\eta, r, 0} \ket{0}, \\
      &\ket{a_1} = \frac{\sqrt{2}}{2} \ket{\eta, r, 1} \ket{0} - \frac{\sqrt{2}}{2}  \ket{\eta, r, 0} \ket{1},\\
      &\ket{a_2} = \frac{i\sqrt{2}}{2} \ket{\eta, r, 1} \ket{0} + \frac{i\sqrt{2}}{2}  \ket{\eta, r, 0} \ket{1}, \\
      &\ket{a_3} =- \ket{\eta, r, 2} \ket{0},
    \end{aligned}
  \end{equation}
  which form a complete basis for a four-dimensional subspace spanned by $\{\ket{\eta, r, 0} \ket{0}, \ket{\eta, r, 0} \ket{1}, \ket{\eta, r, 1} \ket{0}, \ket{\eta, r, 2} \ket{0}\}$. Since all operators are within this subspace, we can restrict to this subspace. Putting ${\ket{a_j}}$ together, we obtain a unitary matrix
  \begin{equation}
    A = \left(\begin{matrix}
      1 & 0 & 0 & 0 \\
      0 & -\frac{\sqrt{2}}{2} & \frac{i\sqrt{2}}{2} & 0 \\
      0 & \frac{\sqrt{2}}{2} & \frac{i\sqrt{2}}{2} & 0 \\
      0 & 0 & 0 & -1
    \end{matrix}\right),
  \end{equation}
then choose a real orthogonal matrix as
  \begin{equation}
    B = \left(\begin{matrix}
    \frac{1}{2} &\frac{1}{2} & \frac{1}{2} & \frac{1}{2}\\
    \frac{1}{2} & -\frac{1}{2}  & \frac{1}{2} & -\frac{1}{2}\\
    \frac{1}{2} &  \frac{1}{2} & -\frac{1}{2} & -\frac{1}{2} \\
    \frac{1}{2} & -\frac{1}{2} & -\frac{1}{2} & \frac{1}{2}
    \end{matrix}\right)
  \end{equation}
  The optimal measurement can then be taken as the projective measurement on the basis given by the rows of $U = B A^{-1}$, which are 
  \begin{equation}
    \begin{aligned}
      &\ket{m_1} = \frac{1}{2} \ket{\hat{\eta}, \hat{r}, 0} \ket{0} + \left(-\frac{\sqrt{2}}{4} + \frac{i\sqrt{2}}{4}\right)\ket{\hat{\eta}, \hat{r}, 0} \ket{1} + \left(\frac{\sqrt{2}}{4} + \frac{i\sqrt{2}}{4}\right)\ket{\hat{\eta}, \hat{r}, 1} \ket{0} -\frac{1}{2} \ket{\hat{\eta}, \hat{r}, 2} \ket{0},\\
      &\ket{m_2} = \frac{1}{2} \ket{\hat{\eta}, \hat{r}, 0} \ket{0} + \left(\frac{\sqrt{2}}{4} + \frac{i\sqrt{2}}{4}\right)\ket{\hat{\eta}, \hat{r}, 0} \ket{1} + \left(-\frac{\sqrt{2}}{4} + \frac{i\sqrt{2}}{4}\right)\ket{\hat{\eta}, \hat{r}, 1} \ket{0} +\frac{1}{2} \ket{\hat{\eta}, \hat{r}, 2} \ket{0},\\
      &\ket{m_3} = \frac{1}{2} \ket{\hat{\eta}, \hat{r}, 0} \ket{0} -\left(\frac{\sqrt{2}}{4} + \frac{i\sqrt{2}}{4}\right)\ket{\hat{\eta}, \hat{r}, 0} \ket{1} + \left(\frac{\sqrt{2}}{4} -\frac{i\sqrt{2}}{4}\right)\ket{\hat{\eta}, \hat{r}, 1} \ket{0} +\frac{1}{2} \ket{\hat{\eta}, \hat{r}, 2} \ket{0},\\
      &\ket{m_4} = \frac{1}{2} \ket{\hat{\eta}, \hat{r}, 0} \ket{0} +\left(\frac{\sqrt{2}}{4} - \frac{i\sqrt{2}}{4}\right)\ket{\hat{\eta}, \hat{r}, 0} \ket{1} - \left(\frac{\sqrt{2}}{4} +\frac{i\sqrt{2}}{4}\right)\ket{\hat{\eta}, \hat{r}, 1} \ket{0} -\frac{1}{2} \ket{\hat{\eta}, \hat{r}, 2} \ket{0},\\
    \end{aligned}
  \end{equation}
  here $\hat{\eta} = \hat{x}_1 + i \hat{x}_2, \hat{r} = \hat{x}_3$ and $\hat{x}_1, \hat{x}_2, \hat{x}_3$ are estimators of $x_1, x_2, x_3$, respectively.
  We can verify that the probabilities of the measurement results are
  \begin{equation}
    \begin{aligned}
      &p_1 = |\inp{m_1}{\eta, r, 0}\ket{0}|^2 = \left|\frac{1}{2}\inp{\hat{\eta}, \hat{r}, 0}{\eta, r, 0} + \left(\frac{\sqrt{2}}{4} - \frac{i\sqrt{2}}{4}\right) \inp{\hat{\eta}, \hat{r}, 1}{\eta, r, 0} - \frac{1}{2} \inp{\hat{\eta}, \hat{r}, 2}{\eta, r, 0}\right|^2,\\
      &p_2 = |\inp{m_2}{\eta, r, 0}\ket{0}|^2 = \left|\frac{1}{2}\inp{\hat{\eta}, \hat{r}, 0}{\eta, r, 0} - \left(\frac{\sqrt{2}}{4} + \frac{i\sqrt{2}}{4}\right) \inp{\hat{\eta}, \hat{r}, 1}{\eta, r, 0} + \frac{1}{2} \inp{\hat{\eta}, \hat{r}, 2}{\eta, r, 0}\right|^2,\\
      &p_3 = |\inp{m_3}{\eta, r, 0}\ket{0}|^2 =  \left|\frac{1}{2}\inp{\hat{\eta}, \hat{r}, 0}{\eta, r, 0} + \left(\frac{\sqrt{2}}{4} + \frac{i\sqrt{2}}{4}\right) \inp{\hat{\eta}, \hat{r}, 1}{\eta, r, 0} + \frac{1}{2} \inp{\hat{\eta}, \hat{r}, 2}{\eta, r, 0}\right|^2,\\
      &p_4 = |\inp{m_4}{\eta, r, 0}\ket{0}|^2 = \left|\frac{1}{2}\inp{\hat{\eta}, \hat{r}, 0}{\eta, r, 0} - \left(\frac{\sqrt{2}}{4} - \frac{i\sqrt{2}}{4}\right) \inp{\hat{\eta}, \hat{r}, 1}{\eta, r, 0} - \frac{1}{2} \inp{\hat{\eta}, \hat{r}, 2}{\eta, r, 0}\right|^2,
    \end{aligned}
  \end{equation}
which gives the classical Fisher information matrix 
  \begin{equation}
    \begin{aligned}
      F_C = \left(\begin{matrix}
        2 e^{2 x_3} & 0 & 0 \\
        0 & 2 e^{-2 x_3} & 0 \\
        0 & 0 & 2
      \end{matrix}\right).
    \end{aligned}
  \end{equation}
For which the tradeoff relation is indeed saturated as $\operatorname{Tr}(F_Q^{-1} F_C) = 2$.
  
We note that in this case the bound can also be saturated without the ancillary system. For example, we can let
  \begin{equation}
    \begin{aligned}
      &|l_{1}\rangle = L_{1}^{\prime} \ket{\eta, r, 0}  =  \ket{\eta, r, 1}, \\
      &|l_{2}\rangle = L_{2}^{\prime} \ket{\eta, r, 0} = i \ket{\eta, r, 1}, \\
      &|l_{3}\rangle = L_{3}^{\prime} \ket{\eta, r, 0}  = - \ket{\eta, r, 2}, \\
      &|l_\perp \rangle = \ket{\eta, r, 3},
    \end{aligned}
  \end{equation}
here $|l_\perp \rangle$ also satisfies $\inp{\eta, r, 0}{l_\perp} = \inp{l_1}{l_\perp} = \inp{l_2}{l_\perp} =0 $ and $\inp{l_\perp}{l_\perp} = 1$.
The optimal  $\{\ket{o_1}, \ket{o_2}, \ket{o_3}\}$ can then be obtained from Eq.~(\ref{construct_beta1}) as (with $\varphi=0$)
  \begin{equation}
    \begin{aligned}
      &\ket{o_1}  = \frac{1}{2} \ket{l_1} + \frac{i}{2} \ket{l_\perp} = \frac{1}{2} \ket{\eta, r, 1} + \frac{i}{2}  \ket{\eta, r, 3}, \\
      &\ket{o_2}  = \frac{i}{2} \ket{l_1} + \frac{1}{2} \ket{l_\perp} = \frac{i}{2} \ket{\eta, r, 1} + \frac{1}{2}  \ket{\eta, r, 3}, \\
      &\ket{o_3}  = \ket{l_3} =- \ket{\eta, r, 2}.
    \end{aligned}
  \end{equation}
%It is easy to see that $\inp{o_1}{o_1} = \inp{o_2}{o_2} = \frac{1}{2}, \inp{o_3}{o_3}=1, \inp{o_1}{o_2} = \inp{o_1}{o_3} = \inp{o_2}{o_3} = 0$.
We then let
  \begin{equation}
    \begin{aligned}
      &\ket{a_0} = \ket{\eta, r, 0},  \\
      &\ket{a_1} = \frac{\sqrt{2}}{2} \ket{\eta, r, 1} + \frac{i\sqrt{2}}{2}  \ket{\eta, r, 3},\\
      &\ket{a_2} = \frac{i\sqrt{2}}{2} \ket{\eta, r, 1} + \frac{\sqrt{2}}{2}  \ket{\eta, r, 3}, \\
      &\ket{a_3} =- \ket{\eta, r, 2},
    \end{aligned}
  \end{equation}
  which form a complete basis for a four-dimensional subspace spanned by $\{\ket{\eta, r, 0}, \ket{\eta, r, 1}, \ket{\eta, r, 2}, \ket{\eta, r, 3}\}$. Again within this subspace, %Then we put $\{\ket{a_j}\}$ together with the basis of $\{\ket{\eta, r, 0}, \ket{\eta, r, 1}, \ket{\eta, r, 2} , \ket{\eta, r, 3}\}$ 
  we can put $\{\ket{a_j}\}$ together to get a unitary matrix 
  \begin{equation}
    A = \left(\begin{matrix}
      1 & 0 & 0 & 0 \\
      0 & \frac{\sqrt{2}}{2} & \frac{i\sqrt{2}}{2} & 0 \\
      0 & 0 & 0 & -1 \\
      0 & \frac{i\sqrt{2}}{2} & \frac{\sqrt{2}}{2} & 0
    \end{matrix}\right),
  \end{equation}
then choose a real orthogonal matrix as
  \begin{equation}
    B = \left(\begin{matrix}
    \frac{1}{2} &\frac{1}{2} & \frac{1}{2} & \frac{1}{2}\\
    \frac{1}{2} & -\frac{1}{2}  & \frac{1}{2} & -\frac{1}{2}\\
    \frac{1}{2} &  \frac{1}{2} & -\frac{1}{2} & -\frac{1}{2} \\
    \frac{1}{2} & -\frac{1}{2} & -\frac{1}{2} & \frac{1}{2}
    \end{matrix}\right).
  \end{equation}
 % to get a unitary $U = B A^{-1}$.
The optimal measurement can then be taken as the projective measurement on the basis given by the rows of $U=B A^{-1}$, which are
  \begin{equation}
    \begin{aligned}
      &\ket{m_1} = \frac{1}{2} \ket{\hat{\eta}, \hat{r}, 0} + \left(\frac{\sqrt{2}}{4} + \frac{i\sqrt{2}}{4}\right)\ket{\hat{\eta}, \hat{r}, 1} -\frac{1}{2} \ket{\hat{\eta}, \hat{r}, 2}+ \left(\frac{\sqrt{2}}{4} + \frac{i\sqrt{2}}{4}\right)\ket{\hat{\eta}, \hat{r}, 3},\\
      &\ket{m_2} =\frac{1}{2} \ket{\hat{\eta}, \hat{r}, 0} - \left(\frac{\sqrt{2}}{4} - \frac{i\sqrt{2}}{4}\right)\ket{\hat{\eta}, \hat{r}, 1} +\frac{1}{2} \ket{\hat{\eta}, \hat{r}, 2}+ \left(\frac{\sqrt{2}}{4} - \frac{i\sqrt{2}}{4}\right)\ket{\hat{\eta}, \hat{r}, 3},\\
      &\ket{m_3} =\frac{1}{2} \ket{\hat{\eta}, \hat{r}, 0} + \left(\frac{\sqrt{2}}{4} - \frac{i\sqrt{2}}{4}\right)\ket{\hat{\eta}, \hat{r}, 1} +\frac{1}{2} \ket{\hat{\eta}, \hat{r}, 2}- \left(\frac{\sqrt{2}}{4} - \frac{i\sqrt{2}}{4}\right)\ket{\hat{\eta}, \hat{r}, 3},\\
      &\ket{m_4} = \frac{1}{2} \ket{\hat{\eta}, \hat{r}, 0} - \left(\frac{\sqrt{2}}{4} + \frac{i\sqrt{2}}{4}\right)\ket{\hat{\eta}, \hat{r}, 1} -\frac{1}{2} \ket{\hat{\eta}, \hat{r}, 2}- \left(\frac{\sqrt{2}}{4} + \frac{i\sqrt{2}}{4}\right)\ket{\hat{\eta}, \hat{r}, 3}.\\
    \end{aligned}
  \end{equation}
  Here $\hat{\eta} = \hat{x}_1 + i \hat{x}_2, \hat{r} = \hat{x}_3$ and $\hat{x}_1, \hat{x}_2, \hat{x}_3$ are estimators of $x_1, x_2, x_3$, respectively.
  
  We can verify that under this projective measurement, the probabilities of the measurement results are 
  {\scriptsize\begin{equation}
    \begin{aligned}
      &p_1 = |\inp{m_1}{\eta, r, 0}|^2= \left|\frac{1}{2}\inp{\hat{\eta}, \hat{r}, 0}{\eta, r, 0} + \left(\frac{\sqrt{2}}{4} - \frac{i\sqrt{2}}{4}\right) \inp{\hat{\eta}, \hat{r}, 1}{\eta, r, 0} - \frac{1}{2} \inp{\hat{\eta}, \hat{r}, 2}{\eta, r, 0} + \left(\frac{\sqrt{2}}{4} - \frac{i\sqrt{2}}{4}\right) \inp{\hat{\eta}, \hat{r}, 3}{\eta, r, 0}\right|^2,\\
      &p_2 = |\inp{m_2}{\eta, r, 0}|^2 = \left|\frac{1}{2}\inp{\hat{\eta}, \hat{r}, 0}{\eta, r, 0} - \left(\frac{\sqrt{2}}{4} + \frac{i\sqrt{2}}{4}\right) \inp{\hat{\eta}, \hat{r}, 1}{\eta, r, 0} + \frac{1}{2} \inp{\hat{\eta}, \hat{r}, 2}{\eta, r, 0} + \left(\frac{\sqrt{2}}{4} + \frac{i\sqrt{2}}{4}\right) \inp{\hat{\eta}, \hat{r}, 3}{\eta, r, 0}\right|^2,\\
      &p_3 = |\inp{m_3}{\eta, r, 0}|^2 =  \left|\frac{1}{2}\inp{\hat{\eta}, \hat{r}, 0}{\eta, r, 0} + \left(\frac{\sqrt{2}}{4} + \frac{i\sqrt{2}}{4}\right) \inp{\hat{\eta}, \hat{r}, 1}{\eta, r, 0} + \frac{1}{2} \inp{\hat{\eta}, \hat{r}, 2}{\eta, r, 0} - \left(\frac{\sqrt{2}}{4} + \frac{i\sqrt{2}}{4}\right) \inp{\hat{\eta}, \hat{r}, 3}{\eta, r, 0}\right|^2,\\
      &p_4 = |\inp{m_4}{\eta, r, 0}|^2 = \left|\frac{1}{2}\inp{\hat{\eta}, \hat{r}, 0}{\eta, r, 0} - \left(\frac{\sqrt{2}}{4} - \frac{i\sqrt{2}}{4}\right) \inp{\hat{\eta}, \hat{r}, 1}{\eta, r, 0} - \frac{1}{2} \inp{\hat{\eta}, \hat{r}, 2}{\eta, r, 0} - \left(\frac{\sqrt{2}}{4} - \frac{i\sqrt{2}}{4}\right) \inp{\hat{\eta}, \hat{r}, 3}{\eta, r, 0}\right|^2,
    \end{aligned}
  \end{equation}}
which gives the classical Fisher information matrix 
  \begin{equation}
    \begin{aligned}
      F_C = \left(\begin{matrix}
        2 e^{2 x_3} & 0 & 0 \\
        0 & 2 e^{-2 x_3} & 0 \\
        0 & 0 & 2
      \end{matrix}\right).
    \end{aligned}
  \end{equation}
This saturates the tradeoff relation with
  \begin{equation}
    \operatorname{Tr}(F_Q^{-1} F_C) = 2.
  \end{equation}

\section{Optimal measurement for simultaneous estimation of range and velocity}
\subsection{Separable photons}\label{sec:opt_separable}
In this section, we explicitly construct the optimal measurement that saturates the Arthurs-Kelly relation $\sigma_{\bar{t}} \sigma_{\bar{\omega}} \geq 1$ for separable photons. 
Recall that the returned single photon state is given by $ \ket{\psi} = \int dt \ \psi(t) \ket{t}$, with
\begin{equation}
  \psi(t) = \left(\frac{2\sigma^2}{\pi}\right)^{1/4} \exp\left\{-(t-\bar{t})^2 \sigma^2 - i \bar{\omega} (t-\bar{t})\right\}.
\end{equation}
The QFIM for simultaneously estimating parameters $\bar{t}$ and $\bar{\omega}$ is 
\begin{equation}
    F_Q=\left(\begin{array}{cc}
4 \sigma^2 & 0 \\
0 & \frac{1}{\sigma^2}
\end{array}\right),
\end{equation}
with the corresponding SLDs provided in Eq.~(\ref{eq:sld_sep}).
We first make a reparametrization,
\begin{equation}
  \left(\begin{matrix}
    \bar{t}^{\prime} \\ \bar{\omega}^{\prime}
  \end{matrix}\right)= F_Q^{1/2} \left(\begin{matrix}
    \bar{t} \\ \bar{\omega}
  \end{matrix}\right)
\end{equation}
under which the SLDs become
\begin{equation}
  \begin{aligned}
    &L_{\bar{t}}^{\prime} = \ket{e_1}\bra{e_2} + \ket{e_2}\bra{e_1},\\
    &L_{\bar{\omega}}^{\prime} = i \ket{e_1}\bra{e_2}-i\ket{e_2}\bra{e_1},
  \end{aligned}
\end{equation}
and $\operatorname{Im}\langle L_{\bar{t}}^{\prime} L_{\bar{\omega}}^{\prime} \rangle = -1$, here \begin{equation}
  \begin{aligned}
    &\ket{e_1} =\ket{\psi(t)}= \int dt  \psi(t) \ket{t}, \\
    &\ket{e_2} = \int dt  2\sigma (t-\bar{t})\psi(t) \ket{t}.\\
\end{aligned}
\end{equation}
We then let
\begin{equation}
  \begin{aligned}
    &\ket{l_1} = L_{\bar{t}}^{\prime} \ket{\psi} = \ket{e_2}, \\
    &\ket{l_2} = L_{\bar{\omega}}^{\prime} \ket{\psi} = -i\ket{e_2}, \\
    &\ket{l_\perp} = \ket{e_3}
  \end{aligned}
\end{equation}
where
\begin{equation}
  \ket{e_3} = \int dt \ e_3(t) \ket{t}
\end{equation}
with
\begin{equation}
  e_3(t) = \frac{1-4(t-\bar{t})^2 \sigma^2}{\sqrt{2}}\psi(t).
\end{equation}
$\ket{l_{\perp}}$ satisfies $\inp{\psi}{l_{\perp}} = \inp{l_1}{l_{\perp}} = \inp{l_2}{l_{\perp}} =0$ and $\inp{l_{\perp}}{l_{\perp}} =1$.
We can then obtain the optimal $\{\ket{o_1}, \ket{o_2}\}$ as
\begin{equation}
  \begin{aligned}
    &\ket{o_1} = \frac{1}{2}\ket{l_1} - \frac{i}{2}\ket{l_\perp} = \frac{1}{2}\ket{e_2}-\frac{i}{2}\ket{e_3},\\
    &\ket{o_2} = -\frac{i}{2}\ket{l_1} + \frac{1}{2}\ket{l_\perp} = -\frac{i}{2}\ket{e_2}+\frac{1}{2}\ket{e_3},
  \end{aligned}
\end{equation}
where $\inp{o_1}{o_1} = \inp{o_2}{o_2}=\frac{1}{2}$, $\inp{o_1}{o_2} = 0$.
We then let
\begin{equation}
  \begin{aligned}
    &\ket{a_0} = \ket{\psi} = \ket{e_1}, \\
    &\ket{a_1} = \frac{\sqrt{2}}{2}\ket{e_2}-\frac{i\sqrt{2}}{2}\ket{e_3},\\
    &\ket{a_2} = -\frac{i\sqrt{2}}{2}\ket{e_2}+\frac{\sqrt{2}}{2}\ket{e_3},\\
  \end{aligned}
\end{equation}
which form a complete basis for the three-dimensional subspace spanned by $\{\ket{e_1},\ket{e_2},\ket{e_3}\}$.
Within this subspace, $\{\ket{a_j}\}$ can be represented as 3-dimensional vectors which can be put together to get a unitary matrix,
\begin{equation}
  A=\left(\begin{matrix}
    1 & 0 & 0 \\
    0 & \frac{\sqrt{2}}{2} & - \frac{i\sqrt{2}}{2} \\
    0 & - \frac{i\sqrt{2}}{2} &  \frac{\sqrt{2}}{2}
  \end{matrix}\right).
\end{equation}
We then choose a real orthogonal matrix,
\begin{equation}
  B=\left(\begin{matrix}
    \frac{1}{2} & \frac{1}{\sqrt{2}} & \frac{1}{2} \\
    \frac{1}{\sqrt{2}} & 0 & -\frac{1}{\sqrt{2}} \\
    \frac{1}{2} & -\frac{1}{\sqrt{2}} & \frac{1}{2}
  \end{matrix}\right),
\end{equation}
and let $U = B A^{-1}$. The optimal measurement can then be taken as the projective measurement on the basis given by the rows of $U$, which are
\begin{equation}
  \begin{aligned}
    &\ket{m_1} = \frac{1}{2} \ket{\hat{e}_1} +\left(\frac{1}{2} - \frac{i\sqrt{2}}{4}\right) \ket{\hat{e}_2} + \left(\frac{\sqrt{2}}{4} - \frac{i}{2}\right) \ket{\hat{e}_3} \\
    &\ket{m_2} = \frac{\sqrt{2}}{2} \ket{\hat{e}_1} + \frac{i}{2} \ket{\hat{e}_2} -\frac{1}{2} \ket{\hat{e}_3} \\
    &\ket{m_3} = \frac{1}{2} \ket{\hat{e}_1} -\left(\frac{1}{2} + \frac{i\sqrt{2}}{4}\right) \ket{\hat{e}_2} + \left(\frac{\sqrt{2}}{4} + \frac{i}{2}\right) \ket{\hat{e}_3} \\
  \end{aligned}
\end{equation}
here
\begin{equation}
  \begin{aligned}
    &\ket{\hat{e}_1} = \int dt \ \hat{\psi}(t) \ket{t}, \\
    &\ket{\hat{e}_2} = \int dt \ 2\sigma (t-\hat{\bar{t}})\hat{\psi}(t) \ket{t},\\
    &\ket{\hat{e}_3} = \int dt \  \frac{1-4(t-\hat{\bar{t}})^2\sigma^2}{\sqrt{2}}\hat{\psi}(t) \ket{t}
  \end{aligned}
\end{equation}
with
\begin{equation}
  \hat{\psi}(t) = \left(\frac{2\sigma^2}{\pi}\right)^{1/4} \exp\left\{-(t-\hat{\bar{t}})^2 \sigma^2 - i \hat{\bar{\omega}} (t-\hat{\bar{t}})\right\}.
\end{equation}
$\hat{\bar{t}}$ and $\hat{\bar{\omega}}$ are estimators of $\bar{t}$ and $\bar{\omega}$, respectively, which need to be updated adaptively. 
When $\hat{\bar{t}}$ and $\hat{\bar{\omega}}$ converge to $\bar{t}$ and $\bar{\omega}$, the classical Fisher information matrix is given by
\begin{equation}
  F_C = \left(\begin{matrix}
    2\sigma^2 & 0 \\  0 & \frac{1}{2\sigma^2}
  \end{matrix}\right).
\end{equation}
We then have $\sigma_{\bar{t}}^2 =\frac{1}{2\sigma^2}$ and $\sigma_{\bar{\omega}}^2 =2\sigma^2$ and \begin{equation}
  \sigma_{\bar{t}} \sigma_{\bar{\omega}} = 1.
\end{equation} 
This shows that the Arthurs-Kelly relation is tight and the constructed measurement is optimal.

\subsection{Entangled photons}\label{sec:opt_entangled}
In this section, we present the optimal measurement that saturates the refined Arthurs-Kelly relation for entangled photons, $\sigma_{\bar{t}}\sigma_{\bar{\omega}} \geq \frac{\sqrt{1-\kappa}}{\sqrt{1+\kappa}}$, for $0\leq \kappa<1$.
Recall that the returned biphoton entangled state is given by $\ket{\Psi} = \int dt \int dt_i \Psi(t,t_i) \ket{t} \ket{t_i}$,
with
\begin{equation}
    \Psi(t,t_i) = \mathcal{N} \exp\{-i \bar{\omega}(t-\bar{t})-i\bar{\omega}_{i0}(t_i-\bar{t}_0) -(t-\bar{t})^2 \sigma^2
    - (t_i-\bar{t}_0)^2 \sigma_{i0}^2 + 2 \kappa (t-\bar{t})(t_i-\bar{t}_0)\sigma\sigma_{i0}\},
\end{equation}
where the normalization factor is given by $\mathcal{N} = \sqrt{\frac{2\sigma \sigma_{i0}}{\pi}}(1-\kappa^2)^{1/4}$.
The QFIM for the simultaneous estimation of $\bar{t}$ and $\bar{\omega}$ is
\begin{equation}
    F_Q=\left(\begin{array}{cc}
4 \sigma^2 & 0 \\
0 & \frac{1}{\sigma^2(1-\kappa^2)}
\end{array}\right),
\end{equation}
with the corresponding SLDs provided explicitly in Eq.~(\ref{eq:sld_ent}).
If we make a reparametrization with
\begin{equation}
  \left(\begin{matrix}
    \bar{t}^{\prime} \\ \bar{\omega}^{\prime}
  \end{matrix}\right)= F_Q^{1/2} \left(\begin{matrix}
    \bar{t} \\ \bar{\omega}
  \end{matrix}\right),
\end{equation}
under which $\tilde{F}_Q = I$, and the SLDs become
\begin{equation}
  \begin{aligned}
    &L_{\bar{t}}^{\prime} = \frac{\sqrt{2(1-\kappa)}}{2} \ket{e_1}\bra{e_2} + \frac{\sqrt{2(1-\kappa)}}{2} \ket{e_2}\bra{e_1}+ \frac{\sqrt{2(1+\kappa)}}{2}  \ket{e_1}\bra{e_3} + \frac{\sqrt{2(1+\kappa)}}{2}  \ket{e_3}\bra{e_1},\\
    &L_{\bar{\omega}}^{\prime} = \frac{i\sqrt{2(1+\kappa)}}{2} \ket{e_1}\bra{e_2} -\frac{i\sqrt{2(1+\kappa)}}{2} \ket{e_2}\bra{e_1}+\frac{i\sqrt{2(1-\kappa)}}{2} \ket{e_1}\bra{e_3} -\frac{i\sqrt{2(1-\kappa)}}{2} \ket{e_3}\bra{e_1}.
  \end{aligned}
\end{equation}
where $\ket{e_j} = \int dt \ \int dt_i \ e_j(t, t_i) \ket{t} \ket{t_i}$, $j = 1, 2,3$, are orthonormal with
\begin{equation}
  \begin{aligned}
    &e_1(t,t_i) = \Psi(t,t_i),\\
    &e_2(t,t_i) = \sqrt{2(1-\kappa)}\left(\sigma (t-\bar{t}) + \sigma_i (t_i-\bar{t}_0)\right) \Psi(t,t_i),\\
    &e_3(t,t_i) = \sqrt{2(1+\kappa)}\left(\sigma (t-\bar{t}) - \sigma_i (t_i-\bar{t}_0)\right) \Psi(t,t_i).
  \end{aligned}
\end{equation}
We then construct the optimal measurement that saturates the relation for $0\leq \kappa<1$.
First, let
\begin{equation}
  \begin{aligned}
    &\ket{l_1} = L_{\bar{t}}^{\prime} \ket{\Psi} = \frac{\sqrt{2(1-\kappa)}}{2} \ket{e_2}+\frac{\sqrt{2(1+\kappa)}}{2}  \ket{e_3}, \\
    &\ket{l_2} = L_{\bar{\omega}}^{\prime} \ket{\Psi} = -\frac{i\sqrt{2(1+\kappa)}}{2} \ket{e_2}-\frac{i\sqrt{2(1-\kappa)}}{2} \ket{e_3}, \\
  \end{aligned}
\end{equation}
The optimal $\{\ket{o_1}, \ket{o_2}\}$ are then 
\begin{equation}
  \begin{aligned}
    &\ket{o_1} = \frac{1+\cos \phi}{2 \cos \phi}\ket{l_1} +\frac{i\sin \phi}{2\cos \phi}\ket{l_2} = \frac{\sqrt{2(1+\kappa)}}{2}  \ket{e_3}, \\
    &\ket{o_2} = -\frac{i\sin \phi}{2\cos \phi}\ket{l_1} + \frac{1+\cos \phi}{2 \cos \phi}\ket{l_2} =-\frac{i\sqrt{2(1+\kappa)}}{2}\ket{e_2},
  \end{aligned}
  \label{eq:radar_optimal_oj}
\end{equation}
here $\sin \phi = -\sqrt{1-\kappa^2}$, $\cos \phi = \kappa$.
In this case $\inp{o_1}{o_1} = \inp{o_2}{o_2}=\frac{1+\kappa}{2}$, $\inp{o_1}{o_2} = 0$.

To get the optimal projective measurement, we let
\begin{equation}
  \begin{aligned}
    &\ket{a_0} = \ket{\psi} = \ket{e_1}, \\
    &\ket{a_1} = \ket{e_3},\\
    &\ket{a_2} = -i\ket{e_2},\\
  \end{aligned}
\end{equation}
which form a basis for the three-dimensional subspace spanned by $\{\ket{e_1},\ket{e_2},\ket{e_3}\}$.
Again within this subspace, we can put $\{\ket{a_j}\}$ together to get a unitary matrix
\begin{equation}
  A=\left(\begin{matrix}
    1 & 0 & 0 \\
    0 & 0 & -i \\
    0 & 1 & 0
  \end{matrix}\right).
\end{equation}
We then choose a real orthogonal matrix,
\begin{equation}
  B=\left(\begin{matrix}
    \frac{1}{2} & \frac{1}{\sqrt{2}} & \frac{1}{2} \\
    \frac{1}{\sqrt{2}} & 0 & -\frac{1}{\sqrt{2}} \\
    \frac{1}{2} & -\frac{1}{\sqrt{2}} & \frac{1}{2}
  \end{matrix}\right),
\end{equation}
and let $U = B A^{-1}$. The optimal measurement can then be obtained as the projective measurement on the basis given by the rows of $U$, which are
\begin{equation}
  \begin{aligned}
    &\ket{m_1} = \frac{1}{2} \ket{\hat{e}_1} - \frac{i}{2} \ket{\hat{e}_2} +  \frac{\sqrt{2}}{2} \ket{\hat{e}_3}, \\
    &\ket{m_2} = \frac{\sqrt{2}}{2} \ket{\hat{e}_1} + \frac{i\sqrt{2}}{2} \ket{\hat{e}_2}, \\
    &\ket{m_3} = \frac{1}{2} \ket{\hat{e}_1} - \frac{i}{2} \ket{\hat{e}_2} - \frac{\sqrt{2}}{2} \ket{\hat{e}_3}, \\
  \end{aligned}
\end{equation}
with
%\begin{widetext}
\begin{equation}
  \begin{aligned}
    &\ket{\hat{e}_1} = \int dt \ \int dt_i \ \hat{\Psi}(t,t_i) \ket{t} \ket{t_i}, \\
    &\ket{\hat{e}_2} = \int dt \ \int dt_i \ \sqrt{2(1-\kappa)}\left(\sigma (t-\hat{\bar{t}}) + \sigma_i (t_i-\bar{t}_0)\right) \hat{\Psi}(t,t_i)  \ket{t} \ket{t_i},\\
    &\ket{\hat{e}_3} = \int dt \ \int dt_i \ \sqrt{2(1+\kappa)}\left(\sigma (t-\hat{\bar{t}}) - \sigma_i (t_i-\bar{t}_0)\right) \hat{\Psi}(t,t_i) \ket{t} \ket{t_i}
  \end{aligned}
\end{equation}
\begin{equation}
  \hat{\Psi}(t,t_i) = (1-\kappa^2)^{1/4}\sqrt{\frac{2\sigma \sigma_i}{\pi}} \exp\{-i \hat{\bar{\omega}}(t-\hat{\bar{t}})-i\bar{\omega}_{i}(t_i-\bar{t}_0) -(t-\hat{\bar{t}})^2 \sigma^2 - (t_i-\bar{t}_0)^2 \sigma_{i}^2 + 2 \kappa (t-\hat{\bar{t}})(t_i-\bar{t}_0)\sigma\sigma_{i}\},
\end{equation}
 %   \end{widetext}
here $\hat{\bar{t}}$ and $\hat{\bar{\omega}}$ are estimators of $\bar{t}$ and $\bar{\omega}$, respectively.
We can verify that the probabilities of the measurement results are
\begin{equation}
  \begin{aligned}
    &p_1 = |\inp{m_1}{\psi}|^2 = \left|\frac{1}{2} \inp{\hat{e}_1}{e_1} +\frac{i}{2} \inp{\hat{e}_2}{e_1} + \frac{\sqrt{2}}{2} \inp{\hat{e}_3}{e_1}\right|^2, \\
    &p_2 = |\inp{m_2}{\psi}|^2 = \left|\frac{\sqrt{2}}{2} \inp{\hat{e}_1}{e_1} - \frac{i\sqrt{2}}{2} \inp{\hat{e}_2}{e_1}\right|^2, \\
    &p_3 = |\inp{m_3}{\psi}|^2 =  \left|\frac{1}{2} \inp{\hat{e}_1}{e_1} +\frac{i}{2} \inp{\hat{e}_2}{e_1} -\frac{\sqrt{2}}{2} \inp{\hat{e}_3}{e_1}\right|^2,
  \end{aligned}
\end{equation}
which gives the classical Fisher information matrix as
\begin{equation}
  F_C = \left(\begin{matrix}
    2\sigma^2(1+\kappa) & 0 \\  0 & \frac{1}{2\sigma^2(1-\kappa)}
  \end{matrix}\right).
\end{equation}
From which we have $\sigma_{\bar{t}}^2 =\frac{1}{2\sigma^2(1+\kappa)}$ and $\sigma_{\bar{\omega}}^2 =2\sigma^2(1-\kappa)$,
%\begin{equation}
  %\end{equation}
which saturates the refined Arthurs-Kelly relation, $\sigma_{\bar{t}} \sigma_{\bar{\omega}} = \frac{\sqrt{1-\kappa}}{\sqrt{1+\kappa}}$. The constructed measurement is thus optimal.

We note that the heuristic measurement proposed by Zhuang et al.~\cite{zhuang2017} and further analyzed by Huang et al.~\cite{huang2021} can be seen as a particular choice of commuting observables $\{O_j\}$ that achieves the minimum error for approximating the reparametrized SLDs $\{L_{\bar{t}}^{'}, L_{\bar{\omega}}^{'}\}$. The detection methods in~\cite{zhuang2017,huang2021} start from an entangled signal–idler biphoton, apply an appropriate biphoton unitary, and then perform intensity measurements of two commuting observables, 
\begin{equation}
    \begin{aligned}
        W_+ &= W \otimes I_i + I_s \otimes W_i = \int d\omega \int d\omega_i (\omega + \omega_i) |\omega\rangle\langle \omega| \otimes |\omega_i\rangle\langle \omega_i|,\\
        T_- &= T\otimes I_i - I_s \otimes T_i = \int dt \int dt_i (t-t_i)|t\rangle\langle t| \otimes |t_i\rangle\langle t_i| ,
    \end{aligned}
\end{equation}
where $W = \int d\omega \omega |\omega\rangle \langle\omega|$, $ W_i = \int d\omega_i \omega_i |\omega_i\rangle \langle\omega_i|$, $T = \int dt t |t\rangle \langle t|$ and $T_i = \int dt_i t_i |t_i\rangle \langle t_i|$. Using the standard time--frequency conjugacy $[T,W]=i$ and $[T_i,W_i]=i$, and the fact that operators acting on different arms commute, one obtains $[T_-,W_+]=0$.
This constitutes a continuous-spectrum joint measurement in the common eigenbasis $\{|\omega_+, t_-\rangle\}$, with eigenvalues $\omega_+ = \omega+\omega_i$ and $t_- = t - t_i$. Following Ref.~\cite{huang2021}, we set $\sigma= \sigma_i$, $\bar{t}_{0} = 0$ and $\bar{\omega}_{i0}  = 0$. With this convention, consider the commuting observables
\begin{equation}
    \begin{aligned}
        O_1 &= \int dt \int dt_i(1+\kappa)\sigma(t - t_i -\bar{t})|t\rangle\langle t| \otimes |t_i\rangle\langle t_i| = (1+\kappa)\sigma\left( T_- - \bar{t}(I_s \otimes I_i)\right), \\
        O_2 &= \int d\omega \int d\omega_i\sqrt{\frac{1+\kappa}{1-\kappa}} \frac{2(\omega+\omega_i)-\bar{\omega}}{2\sigma}|\omega\rangle\langle \omega| \otimes |\omega_i\rangle\langle \omega_i|= \sqrt{\frac{1+\kappa}{1-\kappa}} \frac{2W_+-\bar{\omega}(I_s \otimes I_i)}{2\sigma},
    \end{aligned}
\end{equation}
which gives
\begin{equation}
    \begin{aligned}
        |o_1\rangle &= O_1|\Psi\rangle = \int dt \int dt_i (1+\kappa)\sigma(t_- -\bar{t}) \Psi(t, t_i) |t\rangle |t_i\rangle = \frac{\sqrt{2(1+\kappa)}}{2}|e_3\rangle,\\
        |o_2\rangle &= O_2|\Psi\rangle =-i \int dt \int dt_i \sqrt{1-\kappa^2}\sigma(t_+ -\bar{t}) \Psi(t, t_i) |t\rangle |t_i\rangle  = -\frac{i\sqrt{2(1+\kappa)}}{2}|e_2\rangle.
    \end{aligned}  
\end{equation}
These coincide with the optimal $\{|o_1\rangle, |o_2\rangle\}$ constructed in Eq.~(\ref{eq:radar_optimal_oj}), and saturate the lower bound in Eq.(\ref{eq:minimum_error}). Moreover, $O_1$ depends only on $T_-$ and $O_2$ depends only on $W_+$, hence they commute and share the same eigenbasis $\{|\omega_+, t_-\rangle\}$. This recovers the measurement in Refs.~\cite{zhuang2017,huang2021}. In this case, the measurement has a continuous-spectrum outcome. While our previous construction makes use of the fact that all information is encoded in an effective 3-dimensional subspace, which leads to optimal measurements with 3 outcomes that are sufficient to estimate two parameters.

\section{Demonstration on a cloud-based superconducting quantum computer}
\label{sec:exp_theo_circuit}
In this section, we present demonstrations of two estimation tasks on a cloud superconducting quantum computer, one example for the estimation of two parameters and the other for the estimation of three parameters.

The demonstrations are conducted using the ``ScQ-P136''(``Baiwang'') backend on the ``Quafu'' cloud superconducting quantum computing platform, where the fidelity of single-qubit gate operations exceeds 99\%.
Figure \ref{ScQ_P136} shows the layout of the ScQ-P136 device. 
\begin{figure}[htbp]
  \centering
  \includegraphics[width=0.8\textwidth]{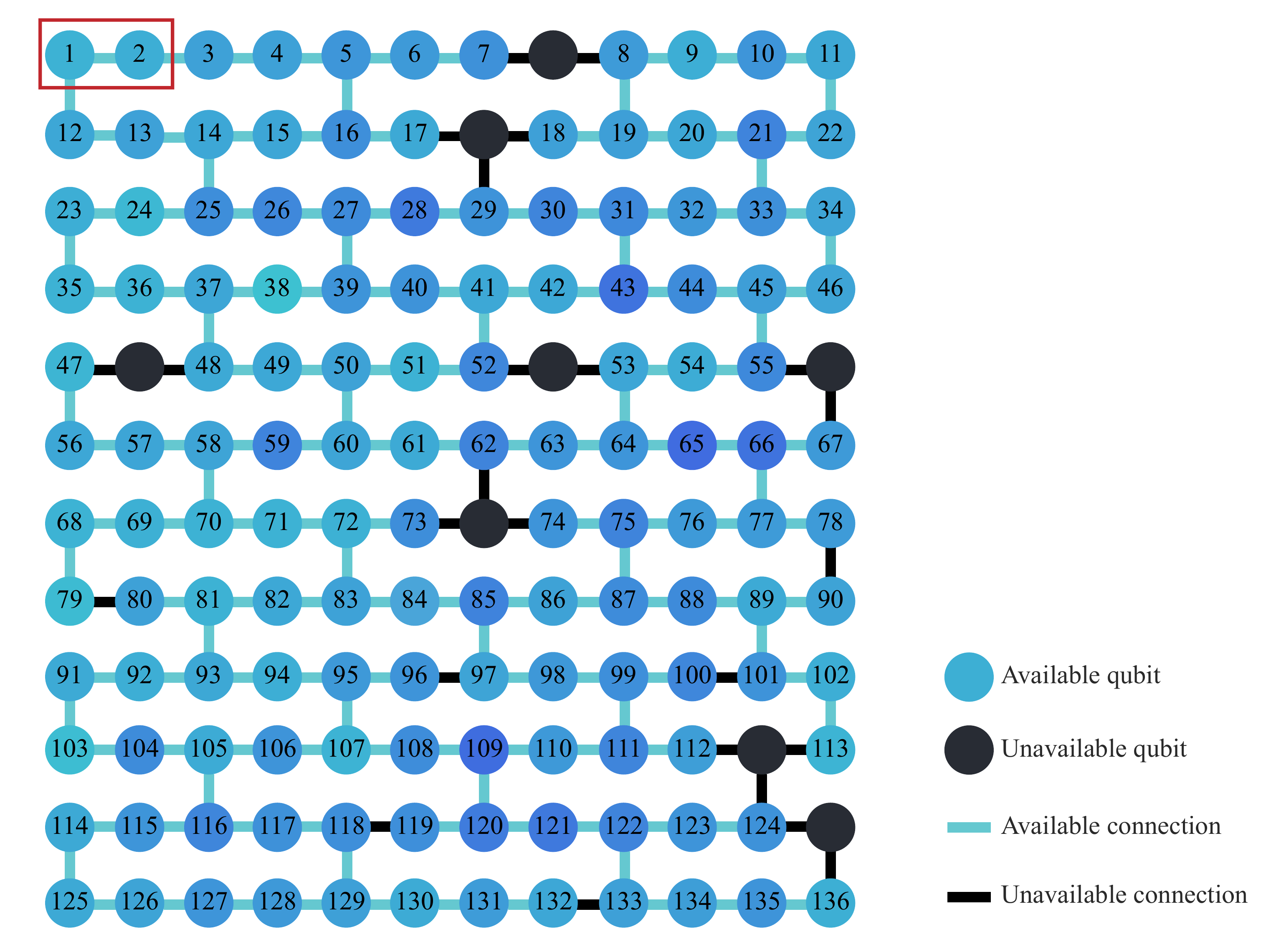}
  \caption{Diagram of ``ScQ-P136''(``Baiwang'') processor on ``Quafu'' cloud quantum computation platform.}
  \label{ScQ_P136}
\end{figure}
In the quantum circuit shown in Fig.~\ref{fig:circuits_2para} and Fig.~\ref{fig:circuits_3para}, qubits $q_0$ and $q_1$ correspond to the physical qubits indexed as $1$ and $2$ on the ScQ-P136, respectively.
Additional details about these qubits are provided in Table \ref{table}.

\begin{table}[htbp]
  \centering % Centres the table on the page, comment out to left-justify
  \begin{tabular}{p{6cm} p{3cm} p{3cm}}
  \toprule % Top horizontal line
  \textbf{Qubit} & 1 & 2\\ 
  \midrule % In-table horizontal line
  Relaxation time $T_1(\mu s)$ & 22.8 & 25 \\ 
  Coherence time $T_2(\mu s)$ & 19.9 & 14.65 \\
  % Anharmonicity $(MHz)$ & 0.294 & 0.291\\
  Qubit frequency $(GHz)$ & 4.524 & 4.652\\
  Readout frequency $(GHz)$ & 6.832 & 6.882\\
  Average fidelity of CNOT gate & 0.985 & 0.984\\
  \bottomrule % Bottom horizontal line
  \end{tabular}
  \caption{The information list of two qubits used in the demonstration.} 
  \label{table} 
  \end{table}

\subsection{Estimation of two parameters}
We first consider the simultaneous estimation of two rotational parameters $(\alpha, \theta)$, encoded in a single-qubit state prepared as 
\begin{equation} \ket{\psi} = R_z(-\alpha) R_y(-\theta) \sigma_x \ket{0}, \end{equation}
where $R_i(\theta) = e^{- \frac{i}{2} \theta \sigma_i}$ are the rotation operators with $i\in\{x,y,z\}$, and $\sigma_{x,y,z}$ denote the Pauli X, Y, and Z gates, respectively.
Explicitly, the state $\ket{\psi}$ can be written as
\begin{equation}
    \ket{\psi} = \left(\begin{matrix}
  e^{\frac{i \alpha}{2}} \sin \frac{\theta}{2} \\ e^{-\frac{i \alpha}{2}} \cos \frac{\theta}{2}
\end{matrix}\right).
\end{equation}
The corresponding SLDs for the parameters $\alpha$ and $\theta$ are given by
\begin{equation}
  L_{\alpha} = \left(\begin{matrix}
    0 & i e^{i \alpha} \sin \theta \\ -i e^{-i \alpha} \sin \theta & 0
  \end{matrix}\right), \quad
  L_{\theta} = \left(\begin{matrix}
    \sin \theta & e^{i \alpha} \cos \theta \\ e^{-i \alpha} \cos \theta & -\sin \theta
  \end{matrix}\right).
\end{equation}
The QFIM can then be obtained as
%\begin{equation}
 $ F_Q = \left(\begin{matrix}
    \sin^2 \theta & 0 \\ 0 & 1
  \end{matrix}\right),$ 
%\end{equation}
%and $C_{\alpha \theta} = $\operatorname{Im} \langle L_{\alpha} L_{\theta} \rangle = -2\sin 2\theta$.
which is not Identity. We thus first make a reparametrization as
\begin{equation}
  \left(\begin{matrix}
    \alpha^{\prime} \\ \theta^{\prime}
  \end{matrix}\right) = F_Q^{\frac{1}{2}} \left(\begin{matrix}
    \alpha \\ \theta
  \end{matrix}\right),
\end{equation}
under which the SLDs become
\begin{equation}
  L_{\alpha}^{\prime} = \frac{L_{\alpha}}{\sqrt{\mathcal{F_{\alpha\alpha}}}} = \left(\begin{matrix}
    0 & i e^{i \alpha} \\ -i e^{-i \alpha} & 0
  \end{matrix}\right), \quad
  L_{\theta}^{\prime} =\frac{L_{\theta}}{\sqrt{\mathcal{F_{\theta\theta}}}} = \left(\begin{matrix}
    \sin \theta & e^{i \alpha} \cos \theta \\ e^{-i \alpha} \cos \theta & -\sin \theta
  \end{matrix}\right).
\end{equation}
%which are the SLDs under the reparametrization
%We have $|\langle L_{\alpha}^{\prime} L_{\theta}^{\prime} \rangle|=1$, where $\operatorname{Im}\langle L_{\alpha}^{\prime} L_{\theta}^{\prime} \rangle=\sin \phi = -1$, $\phi = -\frac{\pi}{2}$.
In this case $\beta=\operatorname{Im}\langle L_{\alpha}^{\prime} L_{\theta}^{\prime} \rangle = -1$, from which we obtained the minimal tradeoff relation as
\begin{equation}
  \operatorname{Tr}(F_Q^{-1} F_C) = 1,
\end{equation}
which coincides with the Gill-Massar bound.

We then provide two optimal measurements that saturate this bound.  

First, we have 
\begin{equation}
  \begin{aligned}
    |l_1\rangle = L_{\alpha}^{\prime} \otimes I |\psi\rangle |0\rangle = \left(\begin{matrix}
      i e^{\frac{i \alpha}{2}} \cos \frac{\theta}{2} \\ 0 \\ -i e^{-\frac{i \alpha}{2}} \sin \frac{\theta}{2} \\ 0
    \end{matrix}\right), \\
    |l_2\rangle = L_{\theta}^{\prime} \otimes I|\psi\rangle |0\rangle = \left(\begin{matrix}
      e^{\frac{i \alpha}{2}} \cos \frac{\theta}{2} \\ 0 \\ - e^{-\frac{i \alpha}{2}} \sin \frac{\theta}{2} \\0
    \end{matrix}\right). \\
  \end{aligned}
\end{equation}
Since $\beta = -1$, we choose
\begin{equation}
  \begin{aligned}
    |l_\perp\rangle = |\psi\rangle |1\rangle = \left(\begin{matrix}
      0 \\ i e^{\frac{i \alpha}{2}} \cos \frac{\theta}{2} \\ 0 \\ -i e^{-\frac{i \alpha}{2}} \sin \frac{\theta}{2}
    \end{matrix}\right),
  \end{aligned}
\end{equation}
which satisfies $\inp{l_1}{l_{\perp}}=\inp{l_2}{l_{\perp}}  = 0$ and $\inp{l_{\perp}}{l_{\perp}}=1$.
With Eq.~(\ref{construct_beta1}) we can then obtain the optimal $\{\ket{o_1}, \ket{o_2}\}$ as (with $\varphi$ taken as $0$)
\begin{equation}
  \begin{aligned}
    \ket{o_1} &= \frac{1}{2} |l_1\rangle - \frac{i}{2} \ket{l_{\perp}}=\frac{1}{2}
    \left(\begin{matrix}
      i e^{\frac{i \alpha}{2}} \cos \frac{\theta}{2} \\ 
      - i e^{\frac{i \alpha}{2}} \sin \frac{\theta}{2} \\ 
      -i e^{-\frac{i \alpha}{2}} \sin \frac{\theta}{2} \\
      -i e^{-\frac{i \alpha}{2}} \cos \frac{\theta}{2}
    \end{matrix}\right),\\
    \ket{o_2} &= -\frac{i}{2} |l_1\rangle + \frac{1}{2} \ket{l_{\perp}}= \frac{1}{2} 
    \left(\begin{matrix}
      e^{\frac{i \alpha}{2}} \cos \frac{\theta}{2} \\ 
      e^{\frac{i \alpha}{2}} \sin \frac{\theta}{2} \\ 
      - e^{-\frac{i \alpha}{2}} \sin \frac{\theta}{2}\\
      e^{-\frac{i \alpha}{2}} \cos \frac{\theta}{2}
    \end{matrix}\right).
  \end{aligned}
\end{equation}
It is easy to compute that $\inp{o_1}{o_1}=\inp{o_2}{o_2} = \frac{1}{2}$, $\inp{o_1}{o_2} = 0$. To get the optimal measurement, we let
\begin{equation}
  \begin{aligned}
    &|a_1\rangle = |\psi\rangle |0\rangle,  \\
    &|a_2\rangle = \frac{|o_1\rangle}{ \sqrt{\inp{o_1}{o_1}}}, \\
    &|a_3\rangle = \frac{|o_2\rangle}{ \sqrt{\inp{o_2}{o_2}}}. \\
  \end{aligned}
\end{equation}
and choose an additional vector $|a_4\rangle=\left(\begin{matrix}
      0 \\ 
       -e^{\frac{i \alpha}{2}} \cos \frac{\theta}{2} \\ 
      0\\
      e^{-\frac{i \alpha}{2}} \sin \frac{\theta}{2}
    \end{matrix}\right)$ to make a complete basis. Put these basis together, we get a unitary matrix 
\begin{equation}
  A = \left(\begin{matrix}
    e^{\frac{i \alpha}{2}} \sin \frac{\theta}{2} & \frac{i e^{\frac{i \alpha}{2}} \cos \frac{\theta}{2}}{\sqrt{2}} & \frac{e^{\frac{i \alpha}{2}} \cos \frac{\theta}{2}}{\sqrt{2}} & 0\\
  0 & -\frac{ i e^{\frac{i \alpha}{2}} \sin \frac{\theta}{2}}{\sqrt{2}} & \frac{e^{\frac{i \alpha}{2}} \sin \frac{\theta}{2}}{\sqrt{2}} &  -e^{\frac{i \alpha}{2}} \cos \frac{\theta}{2}\\
  e^{-\frac{i \alpha}{2}} \cos \frac{\theta}{2} & -\frac{i e^{\frac{-i \alpha}{2}} \sin \frac{\theta}{2}}{\sqrt{2}} & -\frac{e^{-\frac{i \alpha}{2}} \sin \frac{\theta}{2}}{\sqrt{2}} & 0 \\
  0 & -\frac{i e^{-\frac{i \alpha}{2}} \cos \frac{\theta}{2}}{\sqrt{2}} & -\frac{e^{-\frac{i \alpha}{2}} \cos \frac{\theta}{2}}{\sqrt{2}} & e^{-\frac{i \alpha}{2}} \sin \frac{\theta}{2}
  \end{matrix}\right).
\end{equation}
One optimal measurement can be obtained by choosing a real orthogonal matrix as 
\begin{equation}
  B = \left(\begin{matrix}
  \frac{1}{2\sqrt{3}} & -\frac{1}{2} & \frac{1}{\sqrt{6}} & -\frac{1}{\sqrt{2}}\\
  \frac{1}{2} & \frac{1}{2\sqrt{3}}   & -\frac{1}{\sqrt{2}} & - \frac{1}{\sqrt{6}}\\
  \frac{1}{\sqrt{6}} & -\frac{1}{\sqrt{2}} & \frac{1}{2\sqrt{3}}  & -\frac{1}{2} \\
  \frac{1}{\sqrt{2}} & \frac{1}{\sqrt{6}} & \frac{1}{2} & \frac{1}{2\sqrt{3}} 
  \end{matrix}\right).
\end{equation}
The optimal project measurement in this case, denoted as $\mathcal{M}_1^{opt} =\{\ket{m_1},\ket{m_2},\ket{m_3},\ket{m_4}\}$ which corresponds to the rows of $U = B A^{-1}$, is %give the basis for the projective measurement.
{\small
\begin{equation}
  \begin{aligned}
    &\ket{m_1} = \left(\begin{matrix}
      -\frac{1}{12} e^{\frac{i \hat{\alpha}}{2}} \left( ( 2\sqrt{3}+i 3 \sqrt{2})\cos \frac{\hat{\theta}}{2}- 2 \sqrt{3}\sin \frac{\hat{\theta}}{2} \right) \\
      \frac{1}{12} e^{\frac{i \hat{\alpha}}{2}} \left( -6\sqrt{2}\cos \frac{\hat{\theta}}{2}- (2\sqrt{3}- i 3 \sqrt{2})\sin \frac{\hat{\theta}}{2} \right) \\
      \frac{1}{12} e^{-\frac{i \hat{\alpha}}{2}} \left( 2\sqrt{3}\cos \frac{\hat{\theta}}{2}+ (2\sqrt{3}+i 3 \sqrt{2})\sin \frac{\hat{\theta}}{2} \right)\\
      -\frac{1}{12} e^{-\frac{i \hat{\alpha}}{2}} \left( (2\sqrt{3}- i 3 \sqrt{2})\cos \frac{\hat{\theta}}{2}- 6 \sqrt{2}\sin \frac{\hat{\theta}}{2} \right)
   \end{matrix}\right), \quad
    \ket{m_2} = \left(\begin{matrix}
      -\frac{1}{12} e^{\frac{i \hat{\alpha}}{2}} \left( ( 6-i\sqrt{6})\cos \frac{\hat{\theta}}{2}- 6\sin \frac{\hat{\theta}}{2} \right) \\
      \frac{1}{12} e^{\frac{i \hat{\alpha}}{2}} \left( 2\sqrt{6}\cos \frac{\hat{\theta}}{2}- (6+i\sqrt{6})\sin \frac{\hat{\theta}}{2} \right) \\
      \frac{1}{12} e^{-\frac{i \hat{\alpha}}{2}} \left( 6 \cos \frac{\hat{\theta}}{2}+ (6-i \sqrt{6})\sin \frac{\hat{\theta}}{2} \right)\\
      \frac{1}{12} e^{-\frac{i \hat{\alpha}}{2}} \left( (6 + i\sqrt{6})\cos \frac{\hat{\theta}}{2} + 2 \sqrt{6} \sin \frac{\hat{\theta}}{2} \right)
    \end{matrix}\right), \\
    &\ket{m_3} = \left(\begin{matrix}
      \frac{1}{12} e^{\frac{i \hat{\alpha}}{2}} \left( ( \sqrt{6}-i 6)\cos \frac{\hat{\theta}}{2}+ 2\sqrt{6}\sin \frac{\hat{\theta}}{2} \right) \\
      \frac{1}{12} e^{\frac{i \hat{\alpha}}{2}} \left( 6\cos \frac{\hat{\theta}}{2}+ (\sqrt{6}+i 6)\sin \frac{\hat{\theta}}{2} \right) \\
      \frac{1}{12} e^{-\frac{i \hat{\alpha}}{2}} \left( 2\sqrt{6} \cos \frac{\hat{\theta}}{2}- (\sqrt{6}-i 6)\sin \frac{\hat{\theta}}{2} \right)\\
      \frac{1}{12} e^{-\frac{i \hat{\alpha}}{2}} \left( (\sqrt{6} + i6)\cos \frac{\hat{\theta}}{2} - 6 \sin \frac{\hat{\theta}}{2} \right)
   \end{matrix}\right), \quad
    \ket{m_4} = \left(\begin{matrix}
      \frac{1}{12} e^{\frac{i \hat{\alpha}}{2}} \left( ( 3\sqrt{2}+i 2 \sqrt{3})\cos \frac{\hat{\theta}}{2}+ 6 \sqrt{2}\sin \frac{\hat{\theta}}{2} \right) \\
      \frac{1}{12} e^{\frac{i \hat{\alpha}}{2}} \left( -2\sqrt{3}\cos \frac{\hat{\theta}}{2}+ (3\sqrt{2}- i 2 \sqrt{3})\sin \frac{\hat{\theta}}{2} \right) \\
      \frac{1}{12} e^{-\frac{i \hat{\alpha}}{2}} \left( 6\sqrt{2}\cos \frac{\hat{\theta}}{2}- (3\sqrt{2}+i 2 \sqrt{3})\sin \frac{\hat{\theta}}{2} \right)\\
      \frac{1}{12} e^{-\frac{i \hat{\alpha}}{2}} \left( (3\sqrt{2}- i 2 \sqrt{3})\cos \frac{\hat{\theta}}{2}+ 2 \sqrt{3}\sin \frac{\hat{\theta}}{2} \right)
    \end{matrix}\right).
  \end{aligned}
\end{equation}}
here we use $\hat{\alpha}$ and $\hat{\theta}$ to denote the estimated values of $\alpha, \theta$, which needs to be adaptively updated in practice when the values are not known a priory.

By choosing a different $B$ as
\begin{equation}
  B = \left(\begin{matrix}
  \frac{1}{2} &\frac{1}{2} & \frac{1}{2} & -\frac{1}{2}\\
  \frac{1}{2} & -\frac{1}{2}  & \frac{1}{2} & \frac{1}{2}\\
  \frac{1}{2} & -\frac{1}{2} & -\frac{1}{2} & -\frac{1}{2} \\
  \frac{1}{2} & \frac{1}{2} & -\frac{1}{2} & \frac{1}{2}
  \end{matrix}\right).
\end{equation}
we can obtain another optimal measurement $M_2^{opt} = \{\ket{m^{'}_1},\ket{m^{'}_2},\ket{m^{'}_3},\ket{m^{'}_4}\}$ as 
\begin{equation}
  \begin{aligned}
    &\ket{m^{'}_1} = \left(\begin{matrix}
      \frac{1}{2} e^{\frac{i \hat{\alpha}}{2}} \left( \frac{(1+i)\cos \frac{\hat{\theta}}{2}}{\sqrt{2}} + \sin \frac{\hat{\theta}}{2} \right) \\
      \frac{1}{2} e^{\frac{i \hat{\alpha}}{2}} \left( \frac{(1+i)\cos \frac{\hat{\theta}}{2}}{\sqrt{2}} + \sin \frac{\hat{\theta}}{2} \right) \\
      \frac{1}{2} e^{-\frac{i \hat{\alpha}}{2}} \left(\cos \frac{\hat{\theta}}{2} - \frac{(1+i)\sin \frac{\hat{\theta}}{2}}{\sqrt{2}} \right)\\
      \frac{1}{2} e^{-\frac{i \hat{\alpha}}{2}} \left(\cos \frac{\hat{\theta}}{2} - \frac{(1+i)\sin \frac{\hat{\theta}}{2}}{\sqrt{2}} \right)
   \end{matrix}\right), \quad
    \ket{m^{'}_2} = \left(\begin{matrix}
      \frac{1}{2} e^{\frac{i \hat{\alpha}}{2}} \left( \frac{(1-i)\cos \frac{\hat{\theta}}{2}}{\sqrt{2}} + \sin \frac{\hat{\theta}}{2} \right) \\
      -\frac{1}{2} e^{\frac{i \hat{\alpha}}{2}} \left( \frac{(1-i)\cos \frac{\hat{\theta}}{2}}{\sqrt{2}} + \sin \frac{\hat{\theta}}{2} \right) \\
      \frac{1}{2} e^{-\frac{i \hat{\alpha}}{2}} \left(\cos \frac{\hat{\theta}}{2} - \frac{(1-i)\sin \frac{\hat{\theta}}{2}}{\sqrt{2}} \right)\\
      -\frac{1}{2} e^{-\frac{i \hat{\alpha}}{2}} \left(\cos \frac{\hat{\theta}}{2} - \frac{(1-i)\sin \frac{\hat{\theta}}{2}}{\sqrt{2}} \right)
    \end{matrix}\right), \\
    &\ket{m^{'}_3} = \left(\begin{matrix}
      \frac{1}{2} e^{\frac{i \hat{\alpha}}{2}} \left( -\frac{(1+i)\cos \frac{\hat{\theta}}{2}}{\sqrt{2}} + \sin \frac{\hat{\theta}}{2} \right) \\
      \frac{1}{2} e^{\frac{i \hat{\alpha}}{2}} \left( -\frac{(1+i)\cos \frac{\hat{\theta}}{2}}{\sqrt{2}} + \sin \frac{\hat{\theta}}{2} \right) \\
      \frac{1}{2} e^{-\frac{i \hat{\alpha}}{2}} \left(\cos \frac{\hat{\theta}}{2} + \frac{(1+i)\sin \frac{\hat{\theta}}{2}}{\sqrt{2}} \right)\\
      \frac{1}{2} e^{-\frac{i \hat{\alpha}}{2}} \left(\cos \frac{\hat{\theta}}{2} + \frac{(1+i)\sin \frac{\hat{\theta}}{2}}{\sqrt{2}} \right)
   \end{matrix}\right), \quad
    \ket{m^{'}_4} = \left(\begin{matrix}
      -\frac{1}{2} e^{\frac{i \hat{\alpha}}{2}} \left( \frac{(1-i)\cos \frac{\hat{\theta}}{2}}{\sqrt{2}} - \sin \frac{\hat{\theta}}{2} \right) \\
      \frac{1}{2} e^{\frac{i \hat{\alpha}}{2}} \left( \frac{(1-i)\cos \frac{\hat{\theta}}{2}}{\sqrt{2}} - \sin \frac{\hat{\theta}}{2} \right) \\
      \frac{1}{2} e^{-\frac{i \hat{\alpha}}{2}} \left(\cos \frac{\hat{\theta}}{2} + \frac{(1-i)\sin \frac{\hat{\theta}}{2}}{\sqrt{2}} \right)\\
      -\frac{1}{2} e^{-\frac{i \hat{\alpha}}{2}} \left(\cos \frac{\hat{\theta}}{2} + \frac{(1-i)\sin \frac{\hat{\theta}}{2}}{\sqrt{2}} \right)
    \end{matrix}\right).
  \end{aligned}
\end{equation}
where $\hat{\alpha}$ and $\hat{\theta}$ are the adaptively updated parameter estimates.

Figure~\ref{fig:circuits_2para} shows the quantum circuits for the estimation of two parameters, which consist of the preparation of the state and the implementation of the optimal measurements, where Figures~\ref{fig:sub1} corresponds to $\mathcal{M}_1^{opt}$ and Figures~\ref{fig:sub2} corresponds to $\mathcal{M}_2^{opt}$. Here $\mathcal{M}_2^{opt}$ is designed to reduce the circuit depth compared to $\mathcal{M}_1^{opt}$. 
The quantum processor used in our demonstrations performs projective measurements in the $z$ basis. Applying the unitary transformation $U=B A^{-1}$ before the measurement allows the optimal measurement to be implemented through standard projective measurements in the $z$ basis. This two-qubit unitary $U$ is decomposed into elementary gates using the "Qiskit" package~\cite{qiskit}. Specifically, the "TwoQubitBasisDecomposer" class is employed to perform the decomposition with the minimal number of CNOT gates and single-qubit operations implemented through using sequences of $R_z$ and $R_y$ rotations.
The circuit implementing $\mathcal{M}_1^{opt}$ requires 3 CNOT gates and 22 single-qubit rotation gates, whereas the implementation of $\mathcal{M}_2^{opt}$ requires only 2 CNOT gates and 14 single-qubit rotations, which results in a higher fidelity and better performance in the presence of platform noise.

\begin{figure}[hpbt]
  \centering  % Center the figure
    \subfigure[]{
    \centering
    \includegraphics[scale=0.38]{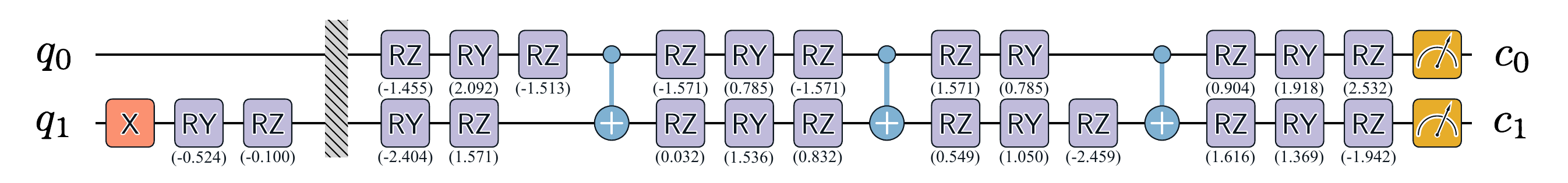}
    \label{fig:sub1}
}
  \hfill
    \subfigure[]{
    \centering
    \includegraphics[scale=0.38]{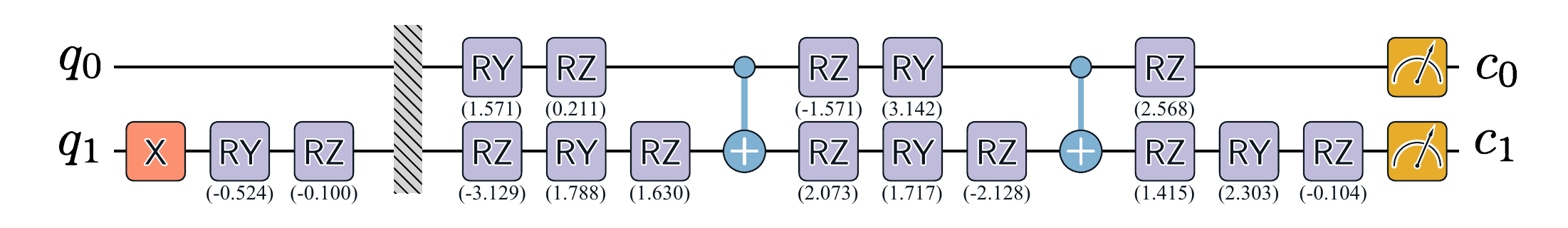}
    \label{fig:sub2}
}
  
  \caption{Quantum circuits used in the two-parameter estimation task.  
  Each circuit is read from left to right. The first qubit is denoted $q_1$ and the second qubit is denoted $q_0$. In (a)(b), $q_0$ acts as the ancilla qubit. The true value of two parameters are $\alpha = 0.1$, $\theta = \frac{\pi}{6}$. The circuit diagram was constructed using PYQUAFU package. (a) Quantum circuit of estimating two parameters with optimal measurement $\mathcal{M}_1^{opt}$ and (b) Quantum circuit of estimating two parameters with optimal measurement $\mathcal{M}_2^{opt}$.}
\label{fig:circuits_2para}
\end{figure}

\subsection{Estimation of three parameters}

We provide another example on the simultaneous estimation of three parameters, $(\alpha_1, \theta_1, \gamma_1)$, encoded in a two-qubit state,
\begin{equation} 
\ket{\psi(\alpha_1, \theta_1, \gamma_1)} = (I_2 \otimes R_z(\alpha_1)), \mathrm{CNOT}, (H R_y(\theta_1) \otimes R_y(\gamma_1)) \ket{00}, 
\end{equation} 
here $H$ denotes the Hadamard gate and CNOT denotes the controlled-NOT gate, with the first qubit serving as the control qubit.

The parameterized two-qubit pure state can be explicitly written as
\begin{equation}
    \ket{\psi} = \left(\begin{matrix}
  \frac{1}{\sqrt{2}} e^{- \frac{i \alpha_1}{2}} \cos \frac{\gamma_1}{2} \left(\cos \frac{\theta_1}{2} -\sin \frac{\theta_1}{2}\right) \\ 
  \frac{1}{\sqrt{2}} e^{\frac{i \alpha_1}{2}} \sin \frac{\gamma_1}{2} \left(\cos \frac{\theta_1}{2} -\sin \frac{\theta_2}{2}\right) \\
  \frac{1}{\sqrt{2}} e^{- \frac{i \alpha_1}{2}} \sin \frac{\gamma_1}{2} \left(\cos \frac{\theta_1}{2} +\sin \frac{\theta_1}{2}\right) \\
  \frac{1}{\sqrt{2}} e^{\frac{i \alpha_1}{2}} \cos \frac{\gamma_1}{2} \left(\cos \frac{\theta_1}{2} +\sin \frac{\theta_1}{2}\right).
\end{matrix}\right)
\end{equation}
The corresponding SLDs, denoted as $L_{\alpha_1}, L_{\theta_1}$ and $L_{\gamma_1}$, can be obtained from which we can get $F_Q$ and $F_{\operatorname{Im}}$ as
\begin{equation}
  F_Q = \left(\begin{matrix}
    \frac{1}{4}(3 + \cos 2 \theta_1 - 2 \cos 2 \gamma_1 \sin^2 \theta_1) &  0 & 0  \\ 0 &1 & 0 \\ 0& 0& 1
  \end{matrix}\right), \quad
  F_{\operatorname{Im}}= \left(\begin{matrix}
    0 &  -\cos \theta_1 \cos \gamma_1& \sin\theta_1 \sin \gamma_1  \\ \cos \theta_1 \cos \gamma_1 & 0 & 0 \\ -\sin\theta_1 \sin \gamma_1& 0& 0
  \end{matrix}\right).
\end{equation}  
Since $F_Q$ is not Identity, we first make a parameterization with
\begin{equation}
  \left(\begin{array}{l}
\alpha_1^{\prime} \\
\theta_1^{\prime}\\
\gamma_1^{\prime}
\end{array}\right)=F_Q^{\frac{1}{2}}\left(\begin{array}{l}
\alpha_1 \\
\theta_1\\
\gamma_1
\end{array}\right),
\end{equation}
under which \begin{equation}
  L_{\alpha_1}^{\prime} = \frac{L_{\alpha_1}}{\sqrt{F_{\alpha_1 \alpha_1}}}; \quad L_{\theta_1}^{\prime} = L_{\theta_1}; \quad L_{\gamma_1}^{\prime} = L_{\gamma_1},
\end{equation} $\tilde{F}_Q = I$, and \begin{equation}
  \tilde{F}_{\operatorname{Im}} = \left(\begin{matrix}
    0 &  -\frac{2\cos \theta_1 \cos \gamma_1}{\sqrt{3+\cos 2\theta_1 - 2 \sin^2 \theta_1 \cos 2\gamma_1}}& \frac{2\sin\theta_1 \sin \gamma_1}{\sqrt{3+\cos 2\theta_1 - 2 \sin^2 \theta_1 \cos 2\gamma_1}}  \\ 
    \frac{2\cos \theta_1 \cos \gamma_1}{\sqrt{3+\cos 2\theta_1 - 2 \sin^2 \theta_1 \cos 2\gamma_1}} & 0 & 0 \\
    -\frac{2\sin\theta_1 \sin \gamma_1}{\sqrt{3+\cos 2\theta_1 - 2 \sin^2 \theta_1 \cos 2\gamma_1}}& 0& 0
  \end{matrix}\right).
\end{equation}
The eigenvalues of $\tilde{F}_{\operatorname{Im}}$ are $\pm i \frac{\sqrt{2+ 2\cos 2\theta_1 \cos 2 \gamma_1}}{\sqrt{3+\cos 2\theta_1 - 2 \sin^2 \theta_1 \cos 2\gamma_1}}$ and 0, which give the tradeoff relation  
\begin{equation}
  \operatorname{Tr}(F_Q^{-1} F_C) \leq \frac{1}{2} \sum_j (1+\sqrt{1-|\lambda_j|^2}) = 2 + \frac{2\left|\cos \theta_1 \sin \gamma_1 \right|}{\sqrt{3+\cos 2\theta_1 - 2 \sin^2 \theta_1 \cos 2\gamma_1}}.
\end{equation}

We then perform another reparametrization
$\left(\begin{array}{l}
\alpha_1^{\prime \prime} \\
\theta_1^{\prime \prime}\\
\gamma_1^{\prime \prime}
\end{array}\right)= P \left(\begin{array}{l}
  \alpha_1^{\prime} \\
  \theta_1^{\prime}\\
  \gamma_1^{\prime}
\end{array}\right)$ with
\begin{equation}
  P = \left(\begin{matrix}
    0 &  -\frac{\sqrt{2} \cos \theta_1 \cos \gamma_1 }{\sqrt{1+\cos 2\theta_1 \cos 2\gamma_1}}  & \frac{\sqrt{2} \sin \theta_1 \sin \gamma_1 }{\sqrt{1+\cos 2\theta_1 \cos 2\gamma_1}}  \\ 
    1 & 0 & 0 \\
    0&  -\frac{\sqrt{2}\left|\sin \theta_1 \sin \gamma_1 \right|}{\sqrt{1+\cos 2\theta_1 \cos 2\gamma_1}} & -\frac{\sqrt{2} \cot \theta_1 \cot \gamma_1\left|\sin \theta_1 \sin \gamma_1 \right|}{\sqrt{1+\cos 2\theta_1 \cos 2\gamma_1}} \\
  \end{matrix}\right),
\end{equation}
to transform $P \tilde{F}_{\operatorname{Im}} P^{T}$ into the block diagonal form as
\begin{equation}
  P \tilde{F}_{\operatorname{Im}} P^{T} = \left(\begin{matrix}
    0 &  -\frac{\sqrt{2+2\cos 2\theta_1 \cos 2\gamma_1}}{\sqrt{3+\cos 2\theta_1 - 2 \sin^2 \theta_1 \cos 2\gamma_1}}& 0  \\ 
    \frac{\sqrt{2+2\cos 2\theta_1 \cos 2\gamma_1}}{\sqrt{3+\cos 2\theta_1 - 2 \sin^2 \theta_1 \cos 2\gamma_1}} & 0 & 0 \\
    0& 0& 0
  \end{matrix}\right).
\end{equation}
After the reparametrization, the SLDs become 
\begin{equation}
  \begin{aligned}
    &L_{\alpha_1}^{\prime \prime} = -\frac{\sqrt{2} \cos \theta_1 \cos \gamma_1 }{\sqrt{1+\cos 2\theta_1 \cos 2\gamma_1}} L_{\theta_1}^{\prime} + \frac{\sqrt{2} \sin \theta_1 \sin \gamma_1 }{\sqrt{1+\cos 2\theta_1 \cos 2\gamma_1}} L_{\gamma_1}^{\prime},\\
    &L_{\theta_1}^{\prime \prime} = L_{\alpha_1}^{\prime},\\
    &L_{\gamma_1}^{\prime \prime} = -\frac{\sqrt{2}\left|\sin \theta_1 \sin \gamma_1 \right|}{\sqrt{1+\cos 2\theta_1 \cos 2\gamma_1}} L_{\theta_1}^{\prime} -\frac{\sqrt{2} \cot \theta_1 \cot \gamma_1\left|\sin \theta_1 \sin \gamma_1 \right|}{\sqrt{1+\cos 2\theta_1 \cos 2\gamma_1}} L_{\gamma_1}^{\prime}.
  \end{aligned}
\end{equation}
To construct optimal measurement, we first let
\begin{equation}
  \begin{aligned}
    &|l_1\rangle = L_{\alpha_1}^{\prime \prime}|\psi\rangle, \\
    &|l_2\rangle = L_{\theta_1}^{\prime \prime}|\psi\rangle, \\
    &|l_3\rangle = L_{\gamma_1}^{\prime \prime}|\psi\rangle.
  \end{aligned}
\end{equation}
Note that ancillary system is not necessary in this case. We then construct the optimal $\{\ket{o_1}, \ket{o_2}, \ket{o_3}\}$ as in Eq.~(\ref{construct_beta_less_1})
\begin{equation}
  \begin{aligned}
    & \left|o_1\right\rangle=\frac{1+\cos \phi}{2 \cos \phi} \ket{l_1} + \frac{i \sin \phi}{2 \cos \phi} \ket{l_2}),\\
    & \left|o_2\right\rangle=- \frac{i \sin \phi}{2 \cos \phi}\ket{l_1} + \frac{1+\cos \phi}{2 \cos \phi} \ket{l_2}, \\
    & \left|o_3\right\rangle = \ket{l_3}.
\end{aligned}
\end{equation}
here $\sin \phi = -\frac{\sqrt{2+2\cos 2\theta_1 \cos 2\gamma_1}}{\sqrt{3+\cos 2\theta_1 - 2 \sin^2 \theta_1 \cos 2\gamma_1}}$, $\cos \phi = \frac{2\left|\cos \theta_1 \sin \gamma_1\right|}{\sqrt{3+\cos 2\theta_1 - 2 \sin^2 \theta_1 \cos 2\gamma_1}}$.
It is easy to see that 
\begin{equation}
  \inp{o_1}{o_1} = \inp{o_2}{o_2} = \frac{1}{2}+\frac{\left|\cos \theta_1 \sin \gamma_1 \right|}{\sqrt{3+\cos 2\theta_1 - 2 \sin^2 \theta_1 \cos 2\gamma_1}}; \quad \inp{o_1}{o_2} = 0.
\end{equation}
To construct the optimal measurement, we let
\begin{equation}
  \begin{aligned}
    &|a_1\rangle = |\psi\rangle, \\
    &|a_2\rangle = \frac{|o_1\rangle}{ \sqrt{\inp{o_1}{o_1}}}, \\
    &|a_3\rangle = \frac{|o_2\rangle}{ \sqrt{\inp{o_2}{o_2}}},
  \end{aligned}
\end{equation}
and add an additional orthonormal vector $|a_4\rangle = \left(\begin{matrix}
  \frac{1}{\sqrt{2}} \sqrt{(1-\sin \theta_1) \cos^2 \frac{\gamma_1}{2}} \\ 
  \frac{1}{2} e^{i \alpha} \tan \frac{\gamma_1}{2} \sqrt{(1-\sin \theta_1)(1+\cos \gamma_1)}  \\
  \frac{\cos \theta_1 \sin \gamma_1}{2\sqrt{(1-\sin \theta_1)(1+\cos \gamma_1)}}\\
  \frac{e^{i \alpha}\cos \theta_1 \cos^2 \frac{\gamma_1}{2}}{2\sqrt{(1-\sin \theta_1)(1+\cos \gamma_1)}}
\end{matrix}\right)$ to form an complete basis.
These vectors can be put together to get a unitary matrix, $A$, as
\begin{equation}
  A = \left(\begin{matrix}
    \ket{a_1} & \ket{a_2} & \ket{a_3} & \ket{a_4}
\end{matrix}\right).
\end{equation}
We then choose a real orthogonal matrix, $B$, as
\begin{equation}
  B = \left(\begin{matrix}
  \frac{1}{2} &\frac{1}{2} & \frac{1}{2} & \frac{1}{2}\\
  \frac{1}{2} & -\frac{1}{2}  & \frac{1}{2} & -\frac{1}{2}\\
  \frac{1}{2} & \frac{1}{2} & -\frac{1}{2} & -\frac{1}{2} \\
  \frac{1}{2} & -\frac{1}{2} & -\frac{1}{2} & \frac{1}{2}
  \end{matrix}\right).
\end{equation}
to get a unitary $U = B A^{-1}$. The optimal measurement can then be obtained as the projective measurement on the basis given by the rows of $U$. We note that the optimal measurement depends on the true value of $\alpha_1$, $\theta_1$ and $\gamma_1$. In practice we will use their estimators, $\hat{\alpha}_1$, $\hat{\theta}_1$ and $\hat{\gamma}_1$ which need to be adaptively updated. 
In our demonstrations, we conduct the optimal measurement directly to approach our theoretical bound with $\hat{\alpha}_1 = \alpha_1$, $\hat{\theta}_1 = \theta_1$ and $\hat{\gamma}_1 = \gamma_1$.
It is straightforward to verify that under this measurement, the minimal tradeoff relation is saturated.

To benchmark the performance of the optimal strategy, we introduce a non-optimal measurement $\mathcal{M} = \{\ket{w_1},\ket{w_2},\ket{w_3},\ket{w_4}\}$ as a comparison.
\begin{equation}
  \begin{aligned}
    &\ket{w_1} = \ket{+}\otimes \ket{+}\\
    &\ket{w_2} = \ket{+}\otimes \ket{-}\\
    &\ket{w_3} = \ket{-}\otimes \ket{+}\\
    &\ket{w_4} = \ket{-}\otimes \ket{-}\\
  \end{aligned}
\end{equation}
where $\ket{+}$ and  $\ket{-}$ are eigenvectors of $\sigma_x$ corresponds to the maximal and minimal eigenvalues respectively.

Fig.~\ref{fig:circuits_3para} presents the quantum circuits designed for the three-parameter estimation task. In particular, Fig.~\ref{fig:sub3} illustrates the circuit implementing the optimal measurement $\mathcal{M}^{opt}$, while Fig.~\ref{fig:sub4} shows the circuit corresponding to the non-optimal measurement $\mathcal{M}$.

The circuit implementing $\mathcal{M}^{opt}$ consists of 3 CNOT gates and 23 single-qubit rotations. In contrast, the circuit for the non-optimal measurement $\mathcal{M}$ adopts a much simpler configuration, requiring only two Hadamard gates. This comparison highlights the advantage of the optimal measurement strategy in achieving the fundamental tradeoff relation.

\begin{figure}[hpbt]
  \centering  % Center the figure
  \subfigure[]{
    \centering
    \includegraphics[scale=0.38]{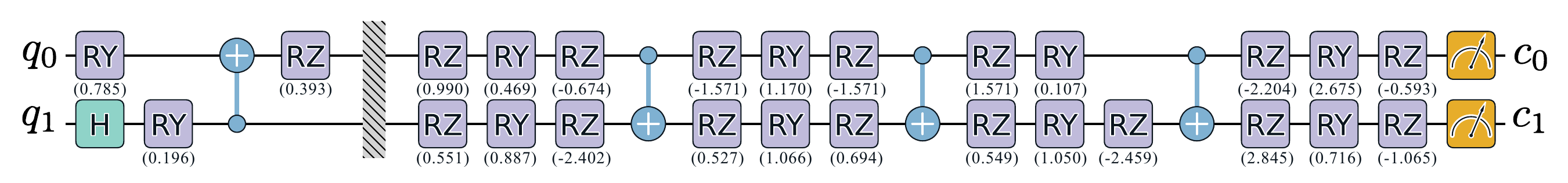}
    \label{fig:sub3}
}
  \hfill
  \subfigure[]{
    \centering
    \includegraphics[scale=0.38]{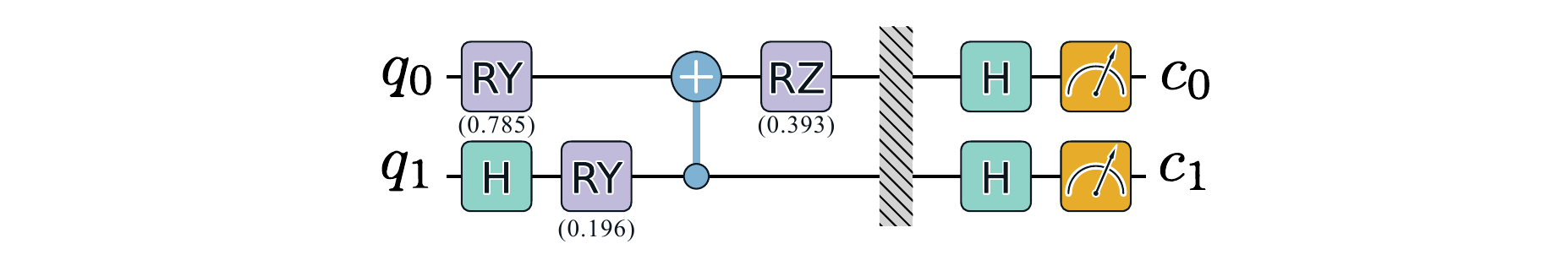}
    \label{fig:sub4}
}
  
  \caption{Quantum circuits used in the three-parameter estimation task. The circuit reads from left to right. In (a)(b), the true value of three parameters are $\alpha_1 = \frac{\pi}{8}$, $\theta_1 = \frac{\pi}{16}$, $\gamma_1 = \frac{\pi}{4}$. The circuit diagram was constructed using PYQUAFU package. (a) Quantum circuit of estimating three parameters with optimal measurement $\mathcal{M}^{opt}$ and (b) Quantum circuit of estimating three parameters with non-optimal measurement $\mathcal{M}$.}
\label{fig:circuits_3para}
\end{figure}

\subsection{Implementation process}

The demonstration begins by applying a sequence of single-qubit rotation gates with unknown parameters to qubits initially prepared in the $|0\rangle$ state. Using our proposed methodology, we design and implement optimal measurement circuits that are theoretically guaranteed to achieve the fundamental precision bound (see Figs.~\ref{fig:circuits_2para} and \ref{fig:circuits_3para} for the circuit implementations). Since the optimal measurement may depend on the true values of the parameters, in this proof-of-principle demonstration we design the circuits using estimators $\hat{x} = x$ fixed at the true parameter values. In practical scenarios where the parameters are unknown, these estimators must be adaptively updated. 

In the first demonstration, the unknown parameters are $x = (\alpha, \theta)$; in the second demonstration, the parameters are $x = (\alpha_1, \theta_1, \gamma_1)$. For each measurement setting (including $\mathcal{M}_1^{opt}$, $\mathcal{M}_2^{opt}$, $\mathcal{M}_{opt}$, and the reference measurement $\mathcal{M}$), we perform $N_{\mathrm{shots}} = 2000$ measurement shots per run. We repeat the full procedure $N = 400$ times, so that each run $l = 1,\dots,N$ produces a set of outcome counts $\{n_i^{(l)}\}$ and the corresponding empirical frequencies $p_i^{\mathrm{meas},(l)} = n_i^{(l)}/N_{\mathrm{shots}}$. From the measurement data, the estimator $\hat{x}^{(l)}$ in run $l$ is obtained by maximum-likelihood estimation. Specifically, we maximize the log-likelihood
\begin{equation}
    \hat{x}^{(l)} = \arg\max_{x} \sum_i p_i^{\mathrm{meas},(l)} \ln p_i(x),
\end{equation}
where $p_i(x)$ denotes the ideal probability of outcome $i$ for the implemented circuit and parameters $x$.

To quantify the estimation performance, we compute the covariance matrix $\mathrm{Cov}(\hat{x})$ from the $N = 400$ independent runs, each based on $N_{\mathrm{shots}} = 2000$ shots. Given the potential for high gate error rates to introduce bias into the estimator, as highlighted in~\cite{Conlon2023}, the covariance matrix is computed as
\begin{equation}
[\mathrm{Cov}(\hat{x})]_{jk} = \frac{1}{N}\sum_{l=1}^{N}(\hat{x}_j^{(l)} - \bar{x}_j) (\hat{x}_k^{(l)} - \bar{x}_k),
\end{equation}
where $\hat{x}_j^{(l)}$ denotes the $j$-th component of the estimator in the $l$-th run, and $\bar{x}_j$ is the sample mean of $\hat{x}_j$ over all $N$ runs.

\end{widetext}


\begin{thebibliography}{99}

\bibitem{Hole82book}
A.~S. Holevo, \textit{Probabilistic and Statistical Aspects of Quantum Theory}
(North-Holland, Amsterdam, 1982).

\bibitem{Hels76book}
Carl~W. Helstrom, \textit{Quantum Detection and Estimation Theory}
(Academic Press, New York, 1976).

\bibitem{Liu2022}
Jing Liu, Mao Zhang, Hongzhen Chen, Lingna Wang, and Haidong Yuan, Optimal
scheme for quantum metrology, Adv. Quantum Tech. \textbf{5}, 2100080
(2022).

\bibitem{Qiushi2023}
Q.~Liu, Z.~Hu, H.~Yuan, and Y.~Yang, Optimal strategies of quantum metrology
with a strict hierarchy, Phys. Rev. Lett. \textbf{130}, 070803 (2023).

\bibitem{Kurdzialek22}
Stanis{\l}aw Kurdzia{\l}ek, Wojciech G{\'o}recki, Francesco Albarelli, and
Rafa{\l} Demkowicz-Dobrza{\'n}ski, Using adaptiveness and causal
superpositions against noise in quantum metrology, Phys. Rev. Lett.
\textbf{131}, 090801 (2023).

\bibitem{MatsumotoThesis}
Keiji Matsumoto, A geometrical approach to quantum estimation theory,
arXiv:2111.09667.

\bibitem{HongzhenPra}
Hongzhen Chen, Yu Chen, and Haidong Yuan, Incompatibility measures in
multiparameter quantum estimation under hierarchical quantum measurements,
Phys. Rev. A \textbf{105}, 062442 (2022).

\bibitem{HongzhenPRL}
H.~Chen, Y.~Chen, and H.~Yuan, Information geometry under hierarchical quantum
measurement, Phys. Rev. Lett. \textbf{128}, 250502 (2022).

\bibitem{GillM00}
R.~D. Gill and S.~Massar, State estimation for large ensembles, Phys. Rev. A
\textbf{61}, 042312 (2000).

\bibitem{HayaM05}
Masahito Hayashi and Keiji Matsumoto, Statistical model with measurement
degree of freedom and quantum physics, in \textit{Asymptotic Theory of Quantum
Statistical Inference: Selected Papers}, edited by Masahito Hayashi
(World Scientific, Singapore, 2005). Original Japanese version was published
in \textit{Surikaiseki Kenkyusho Kokyuroku} 1055, 96--110 (1998).

\bibitem{Zhu2018universally}
Huangjun Zhu and Masahito Hayashi, Universally Fisher-symmetric informationally
complete measurements, Phys. Rev. Lett. \textbf{120}, 030404 (2018).

\bibitem{Lu2021}
Xiao-Ming Lu and Xiaoguang Wang, Incorporating Heisenberg's uncertainty
principle into quantum multiparameter estimation, Phys. Rev. Lett.
\textbf{126}, 120503 (2021).

\bibitem{Suzuki2016}
Jun Suzuki, Explicit formula for the Holevo bound for two-parameter
qubit-state estimation problem, J. Math. Phys. \textbf{57}, 042201
(2016).

\bibitem{Sidhu2021}
J.~S. Sidhu, Y.~Ouyang, E.~T. Campbell, and P.~Kok, Tight bounds on the
simultaneous estimation of incompatible parameters, Phys. Rev. X \textbf{11},
011028 (2021).

\bibitem{Nagaoka1}
H.~Nagaoka, A new approach to Cram{\'e}r--Rao bounds for quantum state
estimation, in \textit{Asymptotic Theory of Quantum Statistical Inference:
Selected Papers}, edited by Masahito Hayashi (World Scientific, Singapore,
2005). Originally published as IEICE Technical Reports No.~89, 228, IT
89--42, 9--14 (1989).

\bibitem{Nagaoka2}
H.~Nagaoka, A generalization of the simultaneous diagonalization of Hermitian
matrices and its relation to quantum estimation theory, in \textit{Asymptotic
Theory of Quantum Statistical Inference: Selected Papers}, edited by Masahito
Hayashi (World Scientific, Singapore, 2005).

\bibitem{Conlon2021}
Lorc{\'a}n~O. Conlon, Jun Suzuki, Ping~Koy Lam, and Syed~M. Assad, Efficient
computation of the Nagaoka--Hayashi bound for multiparameter estimation with
separable measurements, npj Quantum Inf. \textbf{7}, 110 (2021).

\bibitem{ALBARELLI2020126311}
F.~Albarelli, M.~Barbieri, M.~G. Genoni, and I.~Gianani, A perspective on
multiparameter quantum metrology: From theoretical tools to applications in
quantum imaging, Phys. Lett. A \textbf{384}, 126311 (2020).

\bibitem{Federico2021}
Federico Belliardo and Vittorio Giovannetti, Incompatibility in quantum
parameter estimation, New J. Phys. \textbf{23}, 063055 (2021).

\bibitem{Carollo_2019}
Angelo Carollo, Bernardo Spagnolo, Alexander~A. Dubkov, and Davide Valenti,
On quantumness in multi-parameter quantum estimation, J. Stat. Mech. (2019)
094010.

\bibitem{Ragy2016}
S.~Ragy, M.~Jarzyna, and R.~Demkowicz-Dobrza{\'n}ski, Compatibility in
multiparameter quantum metrology, Phys. Rev. A \textbf{94}, 052108 (2016).

\bibitem{Chen_2017}
Yu Chen and Haidong Yuan, Maximal quantum Fisher information matrix, New J.
Phys. \textbf{19}, 063023 (2017).

\bibitem{Liu_2019}
Jing Liu, Haidong Yuan, Xiao-Ming Lu, and Xiaoguang Wang, Quantum Fisher
information matrix and multiparameter estimation, J. Phys. A \textbf{53},
023001 (2020).

\bibitem{ChenHZ2019}
Hongzhen Chen and Haidong Yuan, Optimal joint estimation of multiple Rabi
frequencies, Phys. Rev. A \textbf{99}, 032122 (2019).

\bibitem{Rafal2020}
R.~Demkowicz-Dobrza{\'n}ski, W.~G{\'o}recki, and M.~Gu\c{t}\u{a},
Multi-parameter estimation beyond quantum Fisher information, J. Phys. A
\textbf{53}, 363001 (2020).

\bibitem{Kok2020}
Jasminder~S. Sidhu and Pieter Kok, Geometric perspective on quantum parameter
estimation, AVS Quantum Sci. \textbf{2}, 014701 (2020).

\bibitem{vidrighin2014}
M.~D. Vidrighin, G.~Donati, M.~G. Genoni, X.-M. Jin, W.~S. Kolthammer,
M.~S. Kim, A.~Datta, M.~Barbieri, and I.~A. Walmsley, Joint estimation of
phase and phase diffusion for quantum metrology, Nat. Commun. \textbf{5}, 3532
(2014).

\bibitem{crowley2014}
Philip~J.~D. Crowley, Animesh Datta, Marco Barbieri, and I.~A. Walmsley,
Tradeoff in simultaneous quantum-limited phase and loss estimation in
interferometry, Phys. Rev. A \textbf{89}, 023845 (2014).

\bibitem{Yue2014}
J.-D. Yue, Y.-R. Zhang, and H.~Fan, Quantum-enhanced metrology for multiple
phase estimation with noise, Sci. Rep. \textbf{4}, 5933 (2014).

\bibitem{Zhang2014}
Y.~R. Zhang and H.~Fan, Quantum metrological bounds for vector parameters,
Phys. Rev. A \textbf{90}, 043818 (2014).

\bibitem{Liu2017}
Jing Liu and Haidong Yuan, Control-enhanced multiparameter quantum estimation,
Phys. Rev. A \textbf{96}, 042114 (2017).

\bibitem{Roccia_2017}
Emanuele Roccia, Ilaria Gianani, Luca Mancino, Marco Sbroscia, Fabrizia Somma,
Marco~G. Genoni, and Marco Barbieri, Entangling measurements for
multiparameter estimation with two qubits, Quantum Sci. Technol. \textbf{3},
01LT01 (2017).

\bibitem{e22111197}
Sholeh Razavian, Matteo~G.~A. Paris, and Marco~G. Genoni, On the quantumness
of multiparameter estimation problems for qubit systems, Entropy \textbf{22},
1197 (2020).

\bibitem{Candeloro_2021}
Alessandro Candeloro, Matteo~G.~A. Paris, and Marco~G. Genoni, On the
properties of the asymptotic incompatibility measure in multiparameter quantum
estimation, J. Phys. A \textbf{54}, 485301 (2021).

\bibitem{Koichi2013}
Koichi Yamagata, Akio Fujiwara, and Richard~D. Gill, Quantum local asymptotic
normality based on a new quantum likelihood ratio, Ann. Stat. \textbf{41},
2197 (2013).

\bibitem{Kahn2009}
Jonas Kahn and M\u{a}d\u{a}lin Gu\c{t}\u{a}, Local asymptotic normality for
finite dimensional quantum systems, Commun. Math. Phys. \textbf{289}, 597
(2009).

\bibitem{Yuxiang2019}
Yuxiang Yang, Giulio Chiribella, and Masahito Hayashi, Attaining the ultimate
precision limit in quantum state estimation, Commun. Math. Phys. \textbf{368},
223 (2019).

\bibitem{Hou20minimal}
Z.~Hou, Z.~Zhang, G.-Y. Xiang, C.-F. Li, G.-C. Guo, H.~Chen, L.~Liu, and
H.~Yuan, Minimal tradeoff and ultimate precision limit of multiparameter
quantum magnetometry under the parallel scheme, Phys. Rev. Lett. \textbf{125},
020501 (2020).

\bibitem{HouSuper2021}
Z.~Hou, Y.~Jin, H.~Chen, J.-F. Tang, C.-J. Huang, H.~Yuan, G.-Y. Xiang,
C.-F. Li, and G.-C. Guo, ``Super-Heisenberg'' and Heisenberg scalings achieved
simultaneously in the estimation of a rotating field, Phys. Rev. Lett.
\textbf{126}, 070503 (2021).

\bibitem{Szczykulska2016}
M.~Szczykulska, T.~Baumgratz, and A.~Datta, Multi-parameter quantum metrology,
Adv. Phys. X \textbf{1}, 621 (2016).

\bibitem{FrancescoPRL}
F.~Albarelli, J.~F. Friel, and A.~Datta, Evaluating the Holevo Cram{\'e}r--Rao
bound for multiparameter quantum metrology, Phys. Rev. Lett. \textbf{123},
200503 (2019).

\bibitem{Conlon2023}
Lorc{\'a}n~O. Conlon, Tobias Vogl, Christian~D. Marciniak, Ivan Pogorelov,
Simon~K. Yung, Falk Eilenberger, Dominic~W. Berry, Fabiana~S. Santana, Rainer
Blatt, Thomas Monz, Ping~Koy Lam, and Syed~M. Assad, Approaching optimal
entangling collective measurements on quantum computing platforms, Nat. Phys.
\textbf{19}, 351 (2023).

\bibitem{hu2024control}
Zhiyao Hu, Shilin Wang, Linmu Qiao, Takuya Isogawa, Changhao Li, Yu Yang,
Guoqing Wang, Haidong Yuan, and Paola Cappellaro, Control incompatibility in
multiparameter quantum metrology, arXiv:2411.18896.

\bibitem{imai2026semiclassical}
Satoya Imai, Jing Yang, and Luca Pezz{\`e}, Semiclassical geometric tensor in
multiparameter quantum information, Phys. Rev. Lett. \textbf{136}, 150801
(2026).

\bibitem{Matsumoto_2002}
K.~Matsumoto, A new approach to the Cram{\'e}r--Rao-type bound of the
pure-state model, J. Phys. A \textbf{35}, 3111 (2002).

\bibitem{arthurs1965}
E.~Arthurs and J.~L. Kelly~Jr., On the simultaneous measurement of a pair of
conjugate observables, Bell Syst. Tech. J. \textbf{44}, 725 (1965).

\bibitem{prl}
L.~Wang, H.~Chen, and H.~Yuan, Minimal trade-off and optimal measurement for multiparameter quantum estimation, arXiv:2605.23514.

\bibitem{Cram46}
Harald Cram{\'e}r, \textit{Mathematical Methods of Statistics}
(Princeton University Press, Princeton, NJ, 1946).

\bibitem{Fish22}
R.~A. Fisher, On the mathematical foundations of theoretical statistics,
Philos. Trans. R. Soc. London A \textbf{222}, 309 (1922).

\bibitem{singularqfim1}
P.~Stoica and T.~L. Marzetta, Parameter estimation problems with singular
information matrices, IEEE Trans. Signal Process. \textbf{49}, 87 (2001).

\bibitem{singularqfim2}
Jiayu He and Matteo~G.~A. Paris, Scrambling for precision: Optimizing
multiparameter qubit estimation in the face of sloppiness and incompatibility,
J. Phys. A \textbf{58}, 325301 (2025).

\bibitem{singularqfim3}
Massimo Frigerio and Matteo~G.~A. Paris, Overcoming sloppiness for enhanced
metrology in a Mach--Zehnder interferometer, Int. J. Quantum Inf. \textbf{24},
2540001 (2025).

\bibitem{singularqfim4}
George Mihailescu, Saubhik Sarkar, Abolfazl Bayat, Steve Campbell, and
Andrew~K. Mitchell, Metrological symmetries in singular quantum multi-parameter
estimation, Quantum Sci. Technol. \textbf{11}, 015006 (2025).

\bibitem{singularqfim5}
Y.~Yang, V.~Montenegro, and A.~Bayat, Overcoming quantum metrology singularity
through sequential measurements, Phys. Rev. Lett. \textbf{135}, 010401 (2025).

\bibitem{Luca2017}
L.~Pezz{\`e}, M.~A. Ciampini, N.~Spagnolo, P.~C. Humphreys, A.~Datta,
I.~A. Walmsley, M.~Barbieri, F.~Sciarrino, and A.~Smerzi, Optimal measurements
for simultaneous quantum estimation of multiple phases, Phys. Rev. Lett.
\textbf{119}, 130504 (2017).

\bibitem{Yuan2016}
Haidong Yuan, Sequential feedback scheme outperforms the parallel scheme for
Hamiltonian parameter estimation, Phys. Rev. Lett. \textbf{117}, 160801
(2016).

\bibitem{Arthurs1988}
E.~Arthurs and M.~S. Goodman, Quantum correlations: A generalized Heisenberg
uncertainty relation, Phys. Rev. Lett. \textbf{60}, 2447 (1988).

\bibitem{Ozawa2003}
Masanao Ozawa, Universally valid reformulation of the Heisenberg uncertainty
principle on noise and disturbance in measurement, Phys. Rev. A \textbf{67},
042105 (2003).

\bibitem{OZAWA2004367}
Masanao Ozawa, Uncertainty relations for joint measurements of noncommuting
observables, Phys. Lett. A \textbf{320}, 367 (2004).

\bibitem{OZAWA2004350}
Masanao Ozawa, Uncertainty relations for noise and disturbance in generalized
quantum measurements, Ann. Phys. (Amsterdam) \textbf{311}, 350 (2004).

\bibitem{OZAWA200321}
Masanao Ozawa, Physical content of Heisenberg's uncertainty relation:
Limitation and reformulation, Phys. Lett. A \textbf{318}, 21 (2003).

\bibitem{Ozawa_2014}
Masanao Ozawa, Heisenberg's uncertainty relation: Violation and reformulation,
J. Phys. Conf. Ser. \textbf{504}, 012024 (2014).

\bibitem{Hall2004}
Michael~J.~W. Hall, Prior information: How to circumvent the standard
joint-measurement uncertainty relation, Phys. Rev. A \textbf{69}, 052113
(2004).

\bibitem{Branciard2013}
Cyril Branciard, Error-tradeoff and error-disturbance relations for
incompatible quantum measurements, Proc. Natl. Acad. Sci. U.S.A. \textbf{110},
6742 (2013).

\bibitem{Branciard2014}
Cyril Branciard, Deriving tight error-trade-off relations for approximate joint
measurements of incompatible quantum observables, Phys. Rev. A \textbf{89},
022124 (2014).

\bibitem{2014Error}
M.~Ozawa, Error-disturbance relations in mixed states, arXiv:1404.3388.

\bibitem{Lu2014}
X.-M. Lu, S.~Yu, K.~Fujikawa, and C.~H. Oh, Improved error-tradeoff and
error-disturbance relations in terms of measurement error components, Phys.
Rev. A \textbf{90}, 042113 (2014).

\bibitem{chen2024simultaneous}
Hongzhen Chen, Lingna Wang, and Haidong Yuan, Simultaneous measurement of
multiple incompatible observables and tradeoff in multiparameter quantum
estimation, npj Quantum Inf. \textbf{10}, 98 (2024).

\bibitem{bengtsson2017geometry}
Ingemar Bengtsson and Karol {\.Z}yczkowski, \textit{Geometry of Quantum States:
An Introduction to Quantum Entanglement} (Cambridge University Press, Cambridge, 2017).

\bibitem{provost1980riemannian}
J.~P. Provost and G.~Vallee, Riemannian structure on manifolds of quantum
states, Commun. Math. Phys. \textbf{76}, 289 (1980).

\bibitem{carollo2020geometry}
Angelo Carollo, Davide Valenti, and Bernardo Spagnolo, Geometry of quantum
phase transitions, Phys. Rep. \textbf{838}, 1 (2020).

\bibitem{Berry1984}
Michael~V. Berry, Quantal phase factors accompanying adiabatic changes, Proc.
R. Soc. London, Ser. A \textbf{392}, 45 (1984).

\bibitem{wang2019heisenberg}
W.~Wang, Y.~Wu, Y.~Ma, W.~Cai, L.~Hu, X.~Mu, Y.~Xu, Z.-J. Chen, H.~Wang,
Y.~P. Song, \textit{et al.}, Heisenberg-limited single-mode quantum metrology
in a superconducting circuit, Nat. Commun. \textbf{10}, 4382 (2019).

\bibitem{wang2022quantum}
W.~Wang, Z.-J. Chen, X.~Liu, W.~Cai, Y.~Ma, X.~Mu, X.~Pan, Z.~Hua, L.~Hu,
Y.~Xu, \textit{et al.}, Quantum-enhanced radiometry via approximate quantum
error correction, Nat. Commun. \textbf{13}, 3214 (2022).

\bibitem{ni2025autonomous}
Zhongchu Ni, Ling Hu, Yanyan Cai, Libo Zhang, Jiasheng Mai, Xiaowei Deng, Pan
Zheng, Song Liu, Shi-Biao Zheng, Yuan Xu, \textit{et al.}, Autonomous quantum
error correction beyond break-even and its metrological application,
arXiv:2509.26042.

\bibitem{zhuang2017}
Q.~Zhuang, Z.~Zhang, and J.~H. Shapiro, Entanglement-enhanced lidars for
simultaneous range and velocity measurements, Phys. Rev. A \textbf{96},
040304(R) (2017).

\bibitem{zhuang2021}
Quntao Zhuang, Quantum ranging with Gaussian entanglement, Phys. Rev. Lett.
\textbf{126}, 240501 (2021).

\bibitem{zhuang2022}
Q.~Zhuang and J.~H. Shapiro, Ultimate accuracy limit of quantum
pulse-compression ranging, Phys. Rev. Lett. \textbf{128}, 010501 (2022).

\bibitem{huang2021}
Zixin Huang, Cosmo Lupo, and Pieter Kok, Quantum-limited estimation of range
and velocity, PRX Quantum \textbf{2}, 030303 (2021).

\bibitem{quantumradar2020}
Ricardo Gallego Torrom{\'e}, Nadya Ben Bekhti-Winkel, and Peter Knott,
Introduction to quantum radar, arXiv:2006.14238.

\bibitem{Maccone2020}
Lorenzo Maccone and Changliang Ren, Quantum radar, Phys. Rev. Lett.
\textbf{124}, 200503 (2020).

\bibitem{reichert2022quantum}
Maximilian Reichert, Roberto Di~Candia, Moe~Z. Win, and Mikel Sanz,
Quantum-enhanced Doppler lidar, npj Quantum Inf. \textbf{8}, 147 (2022).

\bibitem{reichert2024heisenberg}
Maximilian Reichert, Quntao Zhuang, and Mikel Sanz, Heisenberg-limited quantum
lidar for joint range and velocity estimation, Phys. Rev. Lett. \textbf{133},
130801 (2024).

\bibitem{li2023entanglement}
Yongqiang Li and Changliang Ren, Entanglement-enhanced quantum strategies for
accurate estimation of multibody-group motion and moving-object
characteristics, Phys. Rev. A \textbf{108}, 062605 (2023).

\bibitem{Quafu1}
Yu-Xin Jin, Hong-Ze Xu, Zheng-An Wang, Wei-Feng Zhuang, Kai-Xuan Huang,
\textit{et al.}, Quafu-RL: The cloud quantum computers based quantum
reinforcement learning, Chin. Phys. B \textbf{33}, 050301 (2024).

\bibitem{Quafu2}
Hong-Ze Xu, Wei-Feng Zhuang, Zheng-An Wang, Kai-Xuan Huang, Yun-Hao Shi,
Wei-Guo Ma, \textit{et al.}, Quafu-Qcover: Explore combinatorial optimization
problems on cloud-based quantum computers, Chin. Phys. B \textbf{33}, 050302
(2024).

\bibitem{Quafu3}
Baqis quafu group, \url{https://quafu.baqis.ac.cn}.

\bibitem{qiskit}
Two Qubit Basis Decomposer,
\url{https://docs.quantum.ibm.com/api/qiskit/qiskit.synthesis.TwoQubitBasisDecomposer}.

\end{thebibliography}
 \end{document}